\documentclass[fleqn,usenatbib, onecolumn]{mnras}

\usepackage{newtxtext,newtxmath}
\usepackage{multirow}
\usepackage{graphicx}
\usepackage{amsmath}

\usepackage{epstopdf}
\usepackage{ulem}
\usepackage[lofdepth,lotdepth]{subfig} 
\usepackage{pdflscape} 

\usepackage[T1]{fontenc}

\DeclareRobustCommand{\VAN}[3]{#2}
\let\VANthebibliography\thebibliography
\def\thebibliography{\DeclareRobustCommand{\VAN}[3]{##3}\VANthebibliography}

\usepackage{graphicx}	
\usepackage{amsmath}	

\usepackage{orcidlink}

\def\apjl{ApJ}
\def\apjs{ApJS}
\def\ao{Appl.~Opt.}
\def\aap{A\&A}

\newcommand{\be}{\begin{equation}}
\newcommand{\ee}{\end{equation}}
\newcommand{\bary}{\begin{eqnarray}}
\newcommand{\eary}{\end{eqnarray}}

\title[Variation of microphysical parameter in GRBs]{Variation of Microphysical Parameters in Reverse-shock Scenario}

\author[Fraija, N. et al.]{
Fraija, N.\orcidlink{0000-0002-0173-6453},$^{1}$\thanks{E-mail: nifraija@astro.unam.mx}
B. Betancourt Kamenetskaia,$^{2}$
Antonio Galv\'an\orcidlink{0000-0001-5193-3693}$^{3}$
and Maria G. Dainotti$^{4,5}$
\\
$^{1}$Instituto de Astronom\' ia, Universidad Nacional Aut\'onoma de M\'exico,\\ Circuito Exterior, C.U., A. Postal 70-264, 04510 M\'exico City, M\'exico\\
$^{2}$Cosmology, Gravity, and Astroparticle Physics Group, Center for Theoretical Physics of the Universe,
Institute for Basic Science (IBS), Daejeon, 34126, Korea\\
$^{3}$Instituto de F\'isica, Universidad Nacional Aut\'onoma de M\'exico,\\ Circuito Exterior, C.U., A. Postal 70-264, 04510 M\'exico City, M\'exico\\
$^{4}$Division of Science, National Astronomical Observatory of Japan, 2-21-1 Osawa, Mitaka, Tokyo 181-8588, Japan \\
$^{5}$The Graduate University for Advanced Studies (SOKENDAI), Shonankokusaimura, Hayama, Miura District, Kanagawa 240-0115, Japan
}

\date{Accepted XXX. Received YYY; in original form ZZZ}

\pubyear{\the\year{}}

\begin{document}
\label{firstpage}
\pagerange{\pageref{firstpage}--\pageref{lastpage}}
\maketitle

\begin{abstract}
Gamma-ray bursts (GRBs), among the most compelling astrophysical phenomena, are potential candidates for exploring the evolution of energy distribution among magnetic fields and particles through multiwavelength observations. The fraction of energy transferred between particles and the magnetic field is governed by microphysical parameters, typically assumed to be constant during relativistic shocks but may in fact vary with time.   In this work, we derive the light curves and closure relations (CRs) of the synchrotron-self Compton (SSC) process from the external reverse shock (RS) with variations of microphysical parameters in a homogeneous and stellar-wind medium. We consider the evolution of the RS in the thick- and thin-shell regimes. We demonstrate that, depending on the microphysical parameters, this process can mimic plateau phases and produce temporal decay indices steeper than those predicted by high-latitude emission alone. The current model is employed to examine the evolution of the spectral and temporal indices of GRBs reported in the Second Fermi-LAT Gamma-ray Burst Catalog (2FLGC) and bursts detected at very high energies, using Markov Chain Monte Carlo (MCMC) simulations.
\end{abstract}


\section{Introduction}

Gamma-ray bursts (GRBs) are among the most energetic phenomena in the universe, characterized by non-repeating gamma-ray pulses, result from the core collapse of massive stars  \citep{1993ApJ...405..273W,1998ApJ...494L..45P, Woosley2006ARA&A} or the merger of binary compact objects (BCO), such as a black hole (BH) - neutron star \citep[NS;][]{1992ApJ...395L..83N}, a BH - BH, or an NS - NS \citep{1992ApJ...392L...9D, 1992Natur.357..472U, 1994MNRAS.270..480T, 2011MNRAS.413.2031M}.  A fast-spinning, newly born, highly magnetized NS (a millisecond magnetar) or a hyperaccreting BH may form as remnants in these scenarios.    The duration of the initial gamma-ray episode ($T_{90}$)\footnote{This duration corresponds to the period during which a burst emits between $5\%$ and $95\%$ of the total observed counts from its prompt emission.} is used to classify GRBs as short (sGRBs), typically associated with BCO mergers, or long (lGRBs), linked to the core collapse of massive stars.  Regardless of the GRB type,  the afterglow, a late-time emission that can include both short- and long-lived components, has been observed across the electromagnetic spectrum, from radio to TeV gamma rays. Long- and short-lived episodes are typically interpreted in the context of synchrotron and synchrotron self-Compton (SSC) mechanisms \citep[e.g., see][]{2001ApJ...559..110Z, 2019ApJ...883..162F}, which are the result of the acceleration of nonthermal electrons in the forward shock \citep[FS;][]{1995ApJ...455L.143S, 1998ApJ...497L..17S, 1999ApJ...513..669K} and reverse shock \citep[RS;][]{2000ApJ...545..807K, 2003ApJ...597..455K, 2016ApJ...818..190F}, respectively.  These shocks result from the outflow encountering the circumburst medium and transferring a portion of its energy to it \citep{1999ApJ...513..669K, 2002ApJ...568..820G, 1998ApJ...497L..17S}.  The energy in these shocks is distributed between the acceleration of particles (electrons and protons) and the amplification of the magnetic field via the microphysical parameters $\varepsilon_{\rm e_j}$ and $\varepsilon_{\rm B_j}$, respectively, with ${\rm j=f}$ for a FS and ${\rm j=r}$ for a RS.\\

Considering a limited understanding of energy transfer from protons to electrons and magnetic fields in relativistic shocks, it is reasonable to infer that the values of the microphysical parameters could be subject to variation.  Variations in these parameters have also been invoked to model multiwavelength afterglow observations \citep[e.g., see][]{2003ApJ...597..459Y, 2003MNRAS.346..905K, 2006A&A...458....7I,2006MNRAS.369..197F, 2006MNRAS.369.2059P, 2005PThPh.114.1317I, 2006MNRAS.370.1946G, 2020ApJ...905..112F}.  For instance, \cite{2005PThPh.114.1317I} suggested that the percentage of energy available to accelerate electrons may change during the early afterglow while the synchrotron radiation was in a fast cooling phase.    \cite{2006MNRAS.370.1946G} proposed that the shallow decay (plateau) phase could be interpreted as resulting from varying microphysical parameters rather than energy injection immediately following the prompt episode. \cite{10.1111/j.1365-2966.2009.15886.x} investigated the evolution of microphysical parameters in a wind bubble environment to replicate the brightenings observed in some GRB afterglows.   In order to explain both the rapid evolution of the synchrotron cooling break and the X-ray plateau exhibited in one of the most powerful long bursts, GRB 160509A, \cite{2020ApJ...905..112F} examined the variation of microphysical parameters. {
The evolution of microphysical parameters is not the only viable framework for explaining features like the X-ray plateau. An alternative explanation involves emission from a structured jet viewed slightly off-axis~\citep{2020MNRAS.492.2847B}. In this model, the shallow plateau arises naturally from the forward shock emission for observers whose line of sight is just outside the jet's energetic core. As the jet decelerates, relativistic beaming decreases, gradually bringing the bright core into the observer's view and producing a plateau phase. Another alternative proposes that the plateau is a signal of the prolonged coasting phase of a jet with a small initial Lorentz factor expanding into a low-density wind environment~\citep{2022NatCo..13.5611D}. In this scenario, the forward shock emission remains nearly constant until the jet begins to decelerate, completely within the framework of the classical fireball model. It was also recently demonstrated in a detailed analysis of GRB 060729 that such behavior is also possible in a constant-density interstellar medium~\citep{2025ApJ...984L..65G}. In this latter scenario, the plateau and subsequent rebrightening are interpreted as the delayed onset of the afterglow from a slow, wide jet that results from early-time interactions between multiple jet components.
}

\cite{Ajello_2019} released the Second Fermi-LAT Gamma-ray Burst Catalog (2FLGC), which corresponded to the bursts detected by the Large Area Telescope (LAT) instrument  onboard the {\it Fermi} satellite from 2008 to 2018. A total of 186 bursts have been identified; 91 exhibit emission within the 30 – 100 MeV energy range (17 of which are exclusively observed in this band), and 169 are discovered above 100 MeV.   This catalog exhibited 86 bursts with much larger emissions than the prompt episode. It described the spectral and temporal features within the 0.1 – 100 GeV energy spectrum, explicitly focusing on the highest-energy photons.  The spectral ($\beta_{\rm L}$) and temporal ($\alpha_{\rm L}$) indices are usually related to the indices of each segment that describe the synchrotron afterglow model through the closure relations (CRs; $F^{\rm syn}_{\rm \nu, f}\propto t^{-\alpha}\nu^{-\beta}$).  The cooling regime, the profile of the surrounding environment, the electron spectral index, and the jet geometry are some of the physical aspects that determine the shape of the CRs.   Several authors have explored the CRs of synchrotron and SSC mechanisms from the FS to explain the spectral and temporal indices of LAT-detected bursts \citep{2019ApJ...883..134T, 2021ApJS..255...13D, 2022ApJ...934..188F, Dainotti2023Galax..11...25D, 2023ApJ...958..126F, 2024MNRAS.527.1884F, 2024MNRAS.527.1674F, 2025ApJ...978...51D, 2024MNRAS.534.3783F}.

\cite{2019ApJ...883..134T} analyzed 59 bursts from 2FLGC, testing the CRs of the synchrotron FS model that evolves in stellar wind environments and interstellar medium (ISM). It was shown that a set of CRs can classify 81\% of the sample. Only one-third of these bursts allow stellar-wind environments and ISM, while the other two-thirds are more partial to ISM. The authors demonstrate that, while the spectral and temporal LAT indices might be satisfied by the CRs of the synchrotron FS model in most situations, a significant fraction of bursts could not be described with these CRs.   They claimed that GRBs that fail to conform to any CR typically exhibit a more gradual temporal decay ($\alpha_L < 1$). This suggested the need for ongoing energy input or alternative physical sources to maintain the observed fluxes.   Finally, they reported that a considerable number of GRBs necessitate an exceedingly minor proportion of the overall energy density present within the magnetic field via the magnetic microphysical parameter ($\varepsilon_B\lesssim  10^{-7}$).   \cite{2021ApJS..255...13D} examined the CRs of the synchrotron model with three specific bursts; GRB 090510A, 090902B, and 160509A. This study revealed that this mechanism lies in a slow-cooling regime, rather than a fast-cooling one, and regardless of whether this afterglow model evolves in a constant-density or a stellar wind medium.   Furthermore, \cite{2023MNRAS.525.1630F} examined the CRs of the SSC afterglow model within the intermediate density profile ($\propto r^{-k}$) with $0 \leq k \leq 2.5$, including the adiabatic/radiative regime and the presence or absence of energy injection. They found that this mechanism evolving in an environment with an intermediate density between stellar wind or constant medium successfully explains a significant number of bursts.  \cite{2023ApJ...958..126F} derived the CRs of the synchrotron and SSC off-axis model with and without energy injection inside a homogeneous and stratified environment. The findings indicate that the most probable scenario for synchrotron emission aligns with the stellar wind, regardless of energy injection. Conversely, the most probable scenario for SSC emission corresponds to a constant density in the absence of energy injection and to the stellar wind when energy injection occurs. Finally, \cite{2024MNRAS.527.1884F} estimated the CRs of the synchrotron and SSC afterglow model evolving in a stratified medium, taking into account the variations of microphysical parameters.  The authors found that the constant-density afterglow scenario was preferred over the stellar wind.\\   

In this manuscript, we derive the SSC  mechanism emitted from the RS region with the evolution of microphysical parameters in the homogeneous and stellar wind medium. Two sets of electron populations characterize the SSC mechanism with a spectral index in the ranges $1<p<2$ and $2 < p$, and RS is analyzed in the thick and thin-shell regime.  We analyze the spectral and temporal indices of the GRBs reported in 2FLGC and those bursts detected by LAT but also with TeV photons.   This manuscript is organized as follows. In Section \S\ref{sec2}, we summarize previous models with variation of microphysical parameters.   In Section \S\ref{sec3}, we derive the SSC mechanism with variation of microphysical parameters from the RS region in the homogeneous and stellar wind medium. In Section \S\ref{sec4}, we introduce \textit{Fermi}-LAT data and the methodical approach with a discussion of the maximum energies, density profile and plateau phase. In Section \S\ref{sec5},  we present a brief summary.

\section{Previous afterglow models with variation of microphysical parameters}\label{sec2}

Variations in the microphysical parameters are required to elucidate microturbulence and the amplification of magnetic fields linked to nonlinear acceleration processes \citep{2003MNRAS.339..881R, 2005PThPh.114.1317I, 2013arXiv1305.3689L,2013MNRAS.428..845L}. These variations also aid in interpreting multiwavelength afterglow observations, which exhibit atypical evolutions in synchrotron and SSC spectra~\citep{2003ApJ...597..459Y, 2006MNRAS.369.2059P}, as well as light curves characterized by features such as a plateau phase \citep{2006A&A...458....7I}, non-standard evolution in spectral breaks \citep{2020ApJ...905..112F}, and bright flashes \citep{2024MNRAS.527.1884F}, among others. 

\cite{2006A&A...458....7I} investigated the temporal evolution of the parameters to elucidate the plateau phase observed in numerous X-ray light curves. They determined that this behavior could be explained if $\epsilon_{\rm e,f}\propto t^{-\alpha_a}$, with $\alpha_a=1/2$. The microphysical parameters were found to be different when \cite{1998ApJ...497..288W} described the radio afterglow observations of GRB 970508 at $\sim$ 1 year, and \cite{1999ApJ...523..177W} at 12 days and \cite{1998ApJ...497..288W}.   \cite{1999ApJ...523..177W} reported values of $\varepsilon_{\rm e,f}=0.12$ and $\varepsilon_{\rm B,f}=0.089$, whereas \cite{1998ApJ...497..288W} found values of $\varepsilon_{\rm e,f}\simeq\varepsilon_{\rm B,f}=0.5$.   \cite{2018ApJ...859..163H} analyzed the early afterglow of GRB 120729A by examining a change of the magnetic parameter represented by a broken power law (BPL) of type $\epsilon_{\rm B,f}\propto t^{-\rm \alpha_b}$, where $\mathrm{\alpha_b}_{1}=-(0.18\pm0.04)$ and $\mathrm{\alpha_b}_{2}=-(0.84\pm0.04)$. \cite{2003ApJ...597..459Y} conducted an analysis of four extensively characterized bursts: GRB 970508, 980329, 980703, and 000926. The authors initially fitted the findings using a model with constant microphysical parameters and observed variability in the values of $\varepsilon_{\rm B,f}$, namely that $0.002<\varepsilon_{\rm B, f}<0.25$.   Subsequently, they suggested the relationship $\epsilon_{\rm B, f}\propto\Gamma^{\alpha_\Gamma}$ with $-2\leq \alpha_\Gamma\leq1$, resulting in a discrepancy of up to an order of magnitude from the parameters of the preceding model. However, the authors observed that the results were not distinctive and therefore no judgments regarding the values could be drawn.  \cite{10.1111/j.1365-2966.2009.15886.x} similarly examined GRB 060206, 070311, and 071010A with a two-region model in which the microphysical parameters evolved as $\epsilon_{\rm j}=\varepsilon_{\rm j}\Gamma^{-\alpha_{\rm i}}$, with ${\rm j\in[\rm e_f,B_f]}$ and ${\rm i\in[1,2]}$ for both regions.  Their model accurately reproduced the observed R-band and X-ray light curves, determining that $\varepsilon_{\rm e_f,0}=0.3$ and $\varepsilon_{\rm B_f}=0.03$ were adequate to fit all GRBs, but varying values of $\alpha_{\rm i}$ were necessary.   In order to fit the X-ray and optical afterglow observations of six bursts,  \cite{2006MNRAS.369.2059P} required the evolution of microphysical parameters such as $\epsilon_{\rm B,f} \propto \Gamma^{-\alpha_b}$ and $\epsilon_{\rm e, f}\propto \Gamma^{-\alpha_e}$. These equipartition parameters displayed a steepening at $\sim$ 1-4 hours after the prompt episode without exhibiting a break in the optical flux.  The power law (PL) indices $\alpha_b$ and $\alpha_e$ were found to be within the ranges of $4.2\leq \alpha_b\leq 7.6$ and $-4.1\leq\alpha_e\leq1.7$, respectively.    \cite{2006MNRAS.369..197F} explored an afterglow model with the variation of microphysical parameters $\epsilon_{\rm e,f}\propto\Gamma^{-\alpha_a}$ and $\epsilon_{\rm B,f}\propto\Gamma^{-\alpha_b}$.    The authors discovered that the X-ray afterglow observations of GRB 050315, GRB 050319, and GRB 050401 could be satisfactorily determined with $0.6\leq \alpha_a \leq 0.7$ and $0.45\leq \alpha_b \leq 1.2$, respectively. This conclusion suggested that the microphysical parameters increased with time. \cite{2006MNRAS.370.1946G} examined an afterglow model in which $\epsilon_{\rm e,f}\propto t^{\alpha_{\rm a}}$ and $\epsilon_{\rm B,f}\propto t^{\alpha_{\rm b}}$ were used to elucidate an observed stage of flattish decay in the X-ray light curves. The authors discovered that this observation could not be replicated without the relationship between the PL indices $\alpha_{\rm e}+\alpha_{\rm B}\sim1-2$. Consequently, they considered linear growth in time of the microphysical parameters as a potential explanation.

The formation of relativistic collisionless shock fronts in a weakly magnetized environment is elucidated through the self-generation of intense small-scale electromagnetic fields that facilitate the transition from the upstream unshocked region to the downstream shocked region. This scenario has been examined by several authors \citep[e.g., see][]{2003MNRAS.339..881R, 2005PThPh.114.1317I, 2013MNRAS.428..845L}, demonstrating how the spectrum undergoes modification when a short magnetic length scale is taken into account.  \cite{2013MNRAS.428..845L} examined the variation of microphysical parameter $\epsilon_{\rm B,f}\propto t^{-\alpha_b}$ and analyzed the influence of the PL index $\alpha_b$ on the synchrotron and SSC spectrum.  \cite{2013MNRAS.435.3009L} modeled some LAT-detected bursts (GRB 090902B, 090323, 090328 and 110731A) using a synchrotron FS model with microphysical variations ($\epsilon_{\rm B,f}\propto t^{-\alpha_b}$). The value of the parameter $\alpha_b$ found in the range of [0.4, 0.5] was related to magnetic field amplification due to microturbulences.  \cite{2020ApJ...905..112F} described the early afterglow observations in GRB 160509A, which exhibited two energetic photons above 10 GeV. The authors demonstrated that the plateau phase exhibited in the X-ray light curve and the atypical synchrotron break could be interpreted in the scenario of variation of microphysical parameters $\epsilon_{\rm e,f}\propto t^{-\rm \alpha_a}$ and $\epsilon_{\rm B,f} \propto t^{-\alpha_b}$ with $\alpha_a \approx 0.3$ and $\alpha_b\approx 1.4$.  \cite{2021MNRAS.504.5685M} analyzed one of the most energetic bursts GRB 190114C in the synchrotron and SSC afterglow scenario in a constant density medium with variation of microphysical parameters. The authors found that for ISM the magnetic parameter lies in the range of $-0.2 < \alpha_b < 0.5$, whereas for the wind environment, it lies in the range $-0.8 <\alpha_b < -0.6$.    \cite{2022ApJ...931L..19S} incorporated a RS component into their FS-plus-RS model to describe the multiwavelength afterglow of GRB 190829A. By including this component, they were able to explain key features of the observed emission, such as the early-time peaks in the X-ray and optical bands. To achieve a self-consistent fit, the authors modeled the RS dynamics by allowing for a rapid decay of the magnetic field in the shocked ejecta after the RS crossed the shell. They found that if the magnetic energy density remained constant, the RS emission would overpredict the late-time radio observations. Recently, \cite{2024MNRAS.527.1884F} derived the CRs of the synchrotron and SSC FS scenario with the variations of microphysical parameters and applied this afterglow model to test the evolution of the spectral and temporal indices of the bursts reported in 2FLGC. They concluded that the constant-density afterglow scenario was favored in comparison to the stellar wind.

\section{Variation of Microphysical parameters in SSC RS}\label{sec3}

{ In standard afterglow models, the microphysical parameters are typically assumed to remain constant during the fireball’s expansion. Nonetheless, the precise microphysical mechanisms occurring in relativistic shocks, such as the energy transfer from protons to electrons and magnetic fields, remain inadequately understood. The microphysical parameters may be variable. Some authors have claimed that the early afterglow would require rapid changes with time of $\epsilon_{\rm e}$ and $\epsilon_{\rm B}$, then these are expected that the temporal evolution terminates at the end of the early afterglow phase (see Section \ref{sec2}).  On the other hand, GRBs occur when the kinetic energy from a relativistic outflow is dissipated and transformed into radiation. The efficiency of this approach is an essential component in any GRB scenario. In the energy injection scenario,  the evolution of microphysical parameters with the blast wave Lorentz factor is considered to be $E(>\Gamma)\propto t^{-e_x}$, $\epsilon_{\rm e_r} \propto\Gamma^{-a}$, and $\epsilon_{\rm B_r}\propto\Gamma^{-b}$, with $e_x$ the parameter for the energy injection \citep[e.g., see][]{2006MNRAS.369..197F, 2006MNRAS.369.2059P}. Although this model is very attractive because these parameters would depend on the properties of the collisionless shock, such as the bulk Lorentz factor of the shock and the particle number density of the environment it
encounters, in some cases it implies very high values of efficiency \citep[e.g., see][]{2006A&A...458....7I, 2006MNRAS.370.1946G}.  This efficiency crisis can be mitigated by considering coherent effects, which would depend on the Lorentz factor of the shock \citep[e.g., see][]{2005PThPh.114.1317I, 2013arXiv1305.3689L,2013MNRAS.428..845L, 2006A&A...458....7I}, and then parameterizing the microphysical parameters as $\epsilon_{\rm e, r}= \varepsilon_{\rm e, r} \left(t/t_0\right)^{-\rm a}$ and $\epsilon_{\rm B, r}=\varepsilon_{\rm B, r} \left(t/t_0\right)^{-\rm b}$ for $t_0 <t< t_{\rm f}$\footnote{The onset and end of the variation of the microphysical parameters are represented by the times $t_0$ and $t_{\rm f}$, respectively.} and $\epsilon_{\rm e, r}= \varepsilon_{\rm e, r}$ ($\epsilon_{\rm B, r}= \varepsilon_{\rm B, r}$), otherwise.  It should be noted that the parameters do not evolve indefinitely but saturate up to a given time ($t_{ f}$) which is presumably determined by the coherence length. For example, the scenario of microturbulence strength and coherence length, where the magnetic parameter evolves with time until a saturation length associated with the compression of the magnetic field \cite{2013arXiv1305.3689L,2013MNRAS.428..845L}. Here, we consider this parameterization of the microphysical parameters. }\\
%
In RS, electrons represented by a PL distribution ($\propto \gamma_e^{-p}\,d\gamma_e$) with the spectral index $p$ are accelerated and mostly cooled by the synchrotron and SSC mechanism. The profile of the synchrotron and SSC light curves is formed by PLs, which are affected by the following factors: the evolution of microphysical parameters $\epsilon_{\rm e, r}= \varepsilon_{\rm e, r} \left(t/t_0\right)^{-\rm a}$ and $\epsilon_{\rm B, r}=\varepsilon_{\rm B, r} \left(t/t_0\right)^{-\rm b}$,\footnote{We will use $t_0=10$ and $10^2$ s, for thick and thin-shell regime, respectively.} respectively, the critical Lorentz factor ($\Gamma_c$), the shock crossing time ($t_{\rm x}$), the burst duration ($T_{90}$), the hardness of the spectral index of the electron distribution ($1< p<2$ and $2 < p$), and the density of the surrounding medium.

\subsection{Evolution in homogeneous environment}

Evolution of the critical Lorentz factor in the surrounding constant density (n) is $\Gamma_c\propto(1+z)^{\frac38} E^{\frac18} n^{-\frac18} T_{90}^{-\frac38}$,\footnote{We will use the convention $Q_{\rm x}=Q/10^{\rm x}$ in c.g.s. units, unless otherwise indicated.} where $z$ corresponds to the redshift and  $E$ the isotropic equivalent kinetic energy, which is 
associated to the isotropic gamma-ray energy ($E_{\rm \gamma}$) and the kinetic efficiency through $\eta=E_{\rm \gamma}/(E+E_{\rm \gamma})$. 
Based on the values of the critical Lorentz factor, the shock crossing time, and the duration of the burst, RS would take place in the thick ($t_{\rm x} \lesssim T_{90}$) or in the thin ($T_{90} < t_{\rm x}$) regime.

\subsubsection{The Thick-shell regime}

 RS in the thick-shell scenario is ultrarelativistic and can significantly decelerate the shell ($\Delta$). In this scenario ($t_{\rm x} \lesssim T_{90}$), after the RS crosses the shell, the magnetic field in the comoving frame varies as $B'\propto t^{-\frac{13+12b}{24}}$,\footnote{Quantities will be designated  as "unprimed" and "primed" when they are measured in the observer and comoving frames, respectively.} and the minimum and cooling breaks of Lorentz factors for an electron distribution with spectral index $1<p<2$ ($2<p$) as $\gamma_{\rm m,r}\propto t^{-\frac{68-21p+96a+24b(2-p)}{96(p-1)}}\,(t^{-\frac{13+48a}{48}})$ and $\gamma_{\rm c,r}\propto t^{\frac{25+48b}{48}}$, respectively. The corresponding spectral breaks and maximum flux of synchrotron radiation evolve as $\nu^{\rm syn}_{\rm m, r}\propto t^{-\frac{21+26p+24(4a+b)}{48(p-1)}}\,(t^{-\frac{73+24(4a+b)}{48}})$ and $\nu^{\rm syn}_{\rm c, r}\propto t^{\frac{1+24b}{16}}$ and $F^{\rm syn}_{\rm max,r}\propto t^{-\frac{47+24b}{48}}$, respectively. { Appendix A provides explicit calculations for the electron Lorentz factors, spectral break, and maximum flux of synchrotron emission.}

The SSC mechanism occurs when the same electron population responsible for synchrotron radiation upscatters these photons to higher energies ($h\nu^{\rm ssc}_{\rm k, r}\simeq\gamma^2_{\rm k, r} h\nu^{\rm syn}_{\rm k, r}$ with $h$ the Planck constant, and ${\rm k=m,\,c}$).  Given the evolution of the synchrotron spectrum together with the electron distribution,  the SSC spectral breaks and the maximum SSC flux are given by

{\small
\begin{eqnarray}\label{SSC_breaks_thick_aft_k0}
h \nu^{\rm ssc}_{\rm m, r} &=&
\begin{cases}
2.1\times 10^{-8}\,{\rm GeV}  \, \left(\frac{1+z}{1.3} \right)^{\frac{137-43p}{48(p-1)}}\,\tilde{g}^{\frac{4}{p-1}}\,\chi_{\rm e,-0.3}^{-4} \varepsilon^{\frac{4}{p-1}}_{\rm e_r,-1} \varepsilon^{\frac{3-p}{2(p-1)}}_{\rm B_r,-3}n^{\frac{7-2p}{4(p-1)}}_{-1}\,\Delta_{11.9}^{\frac{5(25+p)}{48(p-1)}}\,\Gamma^{\frac{4}{p-1}}_{2.7}\,E^{-\frac{1}{4(p-1)}}_{53.5}\cr
\hspace{8.5cm} t^{-\frac{89+5p+24[8a+b(3-p)]}{48(p-1)}}_{1.5}\hspace{2.2cm}{\rm for} \hspace{0.1cm} { 1<p<2 }\cr
6.4\times 10^{-5}\,{\rm GeV}  \,  \left(\frac{1+z}{1.3} \right)^{\frac{17}{16}}\,g^4\,\chi_{\rm e,-0.3}^{-4}\, \,\varepsilon^4_{\rm e_r,-1} \varepsilon^{\frac12}_{\rm B_r,-3}  n^{\frac{3}{4}}_{-1}\,\Gamma^4_{2.7}\Delta_{11.9}^{\frac{45}{16}}\,E^{-\frac{1}{4}}_{53.5}t^{-\frac{33+8(8a+b)}{16}}_{1.5}\hspace{3.8cm}{\rm for} \hspace{0.1cm} {2<p }
\end{cases}
\cr
h \nu^{\rm ssc}_{\rm cut, r}&\simeq& 2.6\times 10\,{\rm GeV}  \,  \left(\frac{1+z}{1.3} \right)^{\frac{17+56b}{16}}\,\varepsilon^{-\frac72}_{\rm B_r,-3} n^{-\frac{9}{4}}_{-1}\,   \left(1+Y_{\rm r}\right)^{-4}\,  \Delta^{\frac{29+56b}{16}}_{11.9}\,E^{-\frac{5}{4}}_{53.5}\,t^{-\frac{33}{16}}_{1.5},\cr
F^{\rm ssc}_{\rm max, r}&\simeq& 2.3\times 10^{-2}\,{\rm mJy}\,\left(\frac{1+z}{1.3} \right)^{\frac{89}{48}}\,g^{-1}\,\chi_{e,-0.3}\,   \varepsilon^{\frac12}_{\rm B_r,-3}\,  \Gamma^{-1}_{2.7}\,n_{-1}\,  \Delta_{11.9}^{\frac{17}{48}} \,d^{-2}_{\rm z,27.7} \,E^{\frac{3}{2}}_{53.5}\,t^{-\frac{41+24b}{48}}_{1.5}. 
\eary
}

For this regime, the shock crossing time and the shell are related by $t_{\rm x}=33\,{\rm s}\,\left(\frac{1+z}{2}\right)\Delta_{12}$.  The parameter $\chi_{\rm e}$ represents the fraction of electrons accelerated in shock~\citep{2006MNRAS.369..197F}, $Y_{\rm r}$, is the Compton parameter,  $\tilde{g}=\frac{2-p}{p-1}$ for $1<p<2$ and $g=\frac{p-2}{p-1}$ for $2<p$. In the following, we use $p=1.95$ for $1<p<2$ and $p=2.25$ for $2<p$, $a=-0.5$ and $b=-0.5$.  For the luminosity distance {\small $d_{\rm z}=(1+z)\frac{c}{H_0}\int^z_0\,\frac{d\tilde{z}}{\sqrt{\Omega_{\rm M}(1+\tilde{z})^3+\Omega_\Lambda}}$}  \citep{1972gcpa.book.....W}, we consider a spatially flat universe $\Lambda$CDM model with the following cosmological parameters: $H_0=69.6\,{\rm km\,s^{-1}\,Mpc^{-1}}$, $\Omega_{\rm M}=0.286$, and $\Omega_\Lambda=0.714$ \citep{2016A&A...594A..13P}.    The new cooling break $\nu^{\rm ssc}_{\rm cut, r}=\nu^{\rm ssc}_{\rm c, r}(t_{\rm x})\,\left(\frac{t}{t_{\rm x}} \right)^{-\frac{33}{16}}$ is derived considering that the fluid expands adiabatically.\\
During this interval, the SSC light curve might have a steeper temporal decay index, because the electrons cannot be reaccelerated inside the RS region, and therefore the emission from this region ceases. Nevertheless, abrupt disappearance is mitigated by emission produced at considerable angles in relation to the jet axis (the phenomenon of angular-time delay). In this case, the SSC flux would evolve as $F^{\rm ssc}_{\nu,r}\propto t^{-(\beta+2)}\nu^{-\beta}$ \citep{2000ApJ...543...66P, 2003ApJ...597..455K}.

The attenuation of up-scattered photons produced by synchrotron radiation results directly from the Klein-Nishina (KN) impact on the SSC spectrum, but the predominance of SSC photons and the cooling of certain injected electrons with varying Lorentz factors are indirect outcomes.  For $\nu^{\rm ssc}_{\rm c, r}<\nu^{\rm ssc}_{\rm m, r}$ and $\nu^{\rm ssc}_{\rm m, r}<\nu^{\rm ssc}_{\rm c, r}$, the spectral breaks in the KN regime are
{\small
\bary
h \nu^{\rm ssc}_{\rm KN, m, r} &=&
\begin{cases}
5.1\,{\rm GeV}\,\left(\frac{1+z}{1.3}\right)^{\frac{122-75p}{96(p-1)}} \tilde{g}^{\frac{1}{p-1}}\,\chi_{e,-0.3}^{-1}\,\varepsilon^{\frac{1}{p-1}}_{\rm e_r,-1}\,\,\varepsilon^{\frac{2-p}{4(p-1)}}_{\rm B_r,-3}\,n^{\frac{5(2-p)}{16(p-1)}}_{-1}\,E^{\frac{p-2}{16(p-1)}}_{53.5}\Delta^{\frac{62+3p}{96(p-1)}}_{11.9}\,\Gamma^{\frac{1}{p-1}}_{2.7}\cr
\hspace{9.6cm} t^{-\frac{26+21p+24[4a+b(2-p)]}{96(p-1)}}_{1.5} \hspace{0.7cm}{\rm for} \hspace{0.1cm} { 1<p<2 }\cr
3.8\times 10\,{\rm GeV}\,\left(\frac{1+z}{1.3}\right)^{-\frac{7}{24}} g\,\chi_{e,-0.3}^{-1}\,\varepsilon_{\rm e_r,-1}\,\Gamma_{2.7}\Delta^{\frac{17}{24}}_{11.9}\,t^{-\frac{17+24a}{24}}_{1.5},\hspace{5.9cm}{\rm for} \hspace{0.1cm} {2<p }
\end{cases}
\cr
h \nu^{\rm ssc}_{\rm KN, c, r}&\simeq& 9.8\times 10^2\,{\rm GeV}\,\left(\frac{1+z}{1.3}\right)^{-\frac{13}{12}} \left(1+Y_{\rm r} \right)^{-1}\,\varepsilon^{-1}_{\rm B_r,-3}\,n^{-\frac{3}{4}}_{-1}\,E^{-\frac{1}{4}}_{53.5}\,\Delta^{-\frac{1}{3}}_{11.9}\,t^{\frac{1+12b}{12}}_{1.5},\,\,
\eary
}

respectively. Requiring the evolution of the SSC spectral breaks and maximum flux (Eq. \ref{SSC_breaks_thick_aft_k0}), we derive the SSC RS light curves evolving in the thick-shell regime and homogeneous medium for two sets of electron spectral indices ($1 < p < 2$ and $2< p$), as listed in Table \ref{table:lc_ssc}. For completeness, we derive synchrotron light curves of RS light curves evolving in the thick-shell regime and homogeneous medium for the same electron spectral indices, as listed in Table \ref{table:lc_syn}.  Furthermore, we derive the CRs of the SSC and the synchrotron mechanisms of the RS in the thick-shell case and homogeneous medium, as listed in Tables \ref{table:cr_ssc} and \ref{table:cr_syn}, respectively.

\subsubsection{The Thin-shell regime}

Because the RS becomes moderately relativistic in the thin-shell phase, it is unable to slow down the shell. For this regime,  the shock crossing time is longer than the duration of the burst ( $T_{90}< t_{\rm x}$), and becomes $t_{\rm x}\propto  
\left(\frac{1+z}{1.3}\right)\, n^{-\frac13}\, E^{\frac13}\, \Gamma^{-\frac83}$. 
During the thin-shell regime, the magnetic field in the RS evolves as $B' \propto t^{-\frac{8+7b}{14}}$.   The minimum value of the electron Lorentz factor and the characteristic synchrotron break for $1<p<2$ ($2< p$ ) evolve as $\gamma_{\rm m, r}\propto t^{-\frac{96-28p+140a+35b(2-p)}{140(p-1)}}\,(t^{-\frac{2+7a}{7}})$ and $\nu^{\rm syn}_{\rm m, r}\propto t^{-\frac{4(7+10p)+5(28a+7b)}{70(p-1)}}\,(t^{-\frac{108+140a+35b}{70}})$, respectively.  Similarly, the evolution of the cooling breaks for the electron Lorentz factor and the synchrotron radiation is $\gamma_{\rm c,r}\propto  t^{\frac{19+35b}{35}}$, and $\nu^{\rm syn}_{\rm c, r}\propto t^{\frac{8+105b}{70}}$, respectively. Finally, the maximum synchrotron flux varies as $F^{\rm syn}_{\rm max,r }\propto t^{-\frac{68+35b}{70}}$ \citep{2000ApJ...545..807K}.\\

In this regime,  the spectral breaks and the maximum flux of the SSC process are given by

{\small
\begin{eqnarray}\label{SSC_breaks_thin_aft_k0}
h \nu^{\rm ssc}_{\rm m, r} &=& \begin{cases}
2.0\times 10^{-9}\,{\rm GeV} \left(\frac{1+z}{1.3}\right)^{\frac{97-29p}{35(p-1)}}\,\tilde{g}^{\frac{4}{p-1}}\,\chi_{\rm e,-0.3}^{-4}\,\varepsilon^{\frac{4}{p-1}}_{\rm e_r,-1} \varepsilon^{\frac{3-p}{2(p-1)}}_{\rm B_r,-3} \, n^{\frac{191-117p}{210(p-1)}}_{-1}\Gamma^{-\frac{286+48p}{105(p-1)}}_{2}\,  E^{\frac{2(3p+31)}{105(p-1)}}_{52.2}\cr
\hspace{9.5cm} t^{-\frac{4(31+3p)+5[56a+7b(3-p)]}{70(p-1)}}_{2.6}\hspace{1.35cm}{\rm for} \hspace{0.1cm} { 1<p<2 }\cr
8.8\times 10^{-8}\,{\rm GeV} \left(\frac{1+z}{2}\right)^{\frac{39}{35}}\,g^4\,\chi_{\rm e,-0.3}^{-4}\,\varepsilon^4_{\rm e_r,-1} \varepsilon^{\frac12}_{\rm B_r,-3}\,n^{-\frac{43}{210}}_{-1}\, \Gamma^{-\frac{382}{105}}_{2}\,  E^{\frac{74}{105}}_{52.2}\,t^{-\frac{148+280a+35b}{70}}_{2.6}\hspace{4.1cm}{\rm for} \hspace{0.1cm} { 2<p }\cr
\end{cases}\cr%
h \nu^{\rm ssc}_{\rm cut, r}&\simeq& 1.9\times 10\,{\rm GeV} \left(\frac{1+z}{1.3}\right)^{\frac{78+245b}{70}}\,\left(1+Y_{\rm r}\right)^{-4}  \varepsilon^{-\frac72}_{\rm B_r,-2}\,n^{-\frac{603+245b}{210}}_{-1} \, \Gamma^{-\frac{2(261+490b)}{105}}_{2}\, E^{-\frac{132-245b}{210}}_{52.2}\, t^{-\frac{74}{35}}_{2.6}\,\cr
F^{\rm ssc}_{\rm max,r} &\simeq&  5.2\times 10^{-9}\,{\rm mJy}\, \left(\frac{1+z}{1.3}\right)^{\frac{62}{35}} \,g^{-1}\,\chi_{\rm e,-1}\,\varepsilon^{\frac12}_{\rm B_r,-3} \, n^{\frac{191}{210}}_{-1} \,\Gamma^{-\frac{181}{105}}_{2}\,d^{-2}_{\rm z,27.7}E^{\frac{167}{105}}_{52.2}\, t^{-\frac{54+35b}{70}}_{2.6}\,.
\eary
}

The cutoff frequency in this regime is $\nu^{\rm ssc}_{\rm cut, r}=\nu^{\rm ssc}_{\rm c, r}(t_{\rm x})\,\left(\frac{t}{t_{\rm x}} \right)^{-\frac{74}{35}}$.   The SSC emission from the RS region could decay faster as a result of the angular-time delay effect. During the emission at high latitudes, the SSC flux evolves equal to the flux mentioned above in the thick-shell regime. \\

The spectral breaks in the KN regime for $\nu^{\rm ssc}_{\rm c, r}<\nu^{\rm ssc}_{\rm m, r}$ and $\nu^{\rm ssc}_{\rm m, r}<\nu^{\rm ssc}_{\rm c, r}$ are
{\small
\bary
h\nu^{\rm ssc}_{\rm KN, m, r}&\simeq&\begin{cases}
1.8\,{\rm GeV}  \,\left(\frac{1+z}{1.3}\right)^{\frac{45-28p}{35(p-1)}}\tilde{g}^{\frac{1}{p-1}}\chi_{e,-0.3}^{-1}\,\varepsilon^{\frac{1}{p-1}}_{\rm e_r,-1}\,\varepsilon^{\frac{2-p}{4(p-1)}}_{\rm B_r,-3}\,n^{\frac{170-133p}{420(p-1)}}_{\rm -1} \,\Gamma^{-\frac{160+7p}{210(p-1)}}_{2}\,E^{\frac{10+7p}{105(p-1)}}_{52.2}\cr
\hspace{8cm}\,t^{-\frac{40+28p+140a+35b(2-p)}{140(p-1)}}_{2.6}\hspace{2.15cm}{\rm for} \hspace{0.1cm} { 1<p<2 }\cr
4.5\times 10\,{\rm GeV}  \,\left(\frac{1+z}{1.3}\right)^{-\frac{11}{35}}g\chi_{e,-0.3}^{-1}\,\varepsilon_{\rm e_r,-1}\,n^{-\frac{8}{35}}_{\rm -1} \,\Gamma^{-\frac{29}{35}}_{2}\,E^{\frac{8}{35}}_{52.2}\,t^{-\frac{24+35a}{35}}_{2.6},\hspace{4.95cm}{\rm for} \hspace{0.1cm} { 2<p }\cr
\end{cases}%
\cr
h\nu^{\rm ssc}_{\rm KN, c, r}&\simeq& 5.8\times 10^{2}\,{\rm GeV} \, \left(\frac{1+z}{1.3}\right)^{-\frac87}\,\left(1+Y_{\rm r}\right)^{-1}\,\varepsilon^{-1}_{\rm B_r,-3}\,n^{-\frac{13}{21}}_{-1} \,\Gamma^{\frac{22}{21}}_{2}\,E^{-\frac{8}{21}}_{\rm  52.2}\,t^{\frac{1+7b}{7}}_{2.6},\,\,\,
\eary
}
respectively. 

Taking into account the evolution of the SSC spectral breaks together the maximum flux derived in Eq. \ref{SSC_breaks_thin_aft_k0}, we compute the SSC light curves for a spectral electron index of $1 < p < 2$ and $2< p$. Table \ref{table:lc_ssc} shows the CRs in the thin-shell regime for each cooling condition in a constant-density medium. For completeness, we derive synchrotron light curves of RS light curves evolving in the thin-shell regime and constant-density medium for the same electron spectral indices, as listed in Table \ref{table:lc_syn}. Furthermore, we derive the CRs of the SSC and the synchrotron mechanisms of the RS in the thin-shell case and homogeneous medium, as listed in Tables \ref{table:cr_ssc} and \ref{table:cr_syn}, respectively.

\subsection{Evolution in stratified environment (stellar wind)}

The evolution of the critical Lorentz factor in the stellar wind ($n(r)=3.0\times 10^{35}\,{\rm cm^{-1}}\,A_{\rm W}\,r^{-\rm 2}$) becomes $\Gamma_c\propto(1+z)^{\frac14} E^{\frac14} A_{\rm W}^{-\frac14} T_{90}^{-\frac14}$, where $A_{\rm W}$ corresponds to the density parameter. For stellar wind, we consider the RS evolving in the thick ($t_{\rm x} \lesssim T_{90}$) and in the thin ($T_{90} < t_{\rm x}$) regime.

\subsubsection{The Thick-shell regime}

RS crosses the shell, and the magnetic field evolves as $B'\propto t^{-\frac{3+2b}{4}}$, and the minimum and cooling electron Lorentz factors for $1<p<2$ ($2< p$) evolve as $\gamma_{\rm m,r}\propto t^{-\frac{16-5p+4[4a+b(2-p)]}{16(p-1)}}\,(t^{-\frac{3+8a}{8}})$ and $\gamma_{\rm c,r}\propto t^{\frac{7+8b}{8}}$, respectively.  The respective synchrotron spectral breaks and the maximum synchrotron flux evolve as $\nu^{\rm syn}_{\rm m, r}\propto t^{-\frac{7+4p+4(4a+b)}{8(p-1)}}\,(t^{-\frac{15+4(4a+b)}{8}})$ and $\nu^{\rm syn}_{\rm c, r}\propto t^{\frac{5+12b}{8}}$ and $F^{\rm syn}_{\rm max,r}\propto t^{-\frac{9+4b}{8}}$, respectively.

When the same population of electrons that generates synchrotron radiation also upscatters them to higher energies, the SSC mechanism results.  By taking into consideration the emergence of the synchrotron spectrum together with the relativistic electrons in the RS region,  the SSC spectral breaks and the maximum SSC flux can be written as

{\small
\begin{eqnarray}\label{SSC_breaks_thick_aft_k2}
h \nu^{\rm ssc}_{\rm m, r} &=& \begin{cases}
5.3\times 10^{-6}\,{\rm GeV}\,\left(\frac{1+z}{1.3}\right)^{\frac{31-9p}{8(p-1)}} \tilde{g}^{\frac{4}{p-1}} \chi_{\rm e,-0.3}^{-4} \varepsilon^{\frac{4}{p-1}}_{\rm e_r,-1} \varepsilon^{\frac{3-p}{2(p-1)}}_{\rm B_r,-3} A^{\frac{7-2p}{2(p-1)}}_{\rm W,-1} \Delta^{\frac{3(5+p)}{8(p-1)}}_{11.9}\,E^{-\frac{4-p}{2(p-1)}}_{53.5} t^{-\frac{23-p+4[8a+b(3-p)]}{8(p-1)}}_{1.5}\hspace{0.7cm}{\rm for} \hspace{0.1cm} { 1<p<2 }\cr
1.1\times 10^{-2}\,{\rm GeV}\, \left(\frac{1+z}{1.3}\right)^{\frac{13}{8}}\,g^{4}\,\chi_{\rm e,-0.3}^{-4}\,\varepsilon^4_{\rm e_r,-1} \varepsilon^{\frac12}_{\rm B_r,-3} \,A^{\frac{3}{2}}_{\rm W, -1}\,\Delta^{\frac{21}{8}}_{11.9}\, \Gamma^{4}_{2.7}\,  E^{-1}_{\rm 53.5}\,t^{-\frac{21+4(8a+b)}{8}}_{1.5}\hspace{2.8cm}{\rm for} \hspace{0.1cm} { 2<p }\cr
\end{cases}\cr %
h \nu^{\rm ssc}_{\rm cut, r}&\simeq& 8.3\times 10^{-8}\,{\rm GeV} \left(\frac{1+z}{1.3}\right)^{\frac{13+28b}{8}}\,\left(1+Y_{\rm r}\right)^{-4}  \varepsilon^{-\frac72}_{\rm B_r,-3}\,A^{-\frac{9}{2}}_{\rm W, -1}\,\Delta^{\frac{37+28b}{8}}_{11.9}\, E_{\rm 53.5}\, t^{-\frac{21}{8}}_{1.5}\,\cr
F^{\rm ssc}_{\rm max,r} &\simeq&  2.8\times 10\,{\rm mJy}\,\left(\frac{1+z}{1.3}\right)^{\frac{19}{8}}\,g^{-1}\,\chi_{\rm e,-0.3}\,  \varepsilon^{\frac12}_{\rm B_r,-3} \, A^{2}_{W,-1}\, \Gamma^{-1}_{2}\,\Delta^{-\frac18}_{11.9}\, d^{-2}_{\rm z,27.7}\,E^{\frac{1}{2}}_{\rm 53.5}\, t^{-\frac{11+4b}{8}}_{1.5}\,.
\eary
}

The new cooling break is predicted to be $\nu^{\rm ssc}_{\rm cut, r}=\nu^{\rm ssc}_{\rm c, r}(t_{\rm x})\,\left(\frac{t}{t_{\rm x}} \right)^{-\frac{21}{8}}$. This is because the fluid expands in an adiabatic approach.  It is possible that the SSC light curve will have a more steep temporal decay index. This is because electrons cannot be reaccelerated within the RS region, and as a result, the emission from this region will stop. However, due to the phenomenon of angular-time delay, the SSC flux would evolve as $F^{\rm ssc}_{\nu,r}\propto t^{-(\beta+2)}\nu^{-\beta}$ \citep{2000ApJ...543...66P, 2003ApJ...597..455K}.

The KN influence on the SSC spectrum is directly responsible for the attenuation of up-scattered photons that are generated by synchrotron radiation.   For $\nu^{\rm ssc}_{\rm c, r}<\nu^{\rm ssc}_{\rm m, r}$ and $\nu^{\rm ssc}_{\rm m, r}<\nu^{\rm ssc}_{\rm c, r}$, the spectral breaks in the KN regime for $1<p<2$ and $2< p$ are
{\small
\bary
h \nu^{\rm ssc}_{\rm KN, m, r}&\simeq&\begin{cases}
5.8\,{\rm GeV}\,\left(\frac{1+z}{1.3}\right)^{\frac{26-15p}{16(p-1)}} \tilde{g}^{\frac{1}{p-1}} \chi_{\rm e,-0.3}^{-1} \varepsilon^{\frac{1}{p-1}}_{\rm e_r,-1}\,\varepsilon^{\frac{2-p}{4(p-1)}}_{\rm B_r,-3} A^{\frac{5(2-p)}{8(p-1)}}_{\rm W,-1} \Delta^{\frac{3(p+2)}{16(p-1)}}_{11.9}\,\Gamma^{\frac{1}{p-1}}_{2.7}\,E^{-\frac{3(2-p)}{8(p-1)}}_{53.5}\cr
\hspace{8.1cm} t^{-\frac{p+10+4[4a+b(2-p)]}{16(p-1)}}_{1.5}\,\hspace{1.7cm}{\rm for} \hspace{0.1cm} { 1<p<2 }\cr
3.8\times 10\,{\rm GeV}\,\left(\frac{1+z}{1.3}\right)^{-\frac14} g\,\chi_{e,-0.3}^{-1}\,\varepsilon_{\rm e_r,-1}\,\Gamma_{2.7}\,\Delta^{\frac34}_{11.9}\,t^{-\frac{3+4a}{4}}_{1.5}\,\hspace{5.5cm}{\rm for} \hspace{0.1cm} { 2<p }
\end{cases}\cr%
h \nu^{\rm ssc}_{\rm KN, c, r}&\simeq& 2.0\,{\rm GeV}\,\left(\frac{1+z}{1.3}\right)^{-\frac32} \left(1+Y_{\rm r} \right)^{-1}\,\varepsilon^{-1}_{\rm B_r,-3}\,A^{-\frac{3}{2}}_{\rm W,-1}\,E^{\frac12}_{53.5}\,t^{\frac{1+2b}{2}}_{1.5},\,\,
\eary
}
respectively.  Using the evolution of the SSC spectral breaks with the maximum SSC flux (Eq. \ref{SSC_breaks_thick_aft_k2}), we estimate and show in Table \ref{table:lc_ssc} the light curves of the SSC mechanism from the RS region when it evolves in the thick-shell regime and the surrounding medium formed by the stellar wind of the progenitor. The temporal PL indices are calculated for an electron distribution with a spectral index in the range of $1 < p < 2$ and $2< p$. For completeness, we derive synchrotron light curves of RS light curves evolving in the thick-shell regime and stellar-wind medium for the same electron spectral indices, as listed in Table \ref{table:lc_syn}.  Furthermore, we derive the CRs of the SSC and the synchrotron mechanisms of the RS in the thick-shell case and the stellar wind medium, as listed in Tables \ref{table:cr_ssc} and \ref{table:cr_syn}, respectively.

\subsubsection{The Thin-shell regime}

The RS, which becomes moderately relativistic in the thin-shell phase, cannot fully decelerate the shell. In this regime, the shock crossing time exceeds the duration of the burst and is defined by $t_{\rm x}\propto  (1+z)\, A_{\rm W}^{-1}\, E\, \Gamma^{-4}$.  In this afterglow wind model, the magnetic field, the minimum and cooling electron Lorentz factors, the synchrotron spectral breaks and the maximum synchrotron flux in terms of the observed time for $1<p<2$ ($2< p$) are $B'\propto t^{-\frac{32+21b}{42}}$, $\gamma_{\rm m,r}\propto t^{-\frac{4(22-7p)+84a+21(2-p)}{84(p-1)}}\,(t^{-\frac{8+21a}{21}})$ and $\gamma_{\rm c,r}\propto t^\frac{6+7b}{7}$, $\nu^{\rm syn}_{\rm m, r}\propto t^{-\frac{2(7+3p)+7(4a+b)}{14(p-1)}}\,(t^{-\frac{26+28a+7b}{14}})$ and $\nu^{\rm syn}_{\rm c, r}\propto t^{\frac{26+63b}{42}}$ and $F^{\rm syn}_{\rm max,r}\propto t^{-\frac{46+21b}{42}}$, respectively. In this regime, the spectral SSC breaks, and the maximum flux can be written as

{\small
\begin{eqnarray}\label{SSC_breaks_thin_aft_k2}
h \nu^{\rm ssc}_{\rm m, r} &=& \begin{cases}
3.7\times 10^{-14}\,{\rm GeV} \left(\frac{1+z}{1.3}\right)^{\frac{86-26p}{21(p-1)}}\,\tilde{g}^{\frac{4}{p-1}}\,\chi_{\rm e,-0.3}^{-4}\,\varepsilon^{\frac{4}{p-1}}_{\rm e_r,-1} \varepsilon^{\frac{3-p}{2(p-1)}}_{\rm B_r,-3} \,A^{\frac{59-53p}{42(p-1)}}_{\rm W,-1}\, \Gamma^{-\frac{2(46+11p)}{21(p-1)}}_{2}\,  E^{\frac{2(8p+1)}{21(p-1)}}_{52.2}\,\cr
\hspace{9.5cm} t^{-\frac{10(13-p)+168a+21b(3-p)}{42(p-1)}}_{2.6}\hspace{1.45cm}{\rm for} \hspace{0.1cm} { 1<p<2 }\cr
2.1\times 10^{-12}\,{\rm GeV} \left(\frac{1+z}{1.3}\right)^{\frac{34}{21}}\,g^4\,\chi_{\rm e,-0.3}^{-4}\,\varepsilon^4_{\rm e_r,-1} \varepsilon^{\frac12}_{\rm B_r,-3}\,A^{-\frac{47}{42}}_{\rm W,-1}  \, \Gamma^{-\frac{136}{21}}_{2}\,  E^{\frac{34}{21}}_{52.2}\,t^{-\frac{110+168a+21b}{42}}_{2.6}\hspace{4.0cm}{\rm for} \hspace{0.1cm} { 2<p }\cr
\end{cases}\cr%
h \nu^{\rm ssc}_{\rm cut, r}&\simeq& 2.2\times 10^{-16}\,{\rm GeV} \left(\frac{1+z}{1.3}\right)^{\frac{68+147b}{42}}\,\left(1+Y_{\rm r}\right)^{-4}  \varepsilon^{-\frac72}_{\rm B_r,-3}\,A^{-\frac{383+147b}{42}}_{\rm W,-1}  \, \Gamma^{-\frac{2(194+147b)}{21}}_{2}\, E^{\frac{236+147b}{42}}_{52.2}\, t^{-\frac{55}{21}}_{2.6}\,\cr
F^{\rm ssc}_{\rm max,r} &\simeq&  5.0\times 10^{-7}\,{\rm mJy}\, \left(\frac{1+z}{1.3}\right)^{\frac{17}{7}} \,g^{-1}\,\chi_{\rm e,-0.3}\,\varepsilon^{\frac12}_{\rm B_r,-3} \, A^{\frac{29}{14}}_{\rm W,-1}\, \,\Gamma^{-\frac{5}{7}}_{2}\,d^{-2}_{\rm z,27.7}E^{\frac{3}{7}}_{52.2}\, t^{-\frac{20+7b}{14}}_{2.6},
\eary
}

where the cutoff frequency in this case is $\nu^{\rm ssc}_{\rm cut, r}=\nu^{\rm ssc}_{\rm c, r}(t_{\rm x})\,\left(\frac{t}{t_{\rm x}} \right)^{-\frac{55}{21}}$.   The SSC emission from the RS region could decay faster as a result of the angular-time delay effect. During emission at high latitudes, the SSC flux evolves to be equal to the flux mentioned above in the thick-shell regime ($F^{\rm ssc}_{\nu,r}\propto t^{-(\beta+2)}\nu^{-\beta}$). \\

For $\nu^{\rm ssc}_{\rm c, r}<\nu^{\rm ssc}_{\rm m, r}$ and $\nu^{\rm ssc}_{\rm m, r}<\nu^{\rm ssc}_{\rm c, r}$, the spectral breaks in the KN regime for $1<p<2$ and $2<p$ are given by
{\small
\bary
h\nu^{\rm ssc}_{\rm KN, m, r}&\simeq& \begin{cases}
1.6\times 10^{-2}\,{\rm GeV} \,\left(\frac{1+z}{1.3}\right)^{\frac{12-7p}{7(p-1)}}\tilde{g}^{\frac{1}{p-1}}\chi_{e,-0.3}^{-1}\,\varepsilon^{\frac{1}{p-1}}_{\rm e_r,-1}\,\varepsilon^{\frac{2-p}{4(p-1)}}_{\rm B_r,-3}\,A^{\frac{22-21p}{28(p-1)}}_{\rm W,-1}E^{-\frac{4-7p}{14(p-1)}}_{52.2}\Gamma^{-\frac{12+7p}{14(p-1)}}_{2}\,t^{-\frac{20+7[4a+b(2-p)]}{28(p-1)}}_{2.6}\hspace{0.4cm}{\rm for} \hspace{0.1cm} { 1<p<2 }\cr
4.4\times 10^{-2}\,{\rm GeV} \,\left(\frac{1+z}{1.3}\right)^{-\frac{2}{7}}g\chi_{e,-0.3}^{-1}\,\varepsilon_{\rm e_r,-1}\,A^{-\frac{5}{7}}_{\rm W,-1} \,\Gamma^{-\frac{13}{7}}_{2}\,E^{\frac{5}{7}}_{52.2}\,t^{-\frac{5+7a}{7}}_{2.6}\hspace{4.9cm}{\rm for} \hspace{0.1cm} { 2<p }\cr
\end{cases}\cr
h\nu^{\rm ssc}_{\rm KN, c, r}&\simeq& 4.2\times 10^{-2}\,{\rm GeV}  \,\left(\frac{1+z}{1.3}\right)^{-\frac{32}{21}} \left(1+Y_{\rm r}\right)^{-1}\,\varepsilon^{-1}_{\rm B_r,-3}\,A^{-\frac{31}{21}}_{\rm W,-1}\,\Gamma^{\frac{2}{21}}_{2}\,E^{\frac{10}{21}}_{52.2}\,t^{\frac{11+21b}{21}}_{2.6},\,\,\,
\eary
}

respectively.

Table \ref{table:lc_ssc} displays the light curves of the SSC emission when the RS lies in the thin-shell case and is decelerated by the stellar wind environment. This derivation is performed considering the SSC spectral breaks and maximum flux (Eq. \ref{SSC_breaks_thin_aft_k2}) and the entire range of the spectral electron index for each cooling condition. For completeness, we derive synchrotron light curves of RS light curves evolving in the thin-shell regime and stellar-wind medium for the same electron spectral indices, as listed in Table \ref{table:lc_syn}. Furthermore, we derive the CRs of the SSC and the synchrotron mechanisms of the RS in the thin-shell case and the stellar wind medium, as listed in Tables \ref{table:cr_ssc} and \ref{table:cr_syn}, respectively.   It is worth noting that for $a=0$ and $b=0$, the spectral breaks and the maximum flux of the standard SSC mechanism are recovered \citep{2025MNRAS.543.1945F}.

\section{Application: Second \textit{Fermi}-LAT Catalog}\label{sec4}

We adopt a similar methodology performed in \cite{2021PASJ...73..970D, 2021ApJS..255...13D, 2020ApJ...903...18S, 2022ApJ...934..188F, 2023MNRAS.525.1630F, 2024MNRAS.527.1884F} for a thorough examination of the CRs of the SSC from the RS region with variation in microphysical parameters.   We consider the evolution of RS in the thick and thin-shell regime for
a homogeneous density medium and stellar wind, and two sets of electron spectral indices $1 < p < 2$ and $2< p$.  Our sample consists of 86 bursts reported in 2FLGC with known redshifts and photons above 100 MeV \citep{2019ApJ...878...52A}.  In the overall scheme of spectral evolution, the photon index is documented in 2FLGC as $\Gamma_{\rm L}=\beta_{\rm L}+1$, where $\beta_{\rm L}$ represents the spectral index reported by the LAT instrument.   The temporal and spectral indices derived from the PL function $F^{\rm ssc}_{\rm \nu, r} \propto t^{-\alpha} \nu^{-\beta}$ are used for the analysis of the CRs.

Figure \ref{fig:MCMC_ISM} and \ref{fig:MCMC_ISM2} show the corner plots from Markov Chain Monte Carlo (MCMC) simulations for the parameters ${\rm a}$ and ${\rm b}$ in our RS model evolving in the thick- and thin-shell regime, respectively. The upper panels correspond to the ISM and the lower panels to the stellar wind. Panels, from left to right, displays the cooling conditions, $ \nu < \nu^{\rm ssc}_{\rm m,r}$, $ \nu^{\rm ssc}_{\rm m,r} < \nu < \nu^{\rm ssc}_{\rm cut,r}$ and  $ \nu^{\rm ssc}_{\rm cut,r} < \nu$. Each histogram shows the marginalized posterior densities with the median values in red lines. Focusing on the last column corresponding to $\nu_{\rm cut,r}^{\rm ssc}<\nu$, we observe that, for all rows, this particular column does not lead to any conclusions as the MCMC algorithm does not converge to a solution. This is because the flux is cut at this frequency range, and hence no observations can be fit. Henceforth, we focus on the first two columns, in which the MCMC is able to find a minimum. Looking at the first row, we notice that, in general, $a$ takes negative values and $b$ positive ones. This means that for our sample of GRBs, the microphysical parameter associated with the acceleration of electrons increases in time, while the magnetic one decreases. By comparing the two columns, we note some variation in the magnitudes. In the first column, we find $(a,b)=(-0.63,0.56)$, while in the second, we have $(a,b)=(-0.6,0.62)$. This suggests that the time evolution of the microphysical parameters in a thick case evolving in ISM is relatively stable across different cooling conditions, with a noticeable increase in $b$ when moving to the second column. Regarding the second row, we observe that $a$ remains negative across both columns, but $b$ shifts from negative in the first column, where we find $(a,b)=(-0.6,-0.34)$, to nearly zero in the second column, with $(a,b)=(-0.74,-0.03)$. This indicates that, unlike in the ISM case, the evolution of the magnetic microphysical parameter in a thick shell wind is not strongly constrained and can vary significantly depending on the fitting conditions. In the case of the third row, we note that $a$ remains negative while $b$ is consistently positive, following the same general trend observed in the thick shell ISM case. However, the best-fit values exhibit more significant variation between the two columns, with $(a,b)=(-0.66,0.72)$ in the first and $(a,b)=(-0.41,-0.28)$ in the second. This suggests that the thin shell ISM environment supports a similar trend of decreasing $\varepsilon_B$ and increasing $\varepsilon_e$, but the precise values are more sensitive to the specific observational constraints. Finally, in the fourth row we observe a consistent negative trend in both parameters. In the first column, we find $(a,b)=(-0.6,-0.47)$, while in the second, the values change to $(a,b)=(-0.72,-0.17)$. This behavior, different from the ISM cases, implies that in a thin shell wind environment, both microphysical parameters decrease over time, with a slightly stronger decrease in $\varepsilon_e$ compared to the first column. Overall, we find that the evolution of $\varepsilon_B$ is highly dependent on the external medium, with the ISM cases favoring an increase over time, while wind environments generally lead to a decrease. Meanwhile, $\varepsilon_e$ consistently decreases across all cases, though the exact magnitude of the PL indices varies depending on the specific cooling conditions.

Figure \ref{fig1:CRsthick} shows the CR of the SSC RS model evolving in a thick-shell regime, for ISM (above) and stellar wind (below),  using the best-fit parameter values ($a$ and $b$) derived from our MCMC analysis and summarized in Table \ref{tab:CR-Results}. Figure \ref{fig2:CRsthin} displays the CR of the SSC RS model evolving in a thin-shell regime, for ISM (above) and stellar wind (below). The uncertainties displayed in this Figure are in the 1$\sigma$ range, and bursts that satisfy the CRs are shown with ellipses in purple; others are shown in gray.   Panels, from left to right, displays the cooling conditions, $ \nu < \nu^{\rm ssc}_{\rm m,r}$, $ \nu^{\rm ssc}_{\rm m,r} < \nu < \nu^{\rm ssc}_{\rm cut,r}$ and  $\nu^{\rm ssc}_{\rm cut,r} < \nu$, for two sets of electron spectral indices $1 < p < 2$ and $2< p$.  Table \ref{tab:CR-Results} summarizes the best-fit values of the parameters $a$ and $b$ together with the coincidences of the bursts reported in 2FLGC. These coincidences and fractions of bursts are shown for each cooling condition of the SSC RS scenario in the thick- and thin-shell case, ISM and stellar wind, and $1 < p < 2$ and $2 < p$.

Overall, we find that, when using a PL description, the CRs of the SSC mechanism from the RS can account for a maximum of 32 occurrences of all the GRBs considered. The analysis shows that an ISM-like environment is preferred over a wind-like medium, as is the thin-shell over the thick-shell case. The most preferred scenario, which accounts for 37.21\% (32 GRBs) of the sample, is a thin shell expanding into a wind environment during the slow-cooling phase $\nu^{\rm ssc}_{\rm m, r} < \nu < \nu^{\rm ssc}_{\rm cut, r}$. The second most desirable scenario, which corresponds to 29.07\% (25 GRBs), is a thin shell in an ISM medium with the same cooling conditions.  Another interesting result is that no GRBs satisfy the CRs when the SSC process lies in the regime $\nu < \nu^{\rm ssc}_{\rm m, r}$, regardless of the surrounding medium or shell type. This suggests that the SSC component from the RS may not play a dominant role at lower frequencies. Finally, for the regime $\nu^{\rm ssc}_{\rm cut,r} < \nu$, only a single GRB (1.16\%) is found to match across all cases, reinforcing the conclusion that in this high-frequency range, the RS SSC contribution is difficult to constrain.

{ Figures \ref{fig:cr_bur_1}, \ref{fig:cr_bur_2}, 
\ref{fig:cr_bur_3} and \ref {fig:cr_bur_4} show the ranges of microphysical parameters ($a$ and $b$) for which the CRs of bursts reported in 2FLGC are satisfied when the SSC model lies in the cooling condition ${\rm \nu_m^{ssc} < \nu_{\rm LAT} < \nu_{\rm cut}^{ssc} }$, and the RS evolves in the thick- and thin-shell regime for ISM and stellar wind. Figure \ref{fig:cr_bur_1} shows the ranges of microphysical parameters in each case for GRB 080916C, 081009, 090323, 090328, 090626, 091003, 091031, 100414A, 100728A and 110428A, Figure \ref{fig:cr_bur_2} for 
GRB 110518A, 110721A, 120316A, 130828A, 131014A, 131029A, 131231A, 140102A,  140110A, 140523A, Figure \ref{fig:cr_bur_3} for GRB 141028A, 141102A, 141222A, 150314A, 150523A, 150627A, 150902A, 160509A, 160521B and 160623A, and Figure \ref{fig:cr_bur_4} for 60816A, 160821A, 160905A, 161109A, 170808B, 171010A, 171102A, 180210A, and 180526A. It is essential to note that there are ranges of parameters that satisfy the closure relations in the case of a constant medium, but not in the stellar wind environment or when the RS evolves in the thick case, but not in the thin case.  For instance, no parameter range fulfills the closure relations in the stellar wind for GRB 080916.}

\subsection{Density profile of surrounding material}

The surrounding material has important implications for the dynamics of the afterglow phase and might help to elucidate the GRB progenitor.  A stratified medium is associated with the stellar wind ejected by the massive stars and therefore with lGRBs \citep{1993ApJ...405..273W,1998ApJ...494L..45P}, and a homogeneous medium with low density is identified with sGRBs \citep{1992ApJ...392L...9D, 1992Natur.357..472U, 1994MNRAS.270..480T, 2011MNRAS.413.2031M}.   Regarding the results summarized in Table \ref{tab:CR-Results}, when comparing the ISM and wind-like environments, we notice that the ISM cases generally show a higher number of coincidences in the thick-shell scenario, whereas the wind-like medium dominates in the thin-shell case. Specifically, the most significant number of matches occurs in the thin-shell wind environment, suggesting that this medium configuration is more compatible with the observed GRBs in the SSC framework. In contrast, the thick-shell wind scenario produces significantly fewer matches, indicating that such conditions are less likely to describe the observed afterglows. 


Table \ref{table:dens_paramet} presents the evolution of the PL density index for the SSC radiation scattered from the RS in the thick and thin case for the cooling conditions, $ \nu < \nu^{\rm ssc}_{\rm m,r}$, $ \nu^{\rm ssc}_{\rm m,r} < \nu < \nu^{\rm ssc}_{\rm cut,r}$ and  $\nu^{\rm ssc}_{\rm cut,r} < \nu$.   It is important to note that as soon as the condition $\nu^{\rm ssc}_{\rm c, r}< \nu$ is satisfied, distinguishing between a stellar wind and a homogeneous environment becomes impossible. Density variations in the stellar-wind environment yield a signature that is more pronounced than that of the homogeneous medium. This signature is more prominent in the thick-shell regime compared to that in the thin-shell regime.  For example, during the cooling condition $\nu^{\rm ssc}_{\rm m, r} < \nu < \nu^{\rm ssc}_{\rm cut, r} $, the SSC flux evolving in the thick case is given by $F^{\rm ssc}_{\rm \nu, r}\propto n^{0.9}(A_{\rm W}^{1.8})$ for $p = 1.9$. Similarly, the evolution in ISM is $F^{\rm ssc}_{\rm \nu, r}\propto n^{0.9}$ and $(n^{0.76})$ for the thick and thin case, respectively.   A comparable analysis may examine an electron population characterized by a spectral index in the range of $ 2 < p$.

The transition radius between stellar wind and ISM, combined with variations in the observed flux, provides information about the progenitor \citep[e.g., see][]{2017ApJ...848...15F,2019ApJ...879L..26F}.  Massive stars located within the ISM will experience a transition at different distances between the stellar wind produced by these progenitors and the homogeneous density constant, leading to a discontinuity or bubble-like structure \citep{1975ApJ...200L.107C, 1977ApJ...218..377W, 2006ApJ...643.1036P}.  In the standard GRB afterglow scenario, the multiwavelength light curves are generated as the electrons accelerated in the external shocks are cooling down by the SSC and synchrotron emission. 
 The profile of the light curves also depends on the surrounding medium (stellar wind and homogeneous), the variation of microphysical parameters (via $a$ and $b$), among others. Therefore, the light curves expected from the stellar wind are different from those of a constant medium, and any transition can be inferred from the light curves, as observed in some GRBs \citep[i.e., ][]{2003ApJ...591L..21D, 2007ApJ...664L...5K, 2009MNRAS.400.1829J, 2020ApJ...900..176L, 2017ApJ...848...15F, 2019ApJ...879L..26F}.   \cite{2019ApJ...883..162F} detailed the non-thermal multiwavelength observations of GRB 160625B during the transition from the stellar wind to ISM. Before $\sim$ 8000 s, these observations were consistent with the evolution of the light curves in the stellar wind and after that with the evolution in the homogeneous medium. The authors located this transition at a distance of 1 parsec from the progenitor.  Table \ref{table:dens_paramet} shows that a transition is predicted to leave a significant signature when the RS evolves under cooling conditions $ \nu < \nu^{\rm ssc}_{\rm m,r}  < \nu^{\rm ssc}_{\rm cut,r}$ and $ \nu^{\rm ssc}_{\rm m,r} < \nu < \nu^{\rm ssc}_{\rm cut,r}$, but not $\nu^{\rm ssc}_{\rm cut,r} < \nu $, neither in the thin nor the thick case. For instance, the SSC flux as a function of the density from stellar wind to ISM changes from  $A_{\rm W}^{-\frac{15 - 14p}{6(p-1)}}$ to $n^{-\frac{15 - 14p}{12(p-1)}}$ and $\propto A_{\rm W}^\frac{11 - 2p}{4}$ to $n^{\frac{11 -2p}{8}}$ in the cooling condition $ \nu < \nu^{\rm ssc}_{\rm m,r}  < \nu^{\rm ssc}_{\rm cut,r}$ and $ \nu^{\rm ssc}_{\rm m,r} < \nu < \nu^{\rm ssc}_{\rm cut,r}$ for $1<p <2$.

\subsection{Hard-spectral index ($1< p < 2$) }

{ The results of the observational analyzes show that the electron spectral index varies between bursts following a normal distribution. Rather than maintaining an absolute standard value, the electron spectral index fluctuates between 1.5 and 3.5 \citep{2002ApJ...581.1248P, 2008MNRAS.388..144R, 2015ApJ...811...83Z, 2006MNRAS.371.1441S, 2008ApJ...672..433S, 2009MNRAS.395..580C, 2019ApJ...883..134T}. \cite{2002ApJ...581.1248P} performed an analysis
using the first spectroscopy catalog of GRBs observed by BATSE\footnote{The Burst and Transient Source Experiment} aboard the \textit{Compton} Gamma Ray Observatory (CGRO) which consisted of 5021 spectra. Requiring several PL configurations of the electron spectral index, they showed via the histograms that, irrespective of the PL configuration, a substantial number of spectra had a spectral index ranging $1<p<2$.   In a large sample of GRB afterglows gathered with the HETE-2, BeppoSAX, and Swift satellites, \cite{2006MNRAS.371.1441S} found no uniform value for the electron spectral index. A significant portion of the spectral data was characterized by spectral indexes ranging between 1 and 2.  In order to model GRB 010222, 020813, and 041006, \cite{2008MNRAS.388..144R} used an electron distribution characterized by a double power law.  For GRB 010222, 020813, and 041006, they found spectral indices of $1.47^{+0.004}_{-0.003}$, $1.40^{+0.007}_{-0.004}$, and $1.29 - 1.32$, respectively.  \cite{2015ApJ...811...83Z} described the early afterglow observations of GRB 091127 with a value of the electron spectral index of $p=1.5\pm 0.01$.  Using spectral and temporal indices reported in 2FLGC, \cite{2019ApJ...883..134T} thoroughly examined the closure relations in 59 randomly chosen LAT-detected bursts. The closure relations of some bursts fulfilled the spectral indices with $1< p <2$, regardless of whether they were characterized with a stratified or constant medium, and regardless of whether they were classified long or short.}

{ In the context of the RS SSC model with a hard electron spectral index $1 < p < 2$, the temporal decay index $\alpha$ of the light curve depends on the evolution of the microphysical parameters $\epsilon_e \propto t^{-a}$ and $\epsilon_B \propto t^{-b}$. Tables~\ref{table:lc_ssc} and~\ref{table:cr_ssc} provide the closure relations for SSC emission in both thick- and thin-shell regimes in homogeneous (ISM) and stellar wind media. When $a = b = 0$, the microphysical parameters remain constant over time \citep[see,][]{2025MNRAS.543.1945F}. In this case, for the thick-shell regime in a homogeneous medium (Table~\ref{table:lc_ssc}), the temporal index $\alpha$ in the slow-cooling regime ($\nu_{\rm m,r}^{\mathrm{ssc}} < \nu < \nu_{\rm cut,r}^{\mathrm{ssc}}$) is given by $\alpha = \frac{5p + 171}{96}$. For $p = 1.95$, this gives $\alpha \approx 1.88$, indicating a rather steep decay. In the stellar wind case, the expression becomes $\alpha = \frac{45 - p}{16}$, which for $p = 1.95$ gives $\alpha \approx 2.69$, a much steeper decay. This benchmark scenario highlights the dependence of the light curve slope on the density profile of the circumburst medium when microphysical parameters are constant. 

To note the effect of the variation of microphysical parameters, we consider the scenario When $a = b = 0.5$, and both $\epsilon_e$ and $\epsilon_B$ decrease with time. In the same thick-shell ISM case, the temporal index becomes $\alpha = \frac{5p + 171 + 192a + 24b(5 - p)}{96}$. Substituting $p = 1.95$, $a = 0.5$, and $b = 0.5$ yields $\alpha \approx 3.26$, significantly steeper than the constant-parameter case. This steepening occurs because the declining microphysical parameters reduce the efficiency of electron acceleration and magnetic field amplification over time. In the stellar wind scenario, the expression is $\alpha = \frac{45 - p + 32a + 4b(5 - p)}{16}$, which gives $\alpha \approx 4.07$ for the same parameters, again steeper than the constant case. Therefore, temporal evolution of microphysical parameters can lead to faster decays in the light curves, particularly in wind-like environments. A similar trend is observed in the thin-shell regime. For ISM, with $a = b = 0$, $\alpha \approx 1.82$ for $p = 1.95$, while with $a = b = 0.5$, $\alpha \approx 3.21$. In the wind case, the values change from $\alpha \approx 2.74$ to $\alpha \approx 4.13$. This behavior illustrates that temporal decrease of microphysical parameters steepens the SSC light curves across all regimes and media, which enables the model to reproduce features of very rapid decays, depending on the specific values of $a$ and $b$.\\
It should be noted that those bursts from the 2FLGC \citep{2019ApJ...878...52A} with spectral and temporal indices of $\alpha_{\rm L}\gtrsim 1.5$ and $\Gamma_{\rm L}\approx 2$, can hardly be described with the CR of the synchrotron FS scenario for $2<p$, and therefore a hard spectral index is required. Although it is interesting to consider a phenomenological scenario with the superposition of synchrotron and SSC emission from RS and FS regions, it is outside the scope of this paper.}

\subsection{Atypical light curve behavior  from RS}

\subsubsection{The Plateau phase}
Typical X-ray afterglows exhibit PL segments $\propto t^{-\alpha}$, including a plateau phase with a shallower decay ($-0.1\lesssim\alpha\lesssim$0.7).  The plateau phase is usually understood through continuous energy injection from a spinning magnetized neutron star \citep[e.g., see][]{zhang2001,2007ApJ...665..599T,dallosso2011,rowlinson2013,rowlinson14,rea15,BeniaminiandMochkovitch2017,Toma2007,2018ApJ...857...95M,Stratta2018,2021ApJ...918...12F}. or a central black hole \citep[e.g., see][]{2005ApJ...635L.133B,  2005ApJ...630L.113K, 2006Sci...311.1127D, 2006ApJ...636L..29P, 2006MNRAS.370L..61P,  2007ApJ...671.1903C, 2017MNRAS.464.4399D, 2019ApJ...872..118B, 2019ApJ...887..254B}, into the circumburst medium.  \cite{2021ApJS..255...13D} examined the occurrence of late-time flattening in LAT data in 2FLGC, providing evidence for the existence of plateaus in light curves.   The authors suggested that the same phenomenological model necessary for interpreting X-ray plateaus could also be applied to characterize GRBs 090510, 090902B, and 160509A, which displayed GeV photons.

Table \ref{table2:a_values} shows the temporal evolution of the SSC flux $F^{\rm ssc}_{\nu,r}\propto t^{-a}$ with $a=a(b,\alpha, p)$ a function of the magnetic parameter $b$, the temporal decay index ($\alpha$) and the spectral index ($\beta$), when the RS lies in the thick- and thin-shell regime for the homogeneous and stellar-wind medium.  This shows the SSC flux for each cooling condition for two ranges of electron populations $1<p<2$ and $2<p$. For example, the SSC flux in the thick case and ISM and evolves as $\frac{-1 - 21p - 4b(p+1) + 16\alpha(p-1)}{32}$ for the cooling conditions $ \nu^{\rm ssc}_{\rm m,r} < \nu < \nu^{\rm ssc}_{\rm cut,r}$ and $1<p <2$. Given $p=1.9$ and $\alpha=0$ (plateau phase), the microphysical parameters satisfy the relation $a=0.94 + 0.385b$, which can be extended for each condition. Therefore, with appropriate values of $a$ and $b$, the SSC emission from the RS could account for the plateau phase in several bursts regardless of the circunstellar medium and regime.  It is important to mention that a late-time flattening in the LAT light curve as exhibited in GRB 090510, 090902B, and 160509A could be interpreted under the thin-shell regime and the cooling condition $ \nu^{\rm ssc}_{\rm m,r} < \nu < \nu^{\rm ssc}_{\rm cut,r}$.
 
\subsubsection{Bursts with temporal index evolution from soft to hard}

The evolution of the temporal index from soft to hard exhibited in distinct multiwavelength light curves has been interpreted in contexts of  energy injection \citep{2006ApJ...641L...5E, 2006MNRAS.370.1946G, 2006MNRAS.366.1357P}, stratified density environment \citep{2006ApJ...640L.139T},  structured jets with different angular distributions \citep{2006ApJ...640L.139T, 2007ApJ...656L..57J, 2017MNRAS.472L..94H},  ejecta with a wide range of Lorentz factors \citep{1998ApJ...496L...1R,2000ApJ...532..286K, 2000ApJ...535L..33S, 2002ApJ...566..712Z, 2015ApJ...806..205D, 2015ApJ...814....1L, 2019ApJ...871..200F}, off-axis emission \citep{2023ApJ...958..126F}, microphysical parameter variations in the FS region \citep{2006MNRAS.369..197F, 2006A&A...458....7I, 2020ApJ...905..112F}, magnetic dissipation \citep{2018ApJ...857...95M}, dust-scattering-driven emission models \citep{2007ApJ...660.1319S} and gravitational microlensing \citep{2021ApJ...921L..30V}.  For instance, \cite{2015ApJ...814....1L} analyzed GRB 100418A, GRB 100901A, GRB 120326A, and GRB 120404A, concluding that the predominant portion of the kinetic energy of the relativistic jet in each burst was transported by slow-moving ejecta. This observation suggests a correlation between the injection rates and the distribution of the Lorentz factor. \\  

Tables \ref{table:lc_ssc} and \ref{table:lc_syn} show that the SSC and synchrotron flux under the cooling conditions $ \nu^{\rm ssc}_{\rm m,r} < \nu < \nu^{\rm ssc}_{\rm cut,r}$ decreases with time for $a=0$ and $b=0$ regardless of whether the RS lies in the thick- and thin-shell regime and the spectral index $1 < p < 2$ and $2<p$. Therefore, if the magnetic microphysical parameter evolves as $b<0$, the temporal index would evolve from soft to hard. A similar situation would occur if the microphysical parameter given to accelerate electrons evolves with $a<0$.

\subsubsection{Explanation for the faster decay than high-latitude emission}

After the peak flux ($t>t_{\rm x}$), when the SSC break exceeds the \textit{Fermi}-LAT band ($\nu^{\rm ssc}_{\rm cut,r} < \nu_{\rm LAT}$), gamma-ray emission from the RS region ceases in the observer frame. However, sudden vanishing is prevented by an angular-time delay (the emission generated at significant angles relative to the jet axis). Therefore, the gamma-ray flux evolves as $F^{\rm ssc}_{\nu,r}\propto t^{-\alpha}\nu^{-\beta}$ with $\alpha=\beta+2$. It is worth noting that afterglow emission with a decay index larger than 2 is difficult to reconcile with synchrotron FS models.

Several physical mechanisms have been used to elucidate the fast decays recorded with \textit{Swift} from gamma-ray bursts during the late prompt and early afterglow phases, which may function in both internal and external shock situations \citep[e.g., see][]{2006ApJ...647.1213O, 2006ApJ...642..354Z, 2006ApJ...646..351L, 2006MNRAS.366.1357P, 2006ApJ...652..482P, 2007MNRAS.377.1464N, 2007ApJ...666.1002Z, 2012A&A...539A...3B, ber13, lu14,li18}. \cite{2006ApJ...647.1213O} presented a temporal and spectral analysis of a sample of 40 X-ray light curves during the early afterglow, which exhibited a steep decay at hundreds of seconds.
They showed that during the faster decay period, the CRs cannot satisfy the theoretical predictions of high-latitude emission ($\beta\sim 1$ and $\alpha\gg3$).  \cite{2012A&A...539A...3B} presented a similar spectral and temporal analysis with a larger sample of 64 X-ray light curves. Similarly, they showed that the temporal and spectral indices of a large fraction of this sample do not fulfill the CRs ($\alpha=\beta+2$) during steep decay.   \cite{lu14} presented a sample of 214 X-ray light curves with established redshifts, categorizing them into four classifications (Gold, Silver, Aluminum, and others). The gold sample exhibited a rapid decline $F^{\rm ssc}_{\nu,r}\propto t^{-\alpha}$ with $3.2\lesssim \alpha\lesssim 8.7$. A similar analysis and conclusions with 101 X-ray light curves classified in Gold, Silver and Bronze  was performed by \cite{li18}.

\subsubsection{Bursts with temporal and spectral indices: $\alpha_{\rm L} > 2$ and $\beta_{\rm L}<1$}

\paragraph{Bursts with spectral index $\beta_{\rm L}\approx 0$.}

Table \ref{tab:beta_eq0} enumerates a selection of bursts included in 2FLGC, exhibiting a spectral index of $\beta_L\pm \delta \beta_L\approx 0$. This sample cannot be well characterized by the synchrotron FS scenario, as the \textit{Fermi}-LAT band must reside below the two spectral breaks ($\nu_{\rm LAT} <\{\nu^{\rm syn}_{\rm m, f}, \nu^{\rm syn}_{\rm c, f}\}$). For instance,  requiring the CRs of the synchrotron FS scenario, \cite{2019ApJ...883..134T} analyzed a large GRB sample reported in 2FLGC and found that they were in the cooling condition $\nu^{\rm syn}_{\rm m, f}<\nu_{\rm LAT}<\nu^{\rm syn}_{\rm c, f}$.  The CRs derived and listed in Tables \ref{table:lc_ssc} and \ref{table:cr_ssc} display that the GRB sample with $\beta_L\pm \delta \beta_L\approx 0$ might be interpreted in the SSC scenario lying the cooling condition ($\nu_{\rm LAT} <\{\nu^{\rm syn}_{\rm m, f}, \nu^{\rm syn}_{\rm c, f}\}$). 
 
Fig.~\ref{fig:SPThick} explores the parameter space of the parameters $a$ and $b$ for the SSC RS model evolving in a thick-shell stellar-wind environment. The panels demonstrate how combinations of these indices, which govern the temporal evolution of microphysical parameters, produce spectral indices $\beta\sim0$. In general, we observe that in the thick-shell regime, achieving a null spectral index requires a large bulk Lorentz factor, as the majority of panels showcase $\Gamma\sim10^3$. We can also note that a significant decay in the magnetic field energy fraction ($b\sim2.0$) is preferred as we have a larger amount of points in the lower panels. We also observe that, while there is little variation in the values of the Lorentz factor, there is a large spread over several orders of magnitude in both the energy and the density parameter, which is especially apparent in the bottom-right panel. In fact, this particular panel is the one that corresponds to the most extreme choice of $(a,b)=(0.5,2.0)$ and is also able to account for the majority of viable points.

Fig.~\ref{fig:SPThick}, analogously, examines the parameter space for the SSC RS model in the thick-shell regime. Here, $\beta\sim0$ is achieved with the smaller Lorentz factor ($\Gamma\sim10^2$). We also consider less extreme parameter combinations. In thin shells, the magnetic field decays more gradually ($b\sim0.5,1.5$), allowing SSC emission to dominate at GeV energies without overwhelming synchrotron contributions. The Figure highlights how smaller temporal variations in $\varepsilon_e$ can flatten the spectrum if paired with a declining $\varepsilon_B$. This contrasts with the thick-shell case, where stronger parameter evolution is required to counteract the shock’s prolonged activity. Another contrasting point with the thick-shell case is that a larger decay in the electron energy fraction ($a=0.5$) is preferred, as we have a larger number of points in the right panels. In fact, for a smaller value of this index ($a=0.1$), we can only account for, at most, two points in the upper left panel, which hints towards the predilection of the thin-shell for rapid decay in the electron energy fraction. Furthermore, while there was a substantial spread in the energy and density parameters in the thick-shell regime, Fig.~\ref{fig:SPThick} shows that the most viable parameters lie in the same order of magnitude with energy around $10^{53}-10^{54}~\rm erg$ and density in $10^{-4}-10^{-3}$.

\subsection{The Evolution of the magnetic parameter}

Based on the results of numerical simulations of mildly magnetized outflows during the very early afterglow, \cite{2004A&A...424..477F} showed that the shock crossing time would be conditional on the value of the magnetization parameter.   Depending on the degree of magnetization of the ejecta, which is defined as the ratio of Poynting to matter energy flux $\sigma={B'}^2_{\rm r}/(4\pi m_pc^2 n\,\Gamma^2)\simeq 8\varepsilon_{\rm B_r}$,\footnote{The term ${\rm m_p}$ corresponds to the proton mass.} the shock crossing time could be somewhat shorter than $T_{90}$.

For $\sigma\ll1$, particle acceleration in the RS is inefficient
and the RS is expected to be weak \citep{2008ApJ...682L...5S, 2011ApJ...726...75S, 2007ApJ...655..973K}, and for $\sigma \gg 1$, the RS is suppressed \citep{2004A&A...424..477F}.  The magnetization parameter in the range of $10^{-1.5}\lesssim \sigma\lesssim 1$ leads to a flow with moderate magnetization, and therefore a bright $\gamma$-ray flare might be expected \citep{2005ApJ...628..315Z, 2004A&A...424..477F, 2007Sci...315.1822M}. Several studies have shown that Poynting flux-dominated models with arbitrary magnetization may explain the high-energy emissions found in the most luminous LAT-detected bursts \citep{2011ApJ...726...90Z, 2014NatPh..10..351U}.  A number within the range of $10^{-1.5}\lesssim \sigma\lesssim 1$ suggests that the ejecta must have spent a considerable amount of Poynting flux during the prompt emission phase, with the internal collision-induced magnetic reconnection and turbulence (ICMART) event being the most plausible mechanism to explain this phenomenon.


\subsection{Maximum synchrotron photons}

During the RS, electrons are accelerated to relativistic energies within the stellar wind and the constant-density medium via Fermi processes. The timescale for an electron with the Lorentz factor $\gamma_e$ to traverse from one side of the shock front to the opposite side is approximately equal to the Larmor scale, expressed as $t'_{\rm L}=\gamma_em_e c/(q_e B')$.   The SSC limit of the RS can be determined by equating the accelerating time scale ($t'_{\rm acc}=\xi 2\pi t'_{\rm L}$) with the cooling time scale ($t'_{\rm syn}=6\pi m_e c/(\sigma_T\gamma_e B'^2)$), where $\xi$ represents the Bohm parameter, $\sigma_T$ denotes the Thomson cross section, ${\rm m_e}$ is the electron mass, $c$ is the speed of light and ${\rm q_e}$ is the elementary charge. The maximum Lorentz factor achievable by electrons becomes $\gamma_{\rm max, r}=\left(3q_e/\xi\sigma_T B'_{\rm r}\right)^{\frac12}$, resulting in the maximum energy emitted by the synchrotron mechanism during both the thick and thin regimes being

{\small
\begin{eqnarray}\label{thick_max}
h \nu^{\rm syn}_{\rm max, r} &=& \begin{cases}
4.9\,{\rm GeV} \left(\frac{1+z}{1.3}\right)^{-1}\, n^{-\frac18}_{-1} E^{\frac{1}{8}}_{53.5} \,t^{-\frac{7}{16}}_{\rm 1.5}\hspace{3.4cm}{\rm for} \hspace{0.1cm} {\rm  k=0 }\cr
2.3\,{\rm GeV} \left(\frac{1+z}{1.3}\right)^{-1}\,A^{-\frac14}_{W,-1} E^{\frac{1}{4}}_{53.5} \,t^{-\frac{3}{8}}_{\rm 1.5}\hspace{3.15cm}{\rm for} \hspace{0.1cm} {\rm  k=2 }\cr
\end{cases},
\eary
}%
and
{\small
\begin{eqnarray}\label{thin_max}
h \nu^{\rm syn}_{\rm max, r} &=& \begin{cases}
1.9\,{\rm GeV} \left(\frac{1+z}{1.3}\right)^{-\frac43}\, n^{-\frac19}_{-1} E^{\frac{1}{9}}_{52.2} \,\Gamma^{\frac19}_{2} \,t^{-\frac{1}{3}}_{\rm 2.6}\hspace{3.2cm}{\rm for} \hspace{0.1cm} {\rm  k=0 }\cr
0.6\,{\rm GeV} \left(\frac{1+z}{1.3}\right)^{-\frac43}\,A^{-\frac13}_{W,-1} E^{\frac{1}{3}}_{52.2} \,\Gamma^{-\frac13}_{2} \,t^{-\frac{1}{3}}_{\rm 2.6}\hspace{2.65cm}{\rm for} \hspace{0.1cm} {\rm  k=2 }\,,\cr
\end{cases},
\eary
}%

respectively.

{  Figure~\ref{fig:Emax} presents with blue points the observed distribution of photons with energies above 100~MeV for GRB 130427A, selected based on a likelihood greater than 90\%. These high-energy photon observations are also compared against the maximum synchrotron photon energies in the thin-shell regime predicted from the RS (see Eqs.~\ref{thick_max} and \ref{thin_max}) with different colored curves. The theoretical curves are shown for RSs evolving in a constant-density medium with densities of $n = 1\,\mathrm{cm}^{-3}$ (upper left panel) and $10^{-2}\,\mathrm{cm}^{-3}$ (upper right panel), as well as in a stellar wind environment with density parameters $A_W = 1$ (lower left panel) and $10^{-2}$ (lower right panel). The modeled values assume an optimistic isotropic energy of $E = 10^{53}\,\mathrm{erg}$ and a bulk Lorentz factor of $\Gamma = 600$. The main conclusion we draw from Figure ~\ref{fig:Emax} is that the RS synchrotron model cannot fully account for all high-energy photons. We also note that, as hinted by the prefactors of Eqs.~\eqref{thick_max} and~\eqref{thin_max}, the maximum synchrotron photon energy is smaller in wind-like environments in comparison to that produced in constant-density media. This would lead to a preference for constant circumburst density in order to explain the observations. However, as the upper panels show, it is not possible to reconcile data with analytical models unless some extreme parameters are chosen for the model. For example, in the $k=0$ case, we would need the combination $(\Gamma E n^{-1})^{1/9}$ to be $20$ times greater than the benchmark parameters chosen to explain the high energy photons. Keeping $\Gamma$ and $E$ of the same order, this would imply a value of $n\sim10^{-11}~\mathrm{cm}^{-3}$ to match the observations, which is highly unrealistic. A similar analysis for the wind scenario leads to the condition $(\Gamma^{-1} E A_W^{-1})^{1/3}>100$, achievable for $A_W\lesssim10^{-6}$ which, while more reasonable than in the constant ISM case, is still unreasonable. This leads to the proposal that a new mechanism, such as SSC from the RS region, should be considered to explain these high-energy observations.}


Figures~\ref{fig_thick:max_en} and~\ref{fig_thin:max_en} present with blue points the observed distribution of photons with energies above 100~MeV for each GRB in the sample, selected based on a likelihood greater than 90\%. These high-energy photon observations are also compared against the maximum synchrotron photon energies predicted from the RS (see Eqs.~\ref{thick_max} and \ref{thin_max}) with different colored curves. Figure~\ref{fig_thick:max_en} shows the results for the thick-shell regime, while Figure~\ref{fig_thin:max_en} focuses on the thin-shell case. In both Figures, theoretical curves are shown for RSs evolving in a constant-density medium with densities of $n = 1\,\mathrm{cm}^{-3}$ and $10^{-2}\,\mathrm{cm}^{-3}$, as well as in a stellar wind environment with density parameters $A_W = 1$ and $10^{-2}$. The modeled values assume an optimistic isotropic energy of $E = 10^{53}\,\mathrm{erg}$ and a bulk Lorentz factor of $\Gamma = 600$. The conclusion we get from Figures \ref{fig_thick:max_en} and~\ref{fig_thin:max_en} is that the RS synchrotron model cannot fully account for high-energy photons in most GRBs. In fact, only in the case of GRB 091127, 110731A and 131108A, could all photons be explained by a synchrotron from the RS region in either the thick- or thin-shell regime. We also note that the synchrotron curves tend to fall below the detected photon energies more often in wind-like environments than in constant-density media. This would lead to a preference for constant circumburst density in order to explain the observations. However, even in the best-case scenario, corresponding to the solid red curves, it is not possible to reconcile data with analytical models. This leads to the proposal that a new mechanism, such as SSC from the RS region, should be considered to explain these high-energy observations.

The maximum energy of synchrotron photons can increase under certain conditions, requiring the magnetic field behind the shock front to diminish over a scale much shorter than the distance high-energy electrons travel before losing half their initial energy. In this instance, given the highest ($B_{\rm w}\sim 1\,{\rm G}$) and smallest ($B_{\rm 0}\sim B_{\rm ISM,-5}$) magnetic fields, and the length of the scale over which the field progressively loses its intensities $L_p\leq x$ \citep[see][]{2012MNRAS.427L..40K},  the behavior of the magnetic field becomes $B(x)=B_{\rm w}(x/L_p)^{-\eta}+ B_0$.  The rate of energy loss of an electron due to synchrotron radiation while traversing behind the shock front can be estimated as {\rm $\frac{d(\gamma_e m_e c^2)}{dt}=-\frac{\gamma_e\sigma_T c}{6\pi}\left[ B_{\rm w}\left(\frac{x}{L_p}\right)^{-\eta}+ B_0\right]^2$}.   In this case, the maximum electron Lorentz factor is $\gamma_{\rm max, r}=\left(3q_e/\xi\sigma_T\,\right)^{\frac12}\, {B}_0^{-\frac12}$, and therefore the Eqs. \ref{thick_max} and \ref{thin_max} are modified by a factor $\frac{B_w}{B_0}$.

{ \cite{1999ApJ...520..641S} considered the forward and reverse external shocks to explore the expected early afterglow at different multiwavelengths.  They found that the reverse shock, at its highest flux, possesses energy similar to that of the GRB itself; however, it exhibits a far lower temperature than the forward shock, resulting in radiation at greatly reduced frequencies. The reverse shock predominates the first optical emission, and an optical flash above the $15^{\rm th}$ magnitude is anticipated alongside the forward shock peak in X-rays or gamma-rays.
Therefore, the expected radiation power of the synchrotron mechanism at RS is very low at MeV-GeV levels, so synchrotron emission can hardly contribute to the emission detected by Fermi - LAT.}

\section{LAT-detected bursts with very-high energy emission}

\subsection{A sample of LAT GRBs}

\subsubsection{GRB 180720B}

GRB 180720B was initially detected by the Gamma-Burst Monitor (GBM) instrument aboard \textit{Fermi} satellite on 20 July 2018 at $T_0=$ 14:21:39.65 Universal Time (UT) \citep{Roberts_2018GCN.22981....1B} and was immediately observed by the Burst Alert Telescope (BAT) onboard \textit{Swift} satellite \citep{Siegel_2018GCN.22981....1B}.   The estimated isotropic energy output in the 50-300 keV band is $E_{\rm \gamma} = (6.0 \pm 0.1) \times 10^{53} \, \rm erg$ for a prompt episode duration of $T_{90}=48.9 \pm 0.4$ seconds. This burst was classified as very luminous due to the prompt episode observed by the GBM, ranking among the top seven brightest occurrences recorded to date \citep{2020A&A...636A..55R}.  The X-ray Telescope (XRT) onboard 
\textit{Swift} satellite detected a luminous afterglow 11 hours post-trigger in the 0.3-10 keV range, observable for up to almost 30 days. The redshift was measured at $z=0.654$ by identifying multiple absorption features, such as Ca II, Mg II, and Fe II \citep{Vreeswijk_2018GCN.22996....1V}.   GRB 180720B was monitored by \textit{Fermi}-LAT between 19 and 700 s, after the trigger time \citep{Bissaldi_2018GCN.22980....1B}. This instrument detected 130 photons with an energy above 100 MeV, being the first and highest photon of 101 MeV and 4.9 GeV at 19.4 s and 142.43 s after the trigger time \citep{2019ApJ...885...29F}.   GRB 180720B burst was detected in very-high-energies (VHEs) by the High Energy Stereoscopic System (HESS) telescopes \citep{Abdalla_2019Natur.575..464A}.  The gamma-ray observations in the energy range of 100 - 440 GeV started 10.1 hours after the trigger time and continued for two hours, uncovering a gamma-ray excess linked to a point-like yet nonsteady source located at $RA=00^{\rm h} 0.2^{\rm min} 0.7.6^{\rm sec}$ and ${\rm Dec} = 0.2^\circ 56' 06''$ (J2000).

\subsubsection{GRB 190114C}

GRB 190114C was initially seen by \textit{Swift}-BAT and \textit{Fermi}-GBM on 14 January 2019 at 20:57:03 UT, classifying it as a long-GRB type with a duration of around 116 s and about 362 s as recorded by \textit{Swift}-BAT and \textit{Fermi}-GBM, respectively \citep{Gropp_2019GCN.23688....1G, Hamburg_2019GCN.23707....1B}.   \cite{2019GCN.23692....1U} identified a source in the Pan-STARRS archive within the vicinity of GRB 190114C, indicating that this source may be the potential host galaxy of this burst. This was validated by NOT, which calculated a redshift of $z=0.4245 \pm 0.0005$ \citep{2019GCN.23695....1S, 2019GCN.23708....1C}. 
The isotropic release of the gamma-ray energy in the 10-1000 keV band during $T_{90}$ was $E_{\rm \gamma} = 2.4 \pm 0.05 \times 10^{53} \, \rm erg$ \citep{MAGIC_2019Natur.575..455M}.  GRB 190114C was monitored by \textit{Fermi}-LAT for almost 100 s, after the trigger time.  This instrument detected 238 photons with an energy above 100 MeV, being the first and the highest photon of 571 MeV and 21.42 GeV at 2.7 s and 21 s after the trigger time \citep{2019ApJ...879L..26F}.  VHE emission in the energy range of 0.2 to 1 TeV began to be detected about one minute after the burst, exceeding 50$\sigma$, revealing an emission component distinct from the synchrotron mechanism. Analysis of muti-wavelength afterglow observation exhibiting a double-bump feature in the spectral energy distribution indicated that the SSC process was the most suitable process responsible for this VHE emission \citep{MAGIC_2019Natur.575..459M}.

\subsubsection{GRB 221009A}

The \textit{Fermi}-GBM instrument detected an extremely bright burst GRB 221009A at 13:16:59 UT on October 9, 2022 \citep{Lesage_2023arXiv230314172L} with coordinates ${\rm RA}=19^{\rm h} 13^{\rm min} 3.48^{\rm sec}$ and ${\rm Dec} = 19^\circ 46' 24.6''$ (J2000). For more than Hundreds of seconds, a variety of instruments also detected this burst in distinct energy ranges \citep{2023ApJ...949L...7F,Williams+23Swift_boat, Dichiara_2022GCN.32632....1D}, including a TeV range by the Large High Altitude Air Shower Observatory \citep[LHAASO;][]{Huang+22LHASO_gcn32677, Cao+23LHAASO}. In fact, this event
emitted more than 5000 photons for almost 2000 s with energies up to  $\sim$ 15 TeV.   The \textit{Fermi}-LAT instrument detected this burst, reporting the highest photon ever observed of 99.3 GeV, which arrived 240 s after the tigger time \citep{Pillera_2022GCN.32658....1P}. This burst exhibited multiple episodes that are described in \cite{Lesage_2023arXiv230314172L}.  Given spectroscopy at optical wavelengths (CaII, CaI and NaID lines), the estimated redshift was $z=0.15095\pm 0.00005$ \citep{deugarte+2222109a_redshift_gcn32648}, determining a luminosity distance of $d_L=2.2\times10^{27}\, {\rm cm}$. \cite{2023ApJ...947L..14Z} showed that the alternative model of proton synchrotron radiation from the RS region could contribute significantly to VHE gamma rays observed in GRB 221009A. \cite{2023ApJ...952L..42L} showed that the variation in the Lorentz factor from the prompt episode to the afterglow is suggestive of a relativistic RS in the thick shell regime  which is consistent with the deceleration time not exceeding
the prompt GRB duration. Additionally, the authors claimed that the abundance of photons provides a stronger indication that the RS is present during the early afterglow.

\subsubsection{Antecedents of RS emission}

\paragraph{GRB 180720B.}

\cite{2024MNRAS.527.1674F} considered the SSC emission from RS and obtained the light curves for a stellar-wind environment. In particular, the LAT observations of GRB 180720B were described with the SSC process evolving in the thick-shell regime. On the other hand, \cite{2024NatAs...8..134A} reported the detection of early photopolarimetric observations, which was interpreted in the RS synchrotron scenario. In addition, the authors required the optical photons as seeds to describe the LAT observation via the SSC model.

\paragraph{GRB 190114C.}  \cite{2019ApJ...879L..26F} analyzed and modeled the \textit{Fermi}-LAT and -GBM data with the function $F(t)\propto (t - t_0)^{\alpha_{\rm L}}\,e^{-\frac{\tau}{t-t_0}}$,\footnote{$\tau$ corresponds to the timescale of flux rise, $t_0$ the initial time and $\alpha_{\rm L}$ the temporal decay index} identifying short- and long-extended-component emission.  Given the best-fit values of the time scale of flux rise and the temporal decay index, the short-component emission was associated with an SSC model from the RS region, and the long-component emission was associated with the standard synchrotron model from the FS region.

\paragraph{GRB 190829A.} \cite{2022MNRAS.512.2337D} modeled the afterglow of GRB 190829A with a two-component scenario consisting of FS and RS emissions. The RS component dominates the early optical, UV, and X-ray light curves while also contributing to the later radio observations. The spectral evolution suggests that the magnetic field strength in the RS region is significantly higher than in the FS, with an estimated $B_{RS}/B_{FS}\sim200$. However, the observed evolution of the spectrum deviates from standard RS models due to a much shallower decline ($\sim t^{-0.4}$ in contrast to the model $\sim t^{-1}$). The explanation would require modifications such as a low initial Lorentz factor and continued energy injection.  Similarly, \cite{2022ApJ...931L..19S} incorporated a RS component into their FS plus RS model to describe the multiwavelength afterglow of GRB 190829A. By including this component, they were able to explain key features of the observed emission, such as the early-time peaks in X-ray and optical bands. To achieve a self-consistent fit, the authors modeled the RS dynamics by allowing for a rapid decay of the magnetic field in the shocked ejecta after the RS crossed the shell. They found that if the magnetic energy density remained constant, the RS emission would overpredict the late-time radio observations.

\paragraph{GRB 221009A.}  \cite{2023ApJ...947L..14Z} showed that proton synchrotron radiation from the RS region could contribute significantly to VHE gamma rays observed in GRB 221009A. \cite{2023ApJ...952L..42L} showed that the variation in the Lorentz factor from the prompt episode to the afterglow is suggestive of a relativistic RS in the thick shell regime  which is consistent with the deceleration time not exceeding
the prompt GRB duration. Additionally, the authors claimed that the abundance of photons provides a stronger indication that the RS is present during the early afterglow. { However, alternative interpretations invoking electron synchrotron emission from the FS have been shown to naturally reproduce the observed GeV–TeV signals without requiring extreme parameters. In particular,~\cite{2025arXiv250317765G} modeled the spreading of an ultra-narrow relativistic jet and its multi-wavelength emission, demonstrating that FS electrons can explain the TeV/GeV emission with more conventional microphysical values. Similar FS-based scenarios have also been discussed by other groups~\citep{2023ApJ...947...53R,2023SciA....9I1405O,2024ApJ...966..141Z} including the possibility of a top-hat jet propagating into a wind-like medium or two-component jet models as an explanation for the VHE emission of GRB 221009A.}

\subsubsection{Analysis}

Figure \ref{fig_vhe:max_en} displays the maximum energy emitted by synchrotron RS emission and all photons detected by the LAT instrument from GRB 190114C and GRB 221009A that were not included in the 2FLGC, but detected in the TeV range by the MAGIC and LHAASO gamma-ray telescopes.  For GRB 190114C, we consider that the RS evolves in a stellar-wind environment, whereas for GRB 221009A a constant-density medium.\footnote{The values used are $\Gamma=600$, $E=10^{54}\,{\rm erg}$, $A_W=1$ and $z=0.42$ GRB 190114C, and $\Gamma=600$, $E=10^{55}\,{\rm erg}$, $n=1\,{\rm cm^{-3}}$ and $z=0.1$ for GRB 221009A.}
For GRB 190114C, we note that the synchrotron emission can only explain two photons, and for GRB 221009A half of these photons. Therefore, SSC emission is needed if we want to interpret the rest of high-energy photons.

Table \ref{table_sample_he} exhibits the values of the spectral and temporal PL indices reported by \textit{Fermi}-LAT \citep{2019Natur.575..459A, 2019ApJ...883..162F, 2019ApJ...885...29F, Pillera_2022GCN.32658....1P} of GRB 180720B, 190114C and 221009A which were detected in the TeV energy range.  To further explore the implications of these indices, Figure~\ref{fig:cr_bur_5} illustrates the possible values of the parameters $a$ and $b$ that can reproduce the LAT observations under the cooling condition $\nu_{\rm m,r}^{\rm ssc}<\nu<\nu_{\rm cut,r}^{\rm ssc}$, assuming a RS that propagates in a thick or thin shell and considering both an ISM and a wind-like environment.  { We want to clarify that we intend to explain only the observed temporal and spectral indices, rather than the lightcurves}. The results indicate that a nonzero choice of $(a,b)$ is necessary to explain the observations.   In addition, a clear inverse proportion between $a$ and $b$ is observed, suggesting a trade-off between these two parameters in the data modeling. While $a$ generally varies within a relatively narrow range, $a\in(-2,1)$, the variation in $b$ is more significant, spanning $b\in(-5,5)$. This difference in variability suggests that the parameter $b$ is more sensitive to the physical conditions that govern the emission, while $a$ remains relatively constrained in different scenarios.

\section{Summary}\label{sec5}
In this work, we have investigated the SSC emission arising from the RS region in GRB afterglows, incorporating the temporal evolution of the microphysical parameters $\epsilon_{e}$ and $\epsilon_{B}$ across both homogeneous and stratified (stellar wind) media. We have derived CRs for the SSC process in the thin- and thick-shell regimes, and considered two distinct ranges of electron spectral indices ($1 < p < 2$ and $ 2 < p$).

We have carried out an analysis between these CRs and the spectral and temporal indices of bursts reported in 2FLGC. We have analyzed LAT-detected bursts documented in 2FLGC and used the MCMC method to determine the best-fit values ($a$ and $b$) that adhere to the CRs under various cooling conditions for the thick- and thin-shell regime.   The quantity and proportion of bursts subsequent to each cooling condition for the SSC RS model progressing in a homogeneous and stellar-wind medium are detailed as follows. Our MCMC simulations reveal that the temporal evolution of the microphysical parameters is not only possible but necessary to match the observed temporal and spectral indices of \textit{Fermi}-LAT GRBs.  The SSC scenario scattered in the RS region and the characteristics of its spectral and temporal indices can be effectively described, given suitable variations in the microphysical parameters, applicable up to the SSC limit. We have shown a sample of 38 bursts with afterglow observations detected in X-rays, optical bands, and radio wavelengths, which required a magnetic microphysical parameter with a value less than $<10^{-7}\,$.  Furthermore, \cite{2019ApJ...883..134T} analyzed the temporal and spectral indices of 59 bursts reported in 2FLGC that required CRs of the standard synchrotron scenario, and found that most of them were well described with a low value of the magnetic microphysical parameter $\varepsilon_B<10^{-7}$. \cite{2009MNRAS.400L..75K, 2010MNRAS.409..226K} adequately described the LAT observations of GRB 080916C, 090510, 090902B using the standard synchrotron afterglow model and an inferred microphysical value of $\epsilon_{\rm B}\sim 10^{-7}$. 

The CRs derived from this evolving SSC model successfully account for up to 32 GRBs in our sample (representing 37.21\% of the total). The most favored configuration corresponds to a thin shell expanding into a stellar-wind medium during the slow-cooling phase ($\nu_{\rm m}^{\rm ssc} < \nu < \nu_{\rm cut}^{\rm ssc}$), followed by the same cooling regime in an ISM environment. Interestingly, no bursts are consistent with the regime $\nu < \nu_{\rm m}^{\rm ssc}$ across any configuration, indicating that the SSC component from the RS predominantly contributes at higher energies (GeV to TeV), and plays a minimal role at lower frequencies. Only a single GRB matched the CRs in the high-frequency regime ($\nu > \nu_{\rm cut}^{\rm ssc}$).

We have also shown that the SSC RS scenario with evolving microphysical parameters can reproduce a variety of atypical light-curve behaviors observed in LAT-detected bursts. Plateau phases ($\alpha \sim 0$) can be achieved through coordinated evolution in $\epsilon_e$ and $\epsilon_B$, while steep decays ($\alpha > 3$), difficult to explain through FS models, are naturally accommodated due to angular time-delay effects. Moreover, transitions in temporal slope from soft to hard, as well as spectral indices $\beta \sim 0$, arise naturally within the SSC framework, particularly in the thin-shell wind scenario.

{ It is worth noting that we have followed the analytical approach proposed in, e.g. \cite{2000ApJ...545..807K,
2003ApJ...597..455K, 2013NewAR..57..141G, 2020ApJ...905..112F}, which shows the evolution of the RS shocked electrons after the shock crossing. The traditional treatment invoking  $\gamma_m$ and  $\gamma_c$ could not be efficient in describing the evolution of the electron spectrum and therefore a more realistic approach of this sevolution is derived in  \cite{2025arXiv250317765G}. 
}

{ Particle-in-cell (PIC) simulations suggest that a subset of electrons may be effectively accelerated during relativistic shocks. In contrast, the remaining electrons establish a relativistic Maxwellian distribution centered at a reduced energy level. \cite{2024ApJ...971...81G} addressed the continuity equation of electrons in energy space, leading to the derivation of multiwavelength afterglows through the integration of various radiative processes. The authors found that a positive correlation exists between the temporal and spectral indices attributed to the cooling of electrons. Furthermore, they reported that thermal electrons lead to concurrent nonmonotonic changes in both the spectral and temporal indices across multiple wavelengths. Consequently, the conclusion may differ from one burst to another, as the quality of the observational data sampling may influence the robustness of the temporal and spectral indices. These results would correspond to a limitation in our SSC scenario.}

Finally, we emphasize that the inferred density profiles from our analysis provide insight into GRB progenitor environments. While ISM-like media are favored in thick-shell scenarios, stellar-wind profiles dominate in thin-shell configurations. The resulting CRs also predict measurable transitions in SSC flux between wind and ISM profiles, which may be detectable in multi-wavelength light curves.

\section*{Acknowledgements}

NF  acknowledges  financial  support  from UNAM-DGAPA-PAPIIT  through  grant  IN112525. BBK is supported by IBS under the project code IBS-R018-D3. M.G.D. acknowledges funding from the AAS Chretienne Fellowship and the MINIATURA2 grant.  AG was supported by Universidad Nacional Autónoma de México Postdoctoral Program (POSDOC).

\section*{Data Availability}

No new data were generated or analysed in support of this research.



\bibliographystyle{mnras}
\bibliography{RS_microp_mnras} 



\clearpage

\appendix

\section{Derivation of SSC scenario}

\subsection{Evolution in homogeneous environment}

\subsubsection{The Thick-shell regime}

The minimum and the cooling Lorentz factors for $1<p<2$ and $2 < p$ are

{\small
\begin{eqnarray}
\gamma_{\rm m, r} &\simeq& \begin{cases}
5.1\times 10\,\left(\frac{1+z}{1.3}\right)^{\frac{68-21p}{96(p-1)}} \tilde{g}^{\frac{1}{p-1}} \chi_{\rm e,-0.3}^{-0.3} \varepsilon^{\frac{1}{p-1}}_{\rm e_r,-1} \varepsilon^{\frac{2-p}{4(p-1)}}_{\rm B_r,-3}\,\Gamma_{2.7}^{\frac{1}{p-1}}\, n^{\frac{8-3p}{16(p-1)}}_{-1}\Delta^{\frac{68-3p}{96(p-1)}}_{11.9}\,E^{-\frac{p}{16(p-1)}}_{53.5} t^{-\frac{68-21p+96a+24b(2-p)}{96(p-1)}}_{1.5}\hspace{1cm}{\rm for} \hspace{0.1cm} { 1<p<2 }\cr
3.8\times 10^2\,\left(\frac{1+z}{1.3}\right)^{\frac{13}{48}} g \chi_{\rm e,-0.3}^{-0.3} \varepsilon_{\rm e_r,-1} \Gamma_{2.7} n^{\frac18}_{-1} \Delta^{\frac{31}{48}}_{11.9} E^{-\frac{1}{8}}_{53.5}t^{-\frac{13+48a}{48}}_{1.5}\hspace{5.65cm}{\rm for} \hspace{0.1cm} {2<p }
\end{cases}\cr
\gamma_{\rm c, r}&\simeq&9.6\times 10^{3}\,\left(\frac{1+z}{1.3}\right)^{-\frac{25}{48}}  \left(1+Y_{\rm r} \right)^{-1}\varepsilon^{-1}_{\rm B_r,-3} n^{-\frac58}_{-1}\Delta^{-\frac{19}{48}}_{11.9} E^{-\frac{3}{8}}_{53.5}\,t^{\frac{25+48b}{48}}_{1.5},
\eary
}
respectively. The characteristic and cooling spectral breaks of the synchrotron model for $1<p<2$ and $2 < p$ are

{\small
\begin{eqnarray}
h \nu^{\rm syn}_{\rm m, r} &\simeq& \begin{cases}
7.9\times 10^{-3}\,{\rm eV}\,\left(\frac{1+z}{1.3}\right)^{\frac{69-22p}{48(p-1)}} \tilde{g}^{\frac{2}{p-1}} \chi_{\rm e,-0.3}^{-2} \varepsilon^{\frac{2}{p-1}}_{\rm e_r,-1} \varepsilon^{\frac{1}{2(p-1)}}_{\rm B_r,-3}\,\Gamma_{2.7}^{\frac{2}{p-1}}\, n^{\frac{6-p}{8(p-1)}}_{-1}\Delta^{\frac{57+8p}{48(p-1)}}_{11.9}\,E^{-\frac{2-p}{8(p-1)}}_{53.5} t^{-\frac{21+26p+24(4a+b)}{48(p-1)}}_{1.5}\hspace{1cm}{\rm for} \hspace{0.1cm} { 1<p<2 }\cr
4.4\times 10^{-1}\,{\rm eV}\,\left(\frac{1+z}{1.3}\right)^{\frac{25}{48}} g^2 \chi_{\rm e,-0.3}^{-2} \varepsilon^2_{\rm e_r,-1} \varepsilon^{\frac12}_{\rm B_r,-3} \Gamma^2_{2.7} n^{\frac12}_{-1} \Delta^{\frac{73}{48}}_{11.9} t^{-\frac{73+24(4a+b)}{48}}_{1.5}\hspace{5.3cm}{\rm for} \hspace{0.1cm} {2<p }
\end{cases}\cr
h\nu^{\rm syn}_{\rm cut, r}&\simeq& 2.8\times 10^2\,{\rm eV}\,\left(\frac{1+z}{1.3}\right)^{\frac{25+72b}{48}}  \left(1+Y_{\rm r} \right)^{-2}\varepsilon^{-\frac32}_{\rm B_r,-3} n^{-1}_{-1}\Delta^{\frac{49+72b}{48}}_{11.9} E^{-\frac{1}{2}}_{53.5}\,t^{-\frac{73}{48}}_{1.5},\cr
F^{\rm syn}_{\rm max,r} &\simeq& 1.8\times 10^5\,{\rm mJy} \,\left(\frac{1+z}{1.3}\right)^{\frac{95}{48}}\chi_{\rm e,-0.3} \varepsilon^{\frac12}_{\rm B_r,-3}\,n^{\frac{1}{4}}_{-1} \, d^{-2}_{\rm z,27.7}\,\Gamma^{-1}_{2.7}\,\Delta^{\frac{11}{48}}_{11.9}\,E^{\frac{5}{4}}_{53.5}\, t^{-\frac{47+24b}{48}}_{1.5}\,,\,\,\,\,\,\,\,\,
\eary
}
respectively.

\subsubsection{The Thin-shell regime}

The minimum and the cooling Lorentz factors for $1<p<2$ and $2 < p$ are

{\small
\begin{eqnarray}
\gamma_{\rm m, r} &\simeq& \begin{cases}
4.6\times 10\,\left(\frac{1+z}{1.3}\right)^{\frac{24-7p}{35(p-1)}} \tilde{g}^{\frac{1}{p-1}} \chi_{\rm e,-0.3}^{-1} \varepsilon^{\frac{1}{p-1}}_{\rm e_r,-1} \varepsilon^{\frac{2-p}{4(p-1)}}_{\rm B_r,-3}\,\Gamma_{2}^{-\frac{174-7p}{210(p-1)}}\, n^{\frac{114-77p}{420(p-1)}}_{-1}\,E^{\frac{24-7p}{105(p-1)}}_{52.2} t^{-\frac{96-28p+140a+35b(2-p)}{140(p-1)}}_{2.6}\hspace{1.2cm}{\rm for} \hspace{0.1cm} { 1<p<2 }\cr
1.2\times 10^2\,\left(\frac{1+z}{1.3}\right)^{\frac27} g \chi_{\rm e,-0.3}^{-1} \varepsilon_{\rm e_r,-1} \Gamma^{-\frac{16}{21}}_{2} n_{-1}^{-\frac{2}{21}} E^{\frac{2}{21}}_{52.2}\,t^{-\frac{2+7a}{7}}_{2.6} \hspace{5.8cm}{\rm for} \hspace{0.1cm} {2<p }
\end{cases}\cr
\gamma_{\rm c, r}&\simeq& 1.5\times 10^4\,\left(\frac{1+z}{1.3}\right)^{-\frac{19}{35}}  \left(1+Y_{\rm r}\right)^{-1}\varepsilon^{-1}_{\rm B_r,-3} n^{-\frac{17}{35}}_{-1}\, \Gamma^{\frac{39}{35}}_{2}E^{-\frac{18}{35}}_{52.2} \,t^{\frac{19+35b}{35}}_{2.6},
\eary
}
respectively. The characteristic and cooling spectral breaks of synchrotron model for $1<p<2$ and $2 < p$ are

{\small
\begin{eqnarray}
h \nu^{\rm syn}_{\rm m, r} &\simeq& \begin{cases}
9.6\times 10^{-4}\,{\rm eV}\,\left(\frac{1+z}{1.3}\right)^{\frac{49-15p}{35(p-1)}} \tilde{g}^{\frac{2}{p-1}} \chi_{\rm e,-0.3}^{-2} \varepsilon^{\frac{2}{p-1}}_{\rm e_r,-1} \varepsilon^{\frac{1}{2(p-1)}}_{\rm B_r,-3}\,\Gamma_{2}^{-\frac{112+55p}{105(p-1)}}\, n^{\frac{77-40p}{210(p-1)}}_{-1}\,E^{\frac{2(7+10p)}{105(p-1)}}_{52.2} t^{-\frac{4(7+10p)+5(28a+7b)}{70(p-1)}}_{2.6}\hspace{1.4cm}{\rm for} \hspace{0.1cm} { 1<p<2 }\cr
6.3\times 10^{-3}\,{\rm eV}\,\left(\frac{1+z}{1.3}\right)^{\frac{19}{35}} g^2\chi_{\rm e,-0.3}^{-2} \varepsilon^2_{\rm e_r,-1} \varepsilon^{\frac12}_{\rm B_r,-3} \Gamma^{-\frac{74}{35}}_{2} n_{-1}^{-\frac{1}{70}} E^{\frac{18}{35}}_{52.2}\,t^{-\frac{108+140a+35b}{70}}_{2.6} \hspace{5.1cm}{\rm for} \hspace{0.1cm} {2<p }
\end{cases}\cr
h\nu^{\rm syn}_{\rm cut, r}&\simeq& 8.6\times 10\,{\rm eV}\,\left(\frac{1+z}{1.3}\right)^{\frac{38+105b}{70}}  \left(1+Y_{\rm r} \right)^{-2}\varepsilon^{-\frac32}_{\rm B_r,-3} n^{-\frac{283+105b}{210}}_{-1}\, \Gamma^{-\frac{4(73+105b)}{105}}_{2}E^{-\frac{32-105b}{210}}_{52.2} \,t^{-\frac{54}{35}}_{2.6},\cr
F^{\rm syn}_{\rm max,r} &\simeq& 2.8\times 10^{3}\,{\rm mJy} \,\left(\frac{1+z}{1.3}\right)^{\frac{69}{35}} \chi_{\rm e,-0.3}\varepsilon^{\frac12}_{\rm B_r,-3}\,n_{-1}^{\frac{37}{210}} \, d^{-2}_{\rm z,27.7}\,\Gamma^{-\frac{167}{105}}_{2}\, E^{\frac{139}{105}}_{52.2}\,t^{-\frac{68+35b}{70}}_{2.6}\,,\,\,\,\,\,\,\,\,
\eary
}

respectively.

\subsection{Evolution in stratified environment (stellar wind)}

\subsubsection{The Thick-shell regime}

The minimum and Lorentz factors for $1<p<2$ and $2 < p$ are

{\small
\begin{eqnarray}
\gamma_{\rm m, r} &\simeq& \begin{cases}
1.2\times 10^2\,\left(\frac{1+z}{1.3}\right)^{\frac{16-5p}{16(p-1)}} \tilde{g}^{\frac{1}{p-1}} \chi_{\rm e,-0.3}^{-1} \varepsilon^{\frac{1}{p-1}}_{\rm e_r,-1} \varepsilon^{\frac{2-p}{4(p-1)}}_{\rm B_r,-3}\,\Gamma_{2.7}^{\frac{1}{p-1}}\, A_{\rm W,-1}^{\frac{8-3p}{8(p-1)}}\Delta^{\frac{8+p}{16(p-1)}}_{11.9}\,E^{-\frac{4-p}{8(p-1)}}_{53.5} t^{-\frac{16-5p+4[4a+b(2-p)]}{16(p-1)}}_{1.5}\hspace{1.4cm}{\rm for} \hspace{0.1cm} { 1<p<2 }\cr
7.9\times 10^2\,\left(\frac{1+z}{1.3}\right)^{\frac38} g \chi_{\rm e,-0.3}^{-1} \varepsilon_{\rm e_r,-1} \Gamma_{2.7} A_{\rm W,-1}^{\frac14} \Delta^{\frac{5}{8}}_{11.9} E^{-\frac{1}{4}}_{53.5}\,t^{-\frac{3+8a}{8}}_{1.5} \hspace{5.65cm}{\rm for} \hspace{0.1cm} {2<p }
\end{cases}\cr
\gamma_{\rm c, r}&\simeq& 4.2\times 10\,\left(\frac{1+z}{1.3}\right)^{-\frac{7}{8}}  \left(1+Y_{\rm r}\right)^{-1}\varepsilon^{-1}_{\rm B_r,-3} A_{\rm W,-1}^{-\frac54}\Delta^{-\frac{1}{8}}_{11.9} E^{\frac{1}{4}}_{53.5}\,t^{\frac{7+8b}{8}}_{1.5},
\eary
}

respectively. The characteristic and cooling spectral breaks of synchrotron radiation for $1<p<2$ and $2 < p$ are  

{\small
\begin{eqnarray}
h \nu^{\rm syn}_{\rm m, r} &\simeq& \begin{cases}
3.7\times 10^{-1}\,{\rm eV}\,\left(\frac{1+z}{1.3}\right)^{\frac{15-4p}{8(p-1)}} \tilde{g}^{\frac{2}{p-1}}\chi_{\rm e,-0.3}^{-2} \varepsilon^{\frac{2}{p-1}}_{\rm e_r,-1} \varepsilon^{\frac{1}{2(p-1)}}_{\rm B_r,-3}\,\Gamma_{2.7}^{\frac{2}{p-1}}\, A_{\rm W,-1}^{\frac{6-p}{4(p-1)}}\Delta^{\frac{7+2p}{8(p-1)}}_{11.9}\,E^{-\frac{4-p}{4(p-1)}}_{53.5} t^{-\frac{7+4p+4(4a+b)}{8(p-1)}}_{1.5}\hspace{1.1cm}{\rm for} \hspace{0.1cm} { 1<p<2 }\cr
1.7\times 10\,{\rm eV}\,\left(\frac{1+z}{1.3}\right)^{\frac78} g^2 \chi_{\rm e,-0.3}^{-2} \varepsilon^2_{\rm e_r,-1} \varepsilon^{\frac12}_{\rm B_r,-3} \Gamma^2_{2.7} A_{\rm W,-1} \Delta^{\frac{11}{8}}_{11.9} E^{-\frac{1}{2}}_{53.5}\,t^{-\frac{15+4(4a+b)}{8}}_{1.5} \hspace{4.95cm}{\rm for} \hspace{0.1cm} {2<p }
\end{cases}\cr
h\nu^{\rm syn}_{\rm cut, r}&\simeq& 4.7\times 10^{-2}\,{\rm eV}\,\left(\frac{1+z}{1.3}\right)^{\frac{7+12b}{8}}  \left(1+Y_{\rm r} \right)^{-2}\varepsilon^{-\frac32}_{\rm B_r,-3} A_{\rm W,-1}^{-2}\Delta^{\frac{19+12b}{8}}_{11.9} E^{\frac{1}{2}}_{53.5}\,t^{-\frac{15}{8}}_{1.5},\cr
F^{\rm syn}_{\rm max,r} &\simeq& 1.6\times 10^{6}\,{\rm mJy} \,\left(\frac{1+z}{1.3}\right)^{\frac{17}{8}} \chi_{\rm e,-0.3}\varepsilon^{\frac12}_{\rm B_r,-3}\, A_{\rm W,-1}^{\frac{1}{2}} \, d^{-2}_{\rm z,27.7}\,\Gamma^{-1}_{2.7}\,\Delta^{\frac18}_{11.9}\, E_{53.5}\,t^{-\frac{9+4b}{8}}_{1.5}   \,,\,\,\,\,\,\,\,\,
\eary
}
respectively.

\subsubsection{The Thin-shell regime}
The minimum and the cooling Lorentz factors for $1<p<2$ and $2 < p$ are

{\small
\begin{eqnarray}
\gamma_{\rm m, r} &\simeq& \begin{cases}
1.3\,\left(\frac{1+z}{1.3}\right)^{\frac{22-7p}{21(p-1)}} \tilde{g}^{\frac{1}{p-1}} \chi_{\rm e,-0.3}^{-1} \varepsilon^{\frac{1}{p-1}}_{\rm e_r,-1} \varepsilon^{\frac{2-p}{4(p-1)}}_{\rm B_r,-3}\,\Gamma_{2}^{-\frac{50+7p}{42(p-1)}}\, A_{\rm W,-1}^{\frac{38-35p}{84(p-1)}}\,E^{\frac{2+7p}{42(p-1)}}_{52.2} t^{-\frac{4(22-7p)+84a+21(2-p)}{84(p-1)}}_{2.6}\hspace{1cm}{\rm for} \hspace{0.1cm} { 1<p<2 }\cr
3.6\,\left(\frac{1+z}{1.3}\right)^{\frac{8}{21}} g \chi_{\rm e,-0.3}^{-1} \varepsilon_{\rm e_r,-1} \Gamma^{-\frac{32}{21}}_{2} A^{-\frac{8}{21}}_{\rm W,-1} E^{\frac{8}{21}}_{52.2}\,t^{-\frac{8+21a}{21}}_{2.6} \hspace{5cm}{\rm for} \hspace{0.1cm} {2<p }
\end{cases}\cr
\gamma_{\rm c, r}&\simeq& 3.6\,{\rm eV}\,\left(\frac{1+z}{1.3}\right)^{-\frac{6}{7}}  \left(1+Y_{\rm r} \right)^{-1}\varepsilon^{-1}_{\rm B_r,-3} A_{\rm W,-1}^{-\frac{8}{7}}\Gamma^{\frac{3}{7}}_{2} E^{\frac{1}{7}}_{52.2}\,t^\frac{6+7b}{7}_{2.6},
\eary
}

respectively. The characteristic and cooling spectral breaks  of synchrotron radiation for $1<p<2$ and $2 < p$ are

{\small
\begin{eqnarray}
h \nu^{\rm syn}_{\rm m, r} &\simeq& \begin{cases}
2.2\times 10^{-5}\,{\rm eV}\,\left(\frac{1+z}{1.3}\right)^{\frac{2(7-2p)}{7(p-1)}} \tilde{g}^{\frac{2}{p-1}} \chi_{\rm e,-0.3}^{-2} \varepsilon^{\frac{2}{p-1}}_{\rm e_r,-1} \varepsilon^{\frac{1}{2(p-1)}}_{\rm B_r,-3}\,\Gamma_{2}^{-\frac{14+5p}{7(p-1)}}\, A_{\rm W,-1}^{\frac{7-6p}{14(p-1)}}\,E^{\frac{3p}{7(p-1)}}_{52.2} t^{-\frac{2(7+3p)+7(4a+b)}{14(p-1)}}_{2.6}\hspace{1.4cm}{\rm for} \hspace{0.1cm} { 1<p<2 }\cr
1.6\times 10^{-4}\,{\rm eV}\,\left(\frac{1+z}{1.3}\right)^{\frac67} g^2 \chi_{\rm e,-0.3}^{-2} \varepsilon^2_{\rm e_r,-1} \varepsilon^{\frac12}_{\rm B_r,-3} \Gamma^{-\frac{24}{7}}_{2} A^{-\frac{5}{14}}_{\rm W,-1} E^{\frac{6}{7}}_{52.2}\,t^{-\frac{26+28a+7b}{14}}_{2.6} \hspace{4.65cm}{\rm for} \hspace{0.1cm} {2<p }
\end{cases}\cr
h\nu^{\rm syn}_{\rm cut, r}&\simeq& 1.1\times 10^{-8}\,{\rm eV}\,\left(\frac{1+z}{1.3}\right)^{\frac{3(4+7b)}{14}}  \left(1+Y_{\rm r} \right)^{-2}\varepsilon^{-\frac32}_{\rm B_r,-3} A_{\rm W,-1}^{-\frac{61+21b}{14}}\Gamma^{-\frac{6(11+7b)}{7}}_{2} E^{\frac{40+21b}{14}}_{52.2}\,t^{-\frac{13}{7}}_{2.6},\cr
F^{\rm syn}_{\rm max,r} &\simeq& 8.1\times 10^{4}\,{\rm mJy} \,\left(\frac{1+z}{1.3}\right)^{\frac{44}{21}} \chi_{\rm e,-0.3}\varepsilon^{\frac12}_{\rm B_r,-3}\, A_{\rm W,-1}^{\frac{17}{42}} \, d^{-2}_{\rm z,27.7}\,\Gamma^{-\frac{29}{21}}_{2}\, E^{\frac{23}{21}}_{52.2}\,t^{-\frac{46+21b}{42}}_{2.6}   \,,\,\,\,\,\,\,\,\,
\eary
}

respectively.

\clearpage
\newpage

\begin{table}[h!]
\centering
\renewcommand{\arraystretch}{1.6}\addtolength{\tabcolsep}{-2pt}
\caption{SSC Light curves ($F^{\rm ssc}_{\rm \nu,r}\propto t^{-\alpha}\nu^{-\beta}$) of the RS model in a thick- and thin-shell case and evolving in ISM and stellar medium.}
\label{table:lc_ssc}
\begin{tabular}{cccccc}
\hline
\hline
     &               & ISM     & ISM      & wind    & wind      \\

     &               & $1<p<2$     & $2<p$       & $1<p<2$     & $2<p$       \\
\hline

 & $\beta$    & $\alpha$    & $\alpha$    & $\alpha$    & $\alpha$    \\
\hline
Thick shell &    &    &    &    &    \\
\hline
$ \nu < \nu^{\rm ssc}_{\rm m,r}$    &    $-\frac13$         &     $-\frac{106-59p+96a+24b(3-2p)}{72(p-1)}$        &      $\frac{1-8a+2b}{6}$       &       $-\frac{28-17p+16a+4b(3-2p)}{12(p-1)}$      &      $\frac{3-8a+2b}{6}$       \\
$ \nu^{\rm ssc}_{\rm m,r} < \nu < \nu^{\rm ssc}_{\rm cut,r}$    &    $\frac{p-1}{2}$           &       $\frac{5p+171+192a+24b(5-p)}{96}$      &       $\frac{99p-17+192a(p-1)+24b(p+1)}{96}$      &      $\frac{45-p+32a+4b(5-p)}{16}$       &      $\frac{21p+1+32a(p-1)+4b(p+1)}{16}$       \\

\hline
Thin shell &  &     &     &    &    \\
\hline
$ \nu < \nu^{\rm ssc}_{\rm m,r}$    &    $-\frac13$         &     $-\frac{143-75p+140a+35b(3-2p)}{105(p-1)}$        &      $\frac{1-20a+5b}{15}$       &       $-\frac{5(31-19p)+84a+21b(3-2p)}{63(p-1)}$      &      $\frac{5-12a+3b}{9}$       \\
$ \nu^{\rm ssc}_{\rm m,r} < \nu < \nu^{\rm ssc}_{\rm cut,r}$    &    $\frac{p-1}{2}$          &       $\frac{4(58+3p)+280a+35b(5-p)}{140}$      &       $\frac{4(37p-10)+280a(p-1)+35b(p+1)}{140}$      &      $\frac{10(25-p)+168a+21b(5-p)}{84}$       &      $\frac{10(11p+1)+168a(p-1)+21b(p+1)}{84}$       \\

\hline
\end{tabular}
\end{table}

\begin{table}
\renewcommand{\arraystretch}{1.6}\addtolength{\tabcolsep}{-5pt}
\caption{Syncrotron light curves ($F^{\rm syn}_{\rm \nu,r}\propto t^{-\alpha}\nu^{-\beta}$) of the RS model in a thick- and thin-shell case and evolving in ISM and stellar medium.}
\label{table:lc_syn}
\begin{tabular}{c c c  c c c}
 \hline \hline
&\hspace{0.5cm}     &\hspace{0.5cm}   ISM &\hspace{0.2cm}   ISM  & \hspace{0.5cm}   wind   & \hspace{0.5cm}  wind \\ 
&\hspace{0.5cm}     &\hspace{0.5cm} $1 < p <2$   &\hspace{0.5cm}   $2 < p$  & \hspace{0.2cm}    $1 < p <2$   & \hspace{0.5cm}  $2 < p$ \\
\hline
 &\hspace{0.5cm} $\beta$    &\hspace{0.5cm} $\alpha$   &\hspace{0.5cm}   $\alpha$  & \hspace{0.2cm}    $\alpha$   & \hspace{0.5cm}  $\alpha$    \\ \hline

Thick shell &\hspace{0.5cm}     &\hspace{0.2cm}    &\hspace{0.5cm}     & \hspace{0.5cm}       & \hspace{0.5cm}  \\ \hline 	
 $ \nu^{\rm syn}_{\rm m, r} < \nu < \nu^{\rm syn}_{\rm cut, r}$   	                & \hspace{0.5cm}  $\frac{p-1}{2}$  &\hspace{0.5cm}  $\frac{115+26p+24(4a+3b)}{96}$	                    &\hspace{0.2cm}  $\frac{45+97p+96a(p-1)}{96}$ &\hspace{0.5cm}  $\frac{25+4p+4(4a+3b)}{16}$ &\hspace{0.5cm}  $\frac{3(1+5p)+16a(p-1)+4b(p+1)}{16}$ \\ 	
 $\nu^{\rm syn}_{\rm cut,r} < \nu$   	                                 & \hspace{0.5cm}  $\frac{p}{2}$     &\hspace{0.5cm}  $\beta+2$	                    &\hspace{0.2cm}  $\beta+2$ &\hspace{0.5cm}  $\beta+2$ &\hspace{0.5cm}  $\beta+2$\\ 
\hline	
Thin shell &\hspace{0.5cm}     &\hspace{0.5cm}    &\hspace{0.5cm}     & \hspace{0.5cm}       & \hspace{0.5cm}  \\ \hline 	
 $ \nu^{\rm syn}_{\rm m, r} < \nu < \nu^{\rm syn}_{\rm cut, r}$   	                & \hspace{0.5cm}  $\frac{p-1}{2}$  &\hspace{0.5cm}  $\frac{4(41+10p)+35(4a+3b)}{140}$	                    &\hspace{0.5cm}  $\frac{4(7+27p)+140a(p-1)+35b(p+1)}{140}$ &\hspace{0.5cm}  $\frac{2(67+9p)+21(4a+3b)}{84}$ &\hspace{0.5cm}  $\frac{2(7+39p)+84a(p-1)+21b(p+1)}{84}$ \\ 	
 $\nu^{\rm syn}_{\rm cut,r} < \nu$   	                                 & \hspace{0.5cm}  $\frac{p}{2}$     &\hspace{0.5cm}  $\beta+2$	                    &\hspace{0.5cm}  $\beta+2$ &\hspace{0.5cm}  $\beta+2$ &\hspace{0.5cm}  $\beta+2$\\

%
%
\hline
\end{tabular}
\end{table}

\begin{table}
\renewcommand{\arraystretch}{1.6}\addtolength{\tabcolsep}{-5pt}
\caption{CRs of SSC of the RS model in a thick- and thin-shell case and evolving in ISM and stellar medium. The relationship between temporal and spectral index ($\alpha=\alpha(\beta)$) is shown for each cooling condition.}
\label{table:cr_ssc}
\begin{tabular}{c c c  c c c}
 \hline \hline
&\hspace{0.1cm}     &\hspace{0.3cm}   ISM &\hspace{0.1cm}   ISM  & \hspace{0.1cm}   wind   & \hspace{0.1cm}  wind \\ 
&\hspace{0.1cm}     &\hspace{0.1cm} $1 < p <2$   &\hspace{0.1cm}   $2 < p$  & \hspace{0.1cm}    $1 < p <2$   & \hspace{0.1cm}  $2 < p$ \\
\hline
 &\hspace{0.1cm} $\beta$    &\hspace{0.1cm} $\alpha(\beta)$   &\hspace{0.1cm}   $\alpha(\beta)$  & \hspace{0.1cm}    $\alpha(\beta)$   & \hspace{0.1cm}  $\alpha(\beta)$    \\ \hline

Thick shell &\hspace{0.3cm}     &\hspace{0.1cm}    &\hspace{0.1cm}     & \hspace{0.1cm}       & \hspace{0.1cm}  \\ \hline 	
 $ \nu < \nu^{\rm ssc}_{\rm m, r}$   	        & \hspace{0.1cm}  $-\frac{1}{3}$           &\hspace{0.1cm}  $\frac{[106-59p+96a+24b(3-2p)]\beta}{24(p-1)}$	                    &\hspace{0.1cm}  $-\frac{(1-8a+2b)\beta}{2}$ &\hspace{0.1cm}  $\frac{[28-17p+16a+4b(3-2p)]\beta}{4(p-1)}$ &\hspace{0.1cm}  $-\frac{(3-8a+2b)\beta}{2}$	\\	
 $ \nu^{\rm ssc}_{\rm m, r} < \nu < \nu^{\rm ssc}_{\rm cut, r}$   	                & \hspace{0.1cm}  $\frac{p-1}{2}$  &\hspace{0.1cm}  $\frac{5\beta+88+96a+24b(2-\beta)}{48}$	                    &\hspace{0.1cm}  $\frac{99\beta+41+192a\beta+24b(\beta+1)}{48}$ &\hspace{0.1cm}  $\frac{22-\beta+16a+4b(2-\beta)}{8}$ &\hspace{0.1cm}  $\frac{21\beta+11+32a\beta+4b(\beta+1)}{8}$ \\ 	
\hline	
Thin shell &\hspace{0.1cm}     &\hspace{0.1cm}    &\hspace{0.1cm}     & \hspace{0.1cm}       & \hspace{0.1cm}  \\ \hline 	
 $ \nu < \nu^{\rm ssc}_{\rm m, r}$   	        & \hspace{0.1cm}  $-\frac{1}{3}$           &\hspace{0.1cm}  $\frac{[143-75p+140a+35b(3-2p)]\beta}{35(p-1)}$	                    &\hspace{0.1cm}  $-\frac{(1-20a+5b)\beta}{5}$ &\hspace{0.1cm}  $\frac{[5(31-19p)+84a+21b(3-2p)]\beta}{21(p-1)}$ &\hspace{0.1cm}  $-\frac{(5-12a+3b)\beta}{3}$	\\	
 $ \nu^{\rm ssc}_{\rm m, r} < \nu < \nu^{\rm ssc}_{\rm cut, r}$   	                & \hspace{0.1cm}  $\frac{p-1}{2}$  &\hspace{0.1cm}  $\frac{2(61+6\beta)+140a+35b(2-\beta)}{70}$	                    &\hspace{0.1cm}  $\frac{2(74\beta+27)+280a\beta+35b(\beta+1)}{70}$ &\hspace{0.1cm}  $\frac{5(12-\beta)+42a+b(21-10\beta)}{21}$ &\hspace{0.1cm}  $\frac{10(11\beta+6)+168a\beta+21b(\beta+1)}{42}$ \\ 	

%
%
\hline
\end{tabular}
\end{table}

\begin{table}
\renewcommand{\arraystretch}{1.6}\addtolength{\tabcolsep}{-5pt}
\caption{CRs of synchrotron radiation of the RS model in a thick- and thin-shell case and evolving in ISM and stellar medium. The relationship between temporal and spectral index ($\alpha=\alpha(\beta)$) is shown for each cooling condition.}
\label{table:cr_syn}
\begin{tabular}{c c c  c c c}
 \hline \hline
&\hspace{0.5cm}     &\hspace{0.5cm}   ISM &\hspace{0.2cm}   ISM  & \hspace{0.5cm}   wind   & \hspace{0.5cm}  wind \\ 
&\hspace{0.5cm}     &\hspace{0.5cm} $1 < p <2$   &\hspace{0.5cm}   $2 < p$  & \hspace{0.2cm}    $1 < p <2$   & \hspace{0.5cm}  $2 < p$ \\
\hline
 &\hspace{0.5cm} $\beta$    &\hspace{0.5cm} $\alpha$   &\hspace{0.5cm}   $\alpha$  & \hspace{0.2cm}    $\alpha$   & \hspace{0.5cm}  $\alpha$    \\ \hline

Thick shell &\hspace{0.5cm}     &\hspace{0.2cm}    &\hspace{0.5cm}     & \hspace{0.5cm}       & \hspace{0.5cm}  \\ \hline 	
 $ \nu^{\rm syn}_{\rm m, r} < \nu < \nu^{\rm syn}_{\rm cut, r}$   	                & \hspace{0.5cm}  $\frac{p-1}{2}$  &\hspace{0.5cm}  $\frac{141+52\beta+24(4a+3b)}{96}$	                    &\hspace{0.2cm}  $\frac{47+73\beta+96a\beta+24(\beta+1)}{48}$ &\hspace{0.5cm}  $\frac{29+8\beta+4(4a+3b)}{16}$ &\hspace{0.5cm}  $\frac{3(3+5\beta)+16a\beta+4b(\beta+1)}{8}$ \\ 	
 $\nu^{\rm syn}_{\rm cut,r} < \nu$   	                                 & \hspace{0.5cm}  $\frac{p}{2}$     &\hspace{0.5cm}  $\beta+2$	                    &\hspace{0.2cm}  $\beta+2$ &\hspace{0.5cm}  $\beta+2$ &\hspace{0.5cm}  $\beta+2$\\ 
\hline	
Thin shell &\hspace{0.5cm}     &\hspace{0.5cm}    &\hspace{0.5cm}     & \hspace{0.5cm}       & \hspace{0.5cm}  \\ \hline 	
 $ \nu^{\rm syn}_{\rm m, r} < \nu < \nu^{\rm syn}_{\rm cut, r}$   	                & \hspace{0.5cm}  $\frac{p-1}{2}$  &\hspace{0.5cm}  $\frac{4(51+20\beta)+35(4a+3b)}{140}$	                    &\hspace{0.5cm}  $\frac{4(17+27\beta)+140a\beta+35b(\beta+1)}{70}$ &\hspace{0.5cm}  $\frac{4(38+9\beta)+21(4a+3b)}{84}$ &\hspace{0.5cm}  $\frac{2(23+39\beta)+84a\beta+21b(\beta+1)}{42}$ \\ 	
 $\nu^{\rm syn}_{\rm cut,r} < \nu$   	                                 & \hspace{0.5cm}  $\frac{p}{2}$     &\hspace{0.5cm}  $\beta+2$	                    &\hspace{0.5cm}  $\beta+2$ &\hspace{0.5cm}  $\beta+2$ &\hspace{0.5cm}  $\beta+2$\\

%
%
\hline
\end{tabular}
\end{table}

\begin{table}
\centering
\caption{The number and percentage of GRBs that adhere to the CRs of the SSC scenario for the RS in thick- and thin-shell regimes. We consider the values of spectral and temporal indices of a SPL and a BPL reported in 2FLGC.}
\label{tab:CR-Results}
\resizebox{\columnwidth}{!}{%
\begin{tabular}{cccccccccc}\cline{1-10}
\multicolumn{2}{c}{} &
  PL &
  $\beta$ &
  \begin{tabular}[c]{@{}c@{}}$\alpha(\beta)$\\ $1 < p < 2$\end{tabular} &
  \begin{tabular}[c]{@{}c@{}}$\alpha(\beta)$\\  $2 < p$\end{tabular} &
  \begin{tabular}[c]{@{}c@{}} a \end{tabular} &
  \begin{tabular}[c]{@{}c@{}} b \end{tabular} &
  \begin{tabular}[c]{@{}c@{}}Coincidences  \end{tabular} &
  \begin{tabular}[c]{@{}c@{}}Percent (\%)  \end{tabular} \\ \cline{1-10} 
\multicolumn{1}{c|}{\multirow{6}{*}{ISM}} & \multirow{3}{*}{Thick}  & $ \nu < \nu^{\rm ssc}_{\rm m, r}$ & $-\frac{1}{3}$ & $\frac{[106-59p+96a+24b(3-2p)]\beta}{24(p-1)}$ & $-\frac{(1-8a+2b)\beta}{2}$ & $-0.63\substack{+0.00 \\ -0.00}$ & $0.56\substack{+0.01 \\ -0.01}$ & 0 & 0.00\% \\
\multicolumn{1}{c|}{}                       &                       & $ \nu^{\rm ssc}_{\rm m, r} < \nu < \nu^{\rm ssc}_{\rm cut, r}$ & $\frac{p-1}{2}$ & $\frac{5\beta+88+96a+24b(2-\beta)}{48}$ & $\frac{99\beta+41+192a\beta+24b(\beta+1)}{48}$ & $-0.60\substack{+0.04 \\ -0.04}$ & $0.62\substack{+0.11 \\ -0.11}$ & 24 & 27.91\% \\
\multicolumn{1}{c|}{}                       &                       & $\nu^{\rm ssc}_{\rm cut,r} < \nu$ & $\frac{p}{2}$ & $\beta+2$ & $\beta+2$ & $0.01\substack{+10.24 \\ -10.21}$ & $0.06\substack{+10.13 \\ -10.23}$ & 1 & 1.16\% \\ \cline{2-10} 
\multicolumn{1}{c|}{}                       & \multirow{3}{*}{Thin} & $ \nu < \nu^{\rm ssc}_{\rm m, r}$ & $-\frac{1}{3}$ & $\frac{[143-75p+140a+35b(3-2p)]\beta}{35(p-1)}$ & $-\frac{(1-20a+5b)\beta}{5}$  & $-0.66\substack{+0.00 \\ -0.00}$ & $0.72\substack{+0.01 \\ -0.01}$ & 0 & 0.00\% \\
\multicolumn{1}{c|}{}                       &                       & $ \nu^{\rm ssc}_{\rm m, r} < \nu < \nu^{\rm ssc}_{\rm cut, r}$ & $\frac{p-1}{2}$ & $\frac{2(61+6\beta)+140a+35b(2-\beta)}{70}$ & $\frac{2(74\beta+27)+280a\beta+35b(\beta+1)}{70}$ & $-0.62\substack{+0.04 \\ -0.05}$ & $0.76\substack{+0.11 \\ -0.11}$ & 25 & 29.07\% \\
\multicolumn{1}{c|}{}                       &                       & $\nu^{\rm ssc}_{\rm cut,r} < \nu$ & $\frac{p}{2}$ & $\beta+2$ & $\beta+2$ & $0.12\substack{+10.09 \\ -10.26}$ & $-0.01\substack{+10.20 \\ -10.23}$ & 1 & 1.16\%\\ \hline
\multicolumn{1}{c|}{\multirow{6}{*}{Wind}}  & \multirow{3}{*}{Thick}  & $ \nu < \nu^{\rm ssc}_{\rm m, r}$ & $-\frac{1}{3}$ & $\frac{[28-17p+16a+4b(3-2p)]\beta}{4(p-1)}$ & $-\frac{(3-8a+2b)\beta}{2}$ & $-0.60\substack{+0.00 \\ -0.00}$ & $-0.34\substack{+0.01 \\ -0.01}$ & 0 & 0.00\% \\
\multicolumn{1}{c|}{}                       &                       & $ \nu^{\rm ssc}_{\rm m, r} < \nu < \nu^{\rm ssc}_{\rm cut, r}$ & $\frac{p-1}{2}$ & $\frac{22-\beta+16a+4b(2-\beta)}{8}$ & $\frac{21\beta+11+32a\beta+4b(\beta+1)}{8}$ & $-0.74\substack{+0.04 \\ -0.04}$ & $-0.03\substack{+0.11 \\ -0.11}$ & 25 & 29.07\%\\
\multicolumn{1}{c|}{}                       &                       & $\nu^{\rm ssc}_{\rm cut,r} < \nu$ & $\frac{p}{2}$ & $\beta+2$ & $\beta+2$ & $-0.09\substack{+10.29 \\ -10.20}$ & $0.09\substack{+10.16 \\ -10.28}$ & 1 & 1.16\% \\ \cline{2-10} 
\multicolumn{1}{c|}{}                       & \multirow{3}{*}{Thin} & $ \nu < \nu^{\rm ssc}_{\rm m, r}$ & $-\frac{1}{3}$ & $\frac{[5(31-19p)+84a+21b(3-2p)]\beta}{21(p-1)}$ & $-\frac{(5-12a+3b)\beta}{3}$ & $-0.63\substack{+0.00 \\ -0.00}$ & $0.56\substack{+0.01 \\ -0.01}$ & 0 & 0.00 \\
\multicolumn{1}{c|}{}                       &                       & $ \nu^{\rm ssc}_{\rm m, r} < \nu < \nu^{\rm ssc}_{\rm cut, r}$ & $\frac{p-1}{2}$ & $\frac{5(12-\beta)+42a+b(21-10\beta)}{21}$ & $\frac{10(11\beta+6)+168a\beta+21b(\beta+1)}{42}$ & $-0.62\substack{+0.04 \\ -0.04}$ & $0.76\substack{+0.11 \\ -0.11}$ & 32 & 37.21\% \\
\multicolumn{1}{c|}{}                       &                       & $\nu^{\rm ssc}_{\rm cut,r} < \nu$ & $\frac{p}{2}$ & $\beta+2$ & $\beta+2$ & $0.09\substack{+10.15 \\ -10.24}$ & $0.07\substack{+10.16 \\ -10.22}$ & 1 & 1.16\% \\ \hline
\end{tabular}%
}
\end{table}

\begin{table}[h!]
\centering \renewcommand{\arraystretch}{1.6}\addtolength{\tabcolsep}{2pt}
\caption{Density parameter evolution in a stellar-wind and constant density medium ($F^{\rm ssc}_{\rm \nu, r} \propto n^{\alpha_{\rm k}} (A_{\rm W}^{\alpha_{\rm k}})$) in the the SSC scenario from the RS.}
\label{table:dens_paramet}
\begin{tabular}{cccccc}
\hline
\hline
     &               & ISM     & ISM      & wind    & wind      \\

     &               & $1<p<2$     & $2<p$       & $1<p<2$     & $2<p$       \\
\hline

 & $\beta$    & $\alpha_{\rm k}$    & $\alpha_{\rm k}$    & $\alpha_{\rm k}$    & $\alpha_{\rm k}$    \\
\hline
Thick shell &    &    &    &    &    \\
\hline
$ \nu < \nu^{\rm ssc}_{\rm m,r}$    &    $-\frac13$         &     $-\frac{15-14p}{12(p-1)}$        &      $\frac{3}{4}$       &       $-\frac{15-14p}{6(p-1)}$      &      $\frac{3}{2}$       \\
$ \nu^{\rm ssc}_{\rm m,r} < \nu < \nu^{\rm ssc}_{\rm cut,r}$    &    $\frac{p-1}{2}$           &       $\frac{11-2p}{8}$      &       $\frac{3p+5}{8}$      &      $\frac{11-2p}{4}$       &      $\frac{3p+5}{4}$       \\

\hline
Thin shell &  &     &     &    &    \\
\hline
$ \nu < \nu^{\rm ssc}_{\rm m,r}$    &    $-\frac13$         &     $-\frac{122-115p}{105(p-1)}$        &      $\frac{44}{45}$       &       $-\frac{164-157p}{63(p-1)}$      &      $\frac{22}{9}$       \\
$ \nu^{\rm ssc}_{\rm m,r} < \nu < \nu^{\rm ssc}_{\rm cut,r}$    &    $\frac{p-1}{2}$          &       $\frac{541-117p}{420}$      &       $\frac{425 -43p}{420}$      &      $\frac{241-53p}{84}$       &      $\frac{221 - 47p}{84}$       \\

\hline
\end{tabular}
\end{table}

\begin{table}[h!]
\centering
\renewcommand{\arraystretch}{1.6}\addtolength{\tabcolsep}{-4pt}
\caption{Values of the microphysical parameter $a$ as a function of $b$, $\alpha$, and $p$  from the SSC RS model evolving in the thick- and thin-shell regime.}
\label{table2:a_values}
\begin{tabular}{cccccc}
\hline
\hline
     &               & ISM     & ISM      & wind    & wind      \\

     &               & $1<p<2$     & $2<p$       & $1<p<2$     & $2<p$       \\
\hline

 & $\beta$    & $a(b, \alpha, p)$    & $a(b, \alpha, p)$    & $a(b, \alpha, p)$    & $a(b, \alpha, p)$    \\
\hline
Thick shell &    &    &    &    &    \\
\hline
$ \nu < \nu^{\rm ssc}_{\rm m,r}$    &    $-\frac13$         &     $\frac{59p - 106 - 72(p-1)\alpha - 24b(3-2p)}{96}$        &      $\frac{1 - 8\alpha + 2b}{8}$       &       $\frac{17p - 28 - 12(p-1)\alpha - 4b(3-2p)}{16}$      &      $\frac{3 - 8\alpha + 2b}{8}$       \\
$ \nu^{\rm ssc}_{\rm m,r} < \nu < \nu^{\rm ssc}_{\rm cut,r}$    &    $\frac{p-1}{2}$           &       $\frac{5p + 171 - 24b(p-5) - 96\alpha}{192}$      &       $\frac{17 - 99p - 24b(p+1) + 96\alpha(p-1)}{192}$      &      $\frac{p - 45 - 4b(5-p) + 16\alpha}{32}$       &      $\frac{-1 - 21p - 4b(p+1) + 16\alpha(p-1)}{32}$       \\

\hline
Thin shell &  &     &     &    &    \\
\hline
$ \nu < \nu^{\rm ssc}_{\rm m,r}$    &    $-\frac13$         &     $\frac{75p - 143 - 105(p-1)\alpha - 35b(3-2p)}{140}$        &      $\frac{1 - 20\alpha + 5b}{20}$       &       $\frac{5(19p - 31) - 84\alpha - 21b(3-2p)}{84}$      &      $\frac{5 - 12\alpha + 3b}{12}$       \\
$ \nu^{\rm ssc}_{\rm m,r} < \nu < \nu^{\rm ssc}_{\rm cut,r}$    &    $\frac{p-1}{2}$          &       $\frac{4(58+3p) - 35b(5-p) - 280\alpha}{280}$      &       $\frac{4(10 - 37p) - 35b(p+1) + 280\alpha(p-1)}{280}$      &      $\frac{10(25-p) - 21b(5-p) - 168\alpha}{168}$       &      $\frac{-10 - 110p - 21b(p+1) + 168\alpha(p-1)}{168}$       \\

\hline
\end{tabular}
\end{table}

\begin{table}
\centering \renewcommand{\arraystretch}{1.5}\addtolength{\tabcolsep}{2pt}

\caption{Sample of 8 bursts with an atypical spectral PL indices with $\beta_{\rm L}=\Gamma_{\rm L}-1$.}
\label{tab:beta_eq0}
\begin{tabular}{lc}
\hline
GRB & $\beta_{\rm L} \pm \delta_{\beta_{\rm L}}$ \\
\hline
090427A	&	$-0.2\pm0.7$		\\
091127	&	$0.2	\pm	0.4$	\\
101014A	&	$0.3	\pm	0.3$	\\
101227B	&	$0.5	\pm	0.5$	\\
150514A	&	$0.1	\pm	0.4$	\\
160422A	&	$0.2	\pm	0.3$    \\
160702A	&	$0.4	\pm	0.4$    \\
160829A	&	$0.3	\pm	0.3$    \\
\hline
\end{tabular}
\end{table}

\begin{table}
\centering \renewcommand{\arraystretch}{1.5}\addtolength{\tabcolsep}{2pt}

\caption{GRBs detected with energetic photons in the TeV energy range.
 }
\label{table_sample_he}
\begin{tabular}{lcc}
\hline
GRB Name & $\alpha_{\rm L} \pm \delta_{\alpha_{\rm L}}$ & $\beta_{\rm L} \pm \delta_{\beta_{\rm L}}$  \\
\hline
180720B	&	$1.90	\pm	0.10$	&	$1.15	\pm	0.10$ \\
190114C	&	$1.10	\pm	0.15$	&	$1.06	\pm	0.30$		\\
221009A	&	$1.32	\pm	0.05$	&	$0.87	\pm	0.04$	         		\\

\hline
\end{tabular}
\end{table}

\clearpage 
\newpage

\begin{landscape}
\begin{figure*}
\centering
    \subfloat{\includegraphics[scale=0.55]{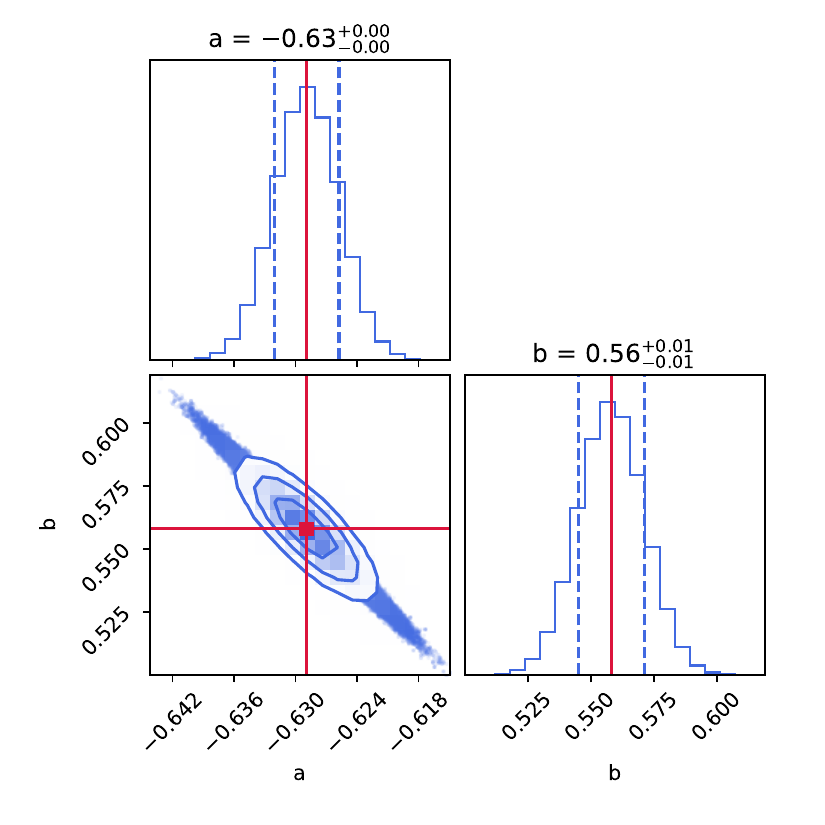}} \label{fig:MCMC_K0_C1}
    \subfloat{\includegraphics[scale=0.55]{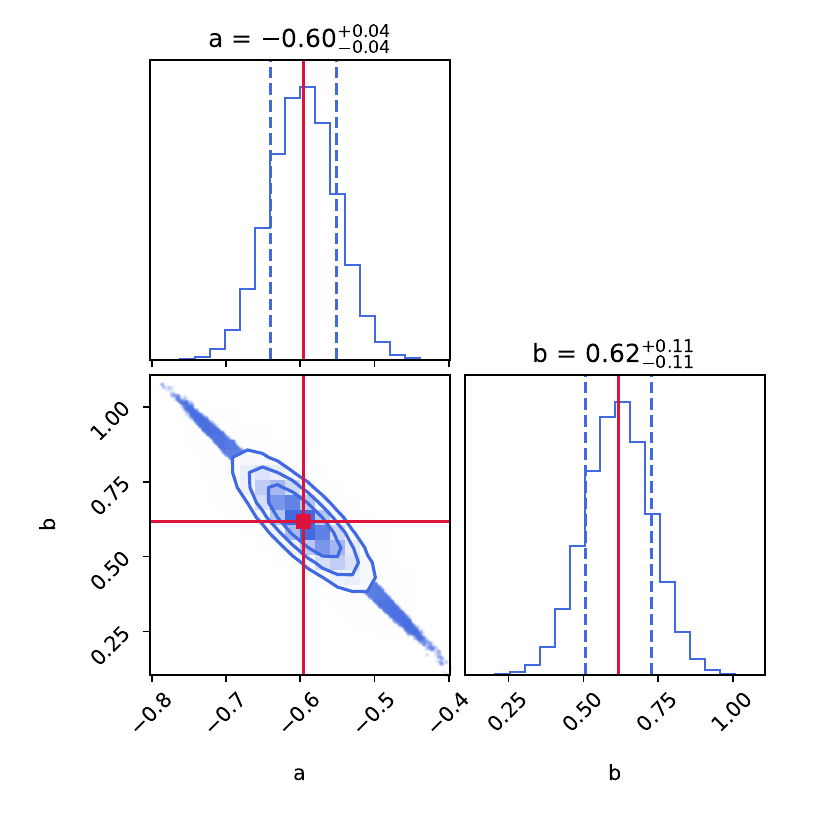}} \label{fig:MCMC_K0_C2}
    \subfloat{\includegraphics[scale=0.55]{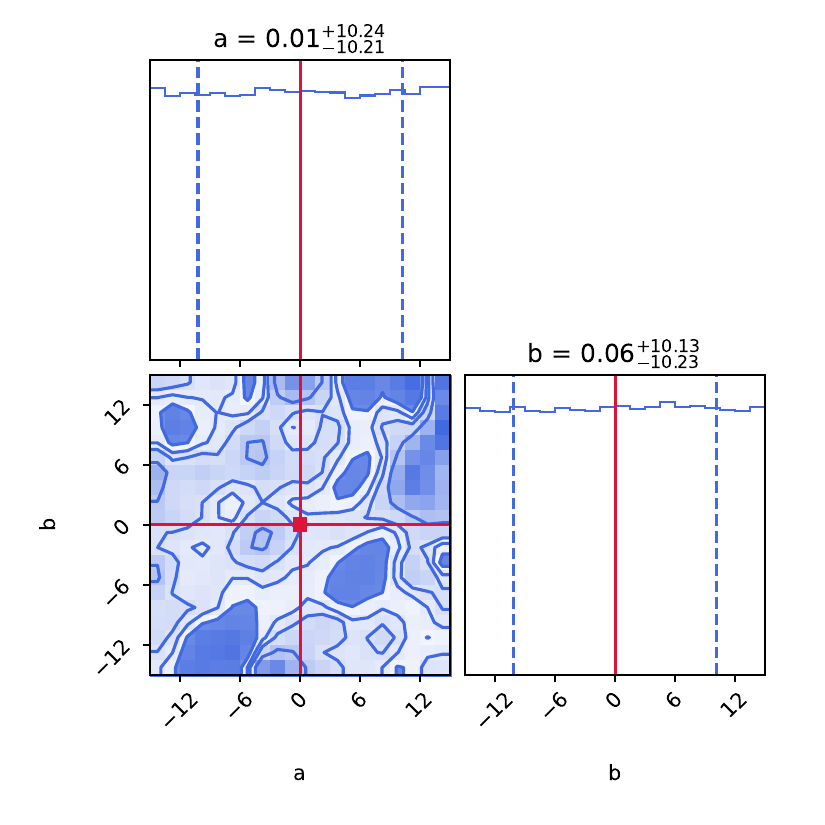}} \label{fig:MCMC_K0_C3}

  \vspace{1em}

    \subfloat{\includegraphics[scale=0.55]{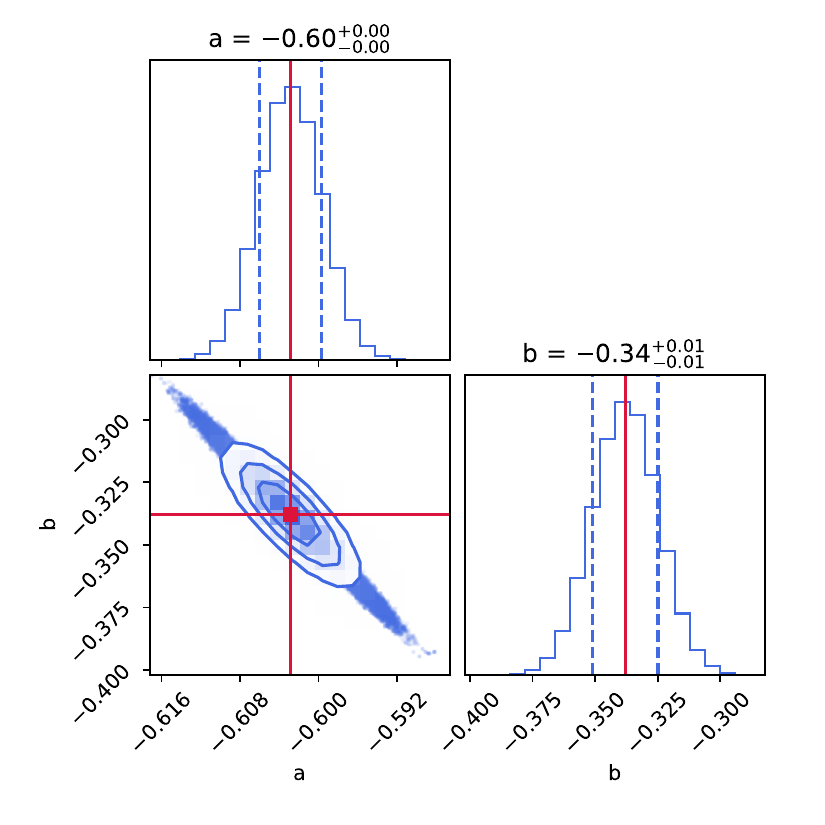}} \label{fig:MCMC_K2_C1}
    \subfloat{\includegraphics[scale=0.55]{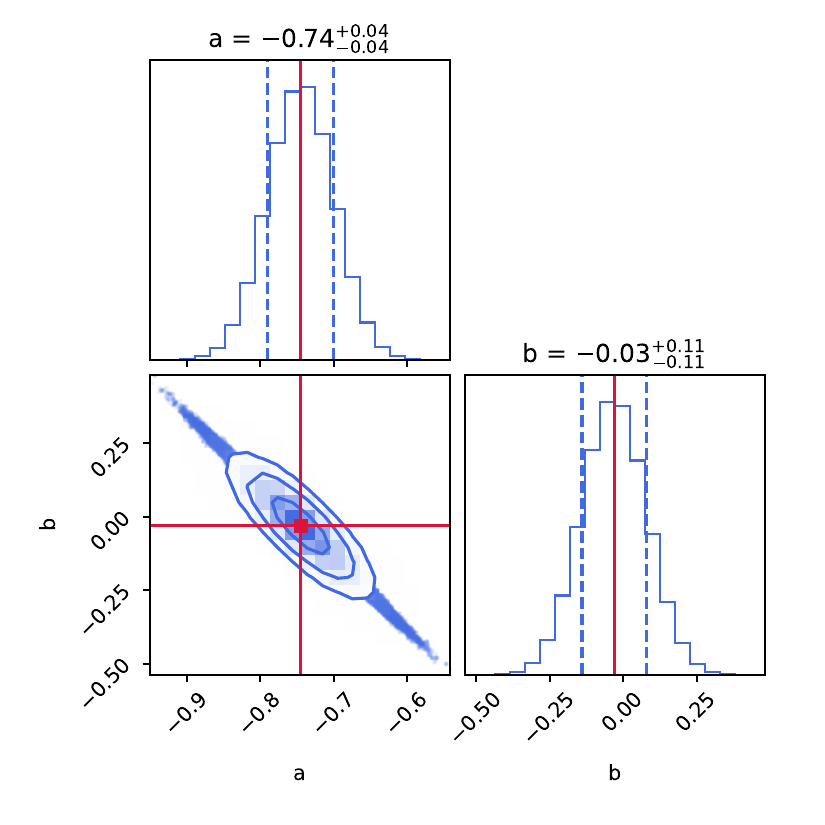}} \label{fig:MCMC_K2_C2}
    \subfloat{\includegraphics[scale=0.55]{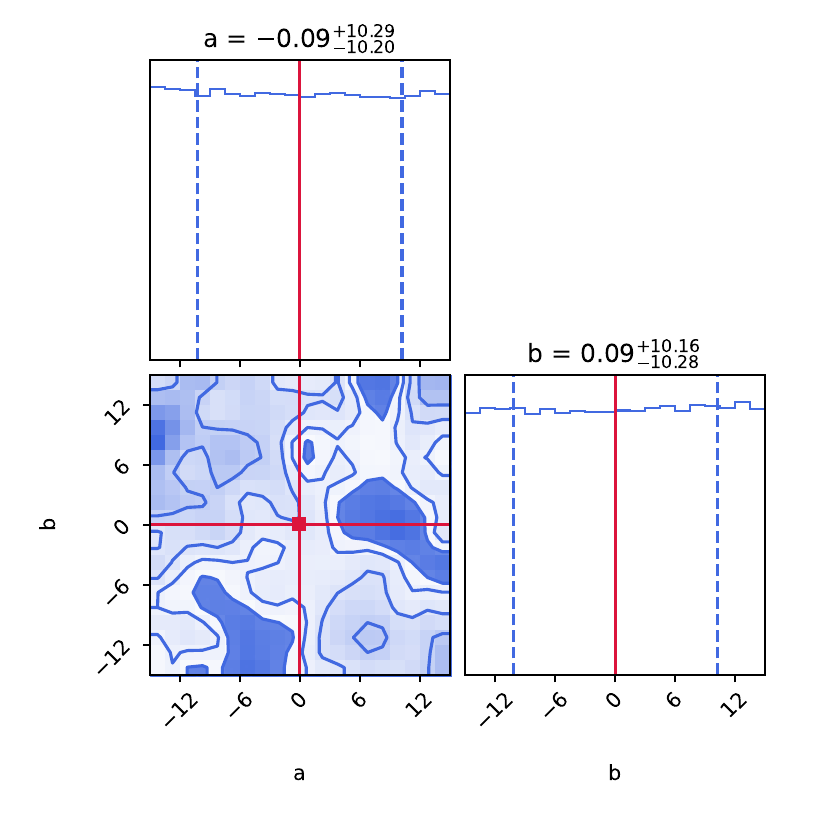}} \label{fig:MCMC_K2_C3}

  \caption{Corner plots resulting for the MCMC process to maximize the CRs reported in Table \ref{table:cr_ssc}. 
  The first column corresponds to $ \nu < \nu^{\rm ssc}_{\rm m, r}$, the second to $ \nu^{\rm ssc}_{\rm m, r} < \nu < \nu^{\rm ssc}_{\rm cut, r}$, 
  and the third to $\nu^{\rm ssc}_{\rm cut,r} < \nu$. 
  We consider the SSC RS evolving thick-shell regime for a constant interstellar medium (upper), 
  and a wind-like medium (lower).}
  \label{fig:MCMC_ISM}
\end{figure*}
\end{landscape}

\begin{landscape}
\begin{figure*}
\centering
    \subfloat{\includegraphics[scale=0.55]{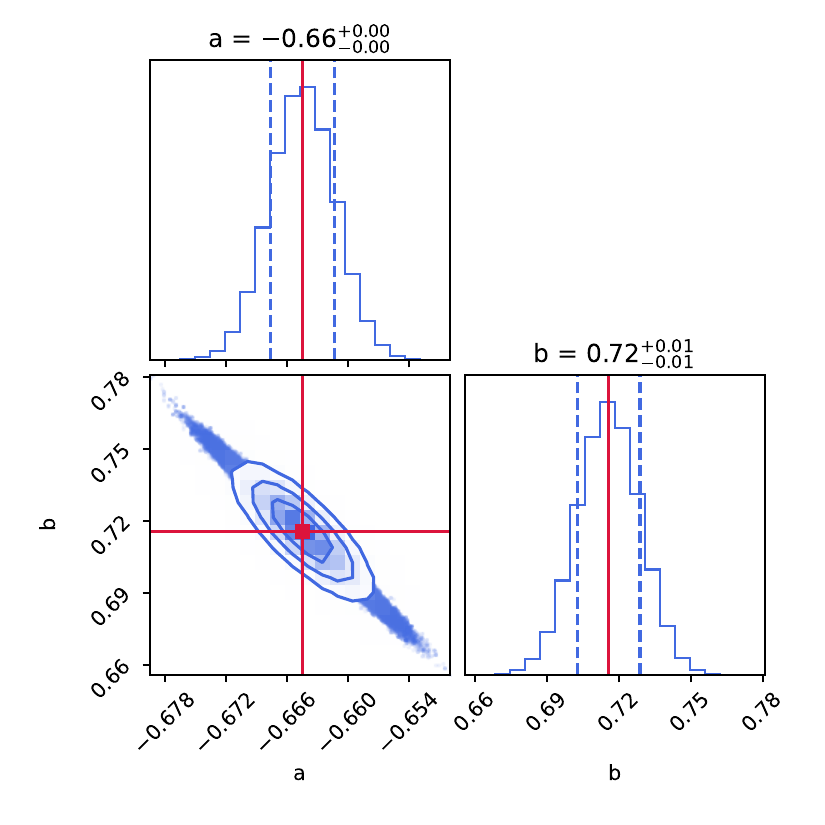}} \label{fig:MCMC_K0_C4}
    \subfloat{\includegraphics[scale=0.55]{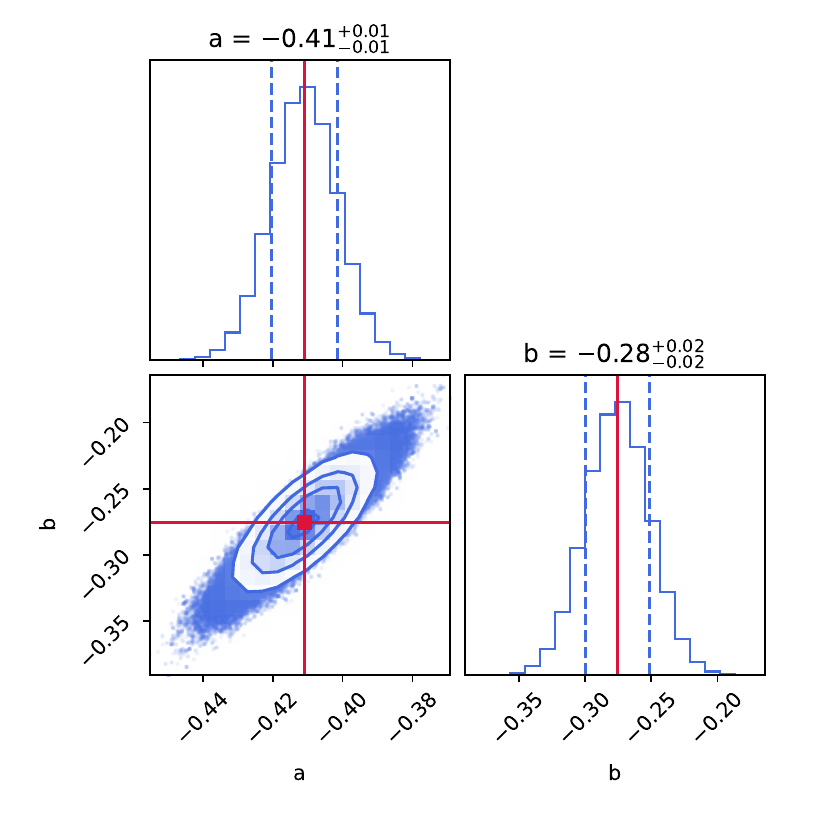}} \label{fig:MCMC_K0_C5}
    \subfloat{\includegraphics[scale=0.55]{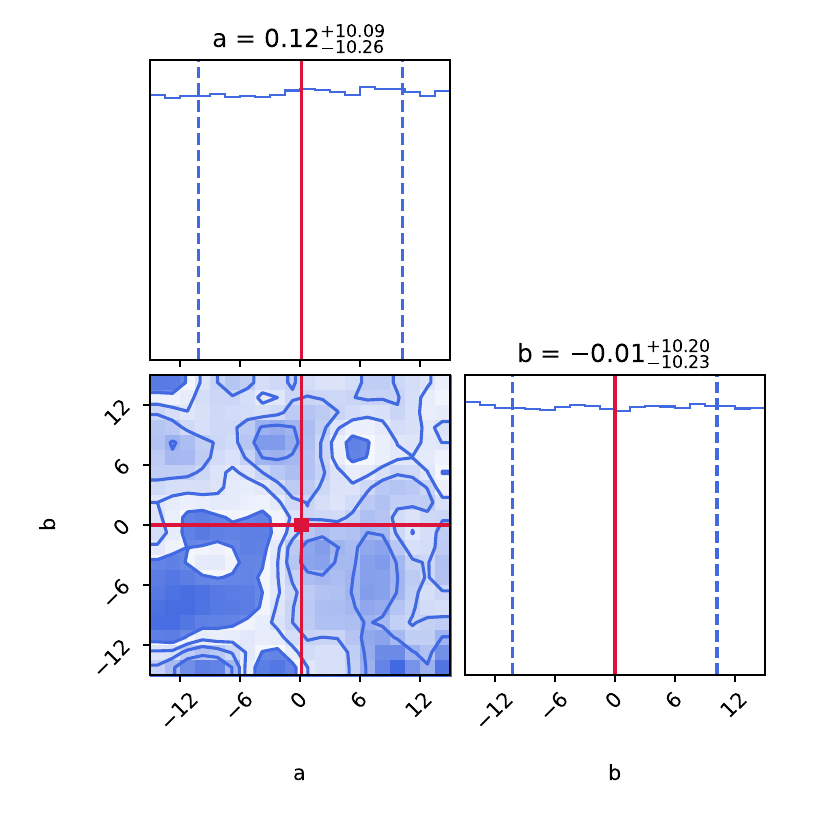}} \label{fig:MCMC_K0_C6}

  \vspace{1em}

    \subfloat{\includegraphics[scale=0.55]{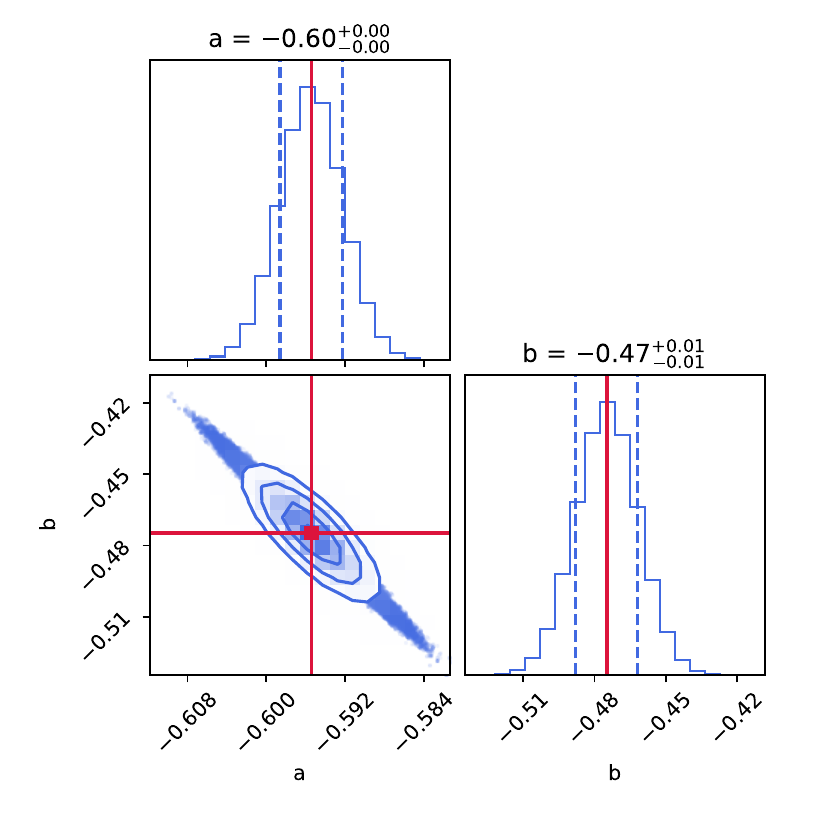}} \label{fig:MCMC_K2_C4}
    \subfloat{\includegraphics[scale=0.55]{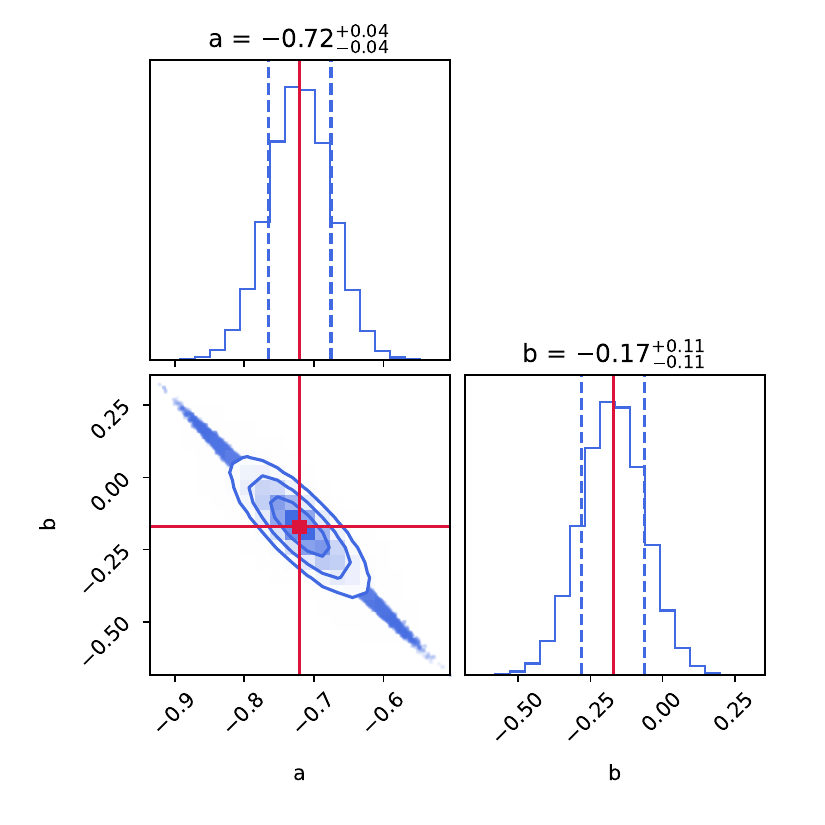}} \label{fig:MCMC_K2_C5}
    \subfloat{\includegraphics[scale=0.55]{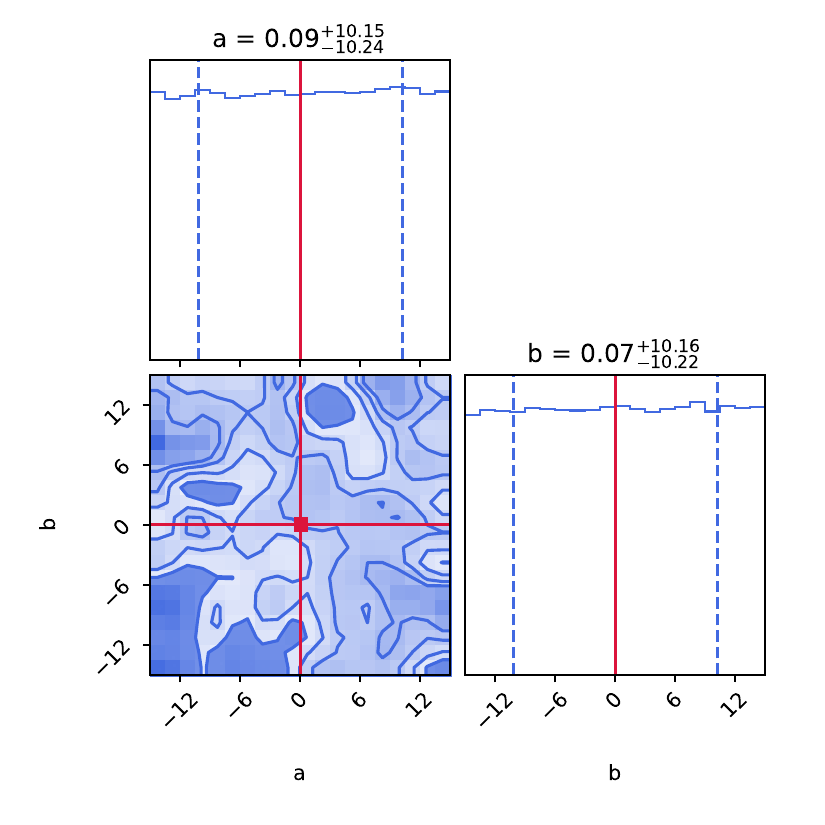}} \label{fig:MCMC_K2_C6}

  \caption{The same as \ref{fig:MCMC_ISM}, but for the thin-shell regime.}
  \label{fig:MCMC_ISM2}
\end{figure*}
\end{landscape}

\begin{figure}
{ \centering
\resizebox*{\textwidth}{0.6\textheight}
{\includegraphics{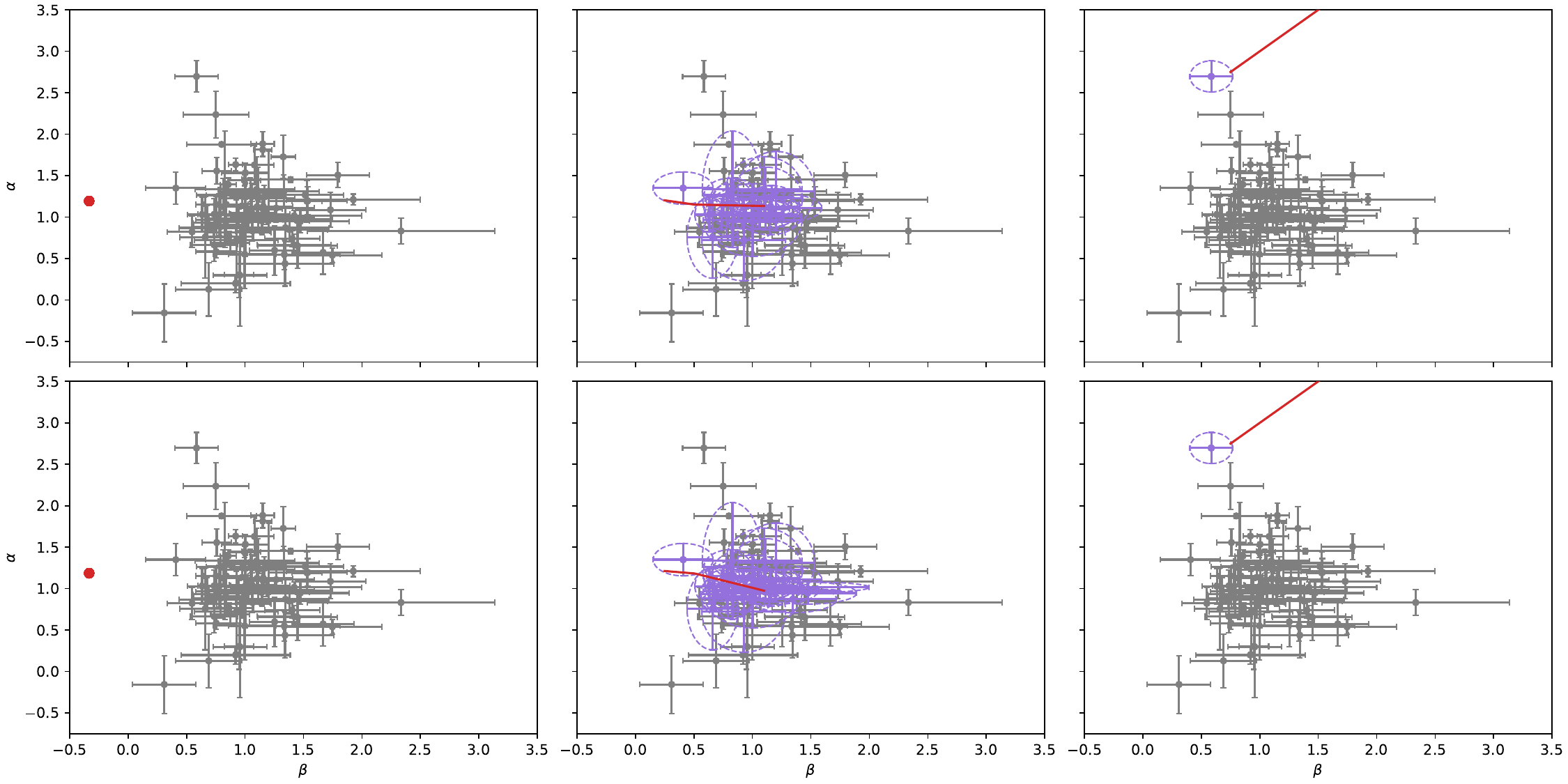}
} 
\caption{These panels show the CRs of SSC RS emission evolving in a thick-shell regime for ISM (above) and stellar wind (below). From left to right panels correspond to the cooling condition 
$ \nu < \nu^{\rm ssc}_{\rm m,r}$, $ \nu^{\rm ssc}_{\rm m,r} < \nu < \nu^{\rm ssc}_{\rm cut,r}$ and  $ \nu^{\rm ssc}_{\rm cut,r} < \nu$.  The purple ellipses are displayed when the confidence regions are within 1$\sigma$ error margins. The red lines and dots represent the estimated CRs.}\label{fig1:CRsthick}
}
\end{figure}

\begin{figure}[h!]
{ \centering
\resizebox*{\textwidth}{0.6\textheight}
{\includegraphics{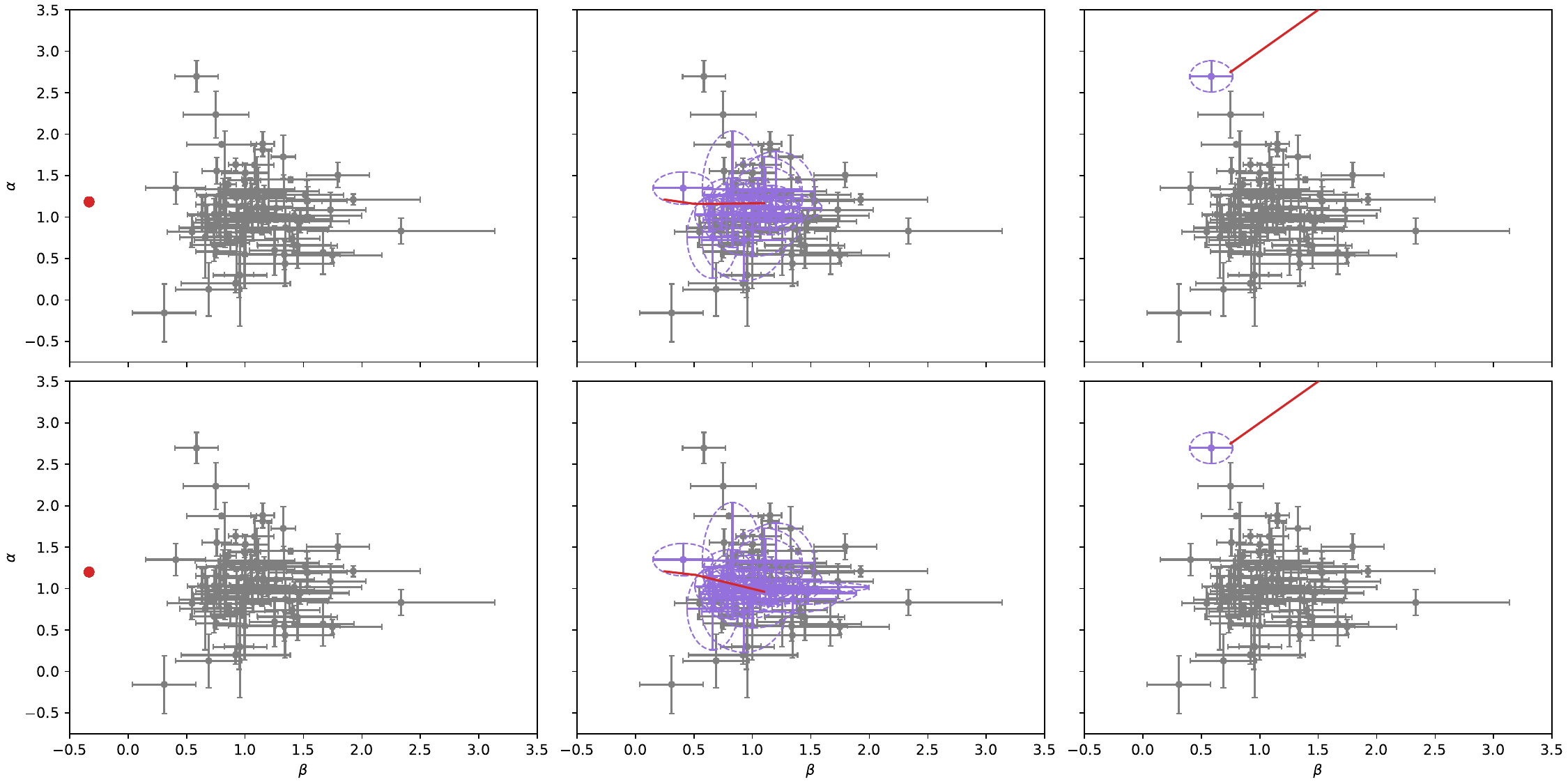}
}
\caption{The same as Figure \ref{fig1:CRsthick}, but for the RS evolving in the thin-shell regime.}\label{fig2:CRsthin}
}
 \end{figure}

\begin{figure}
{ \centering
\resizebox*{\textwidth}{0.9\textheight}
{\includegraphics{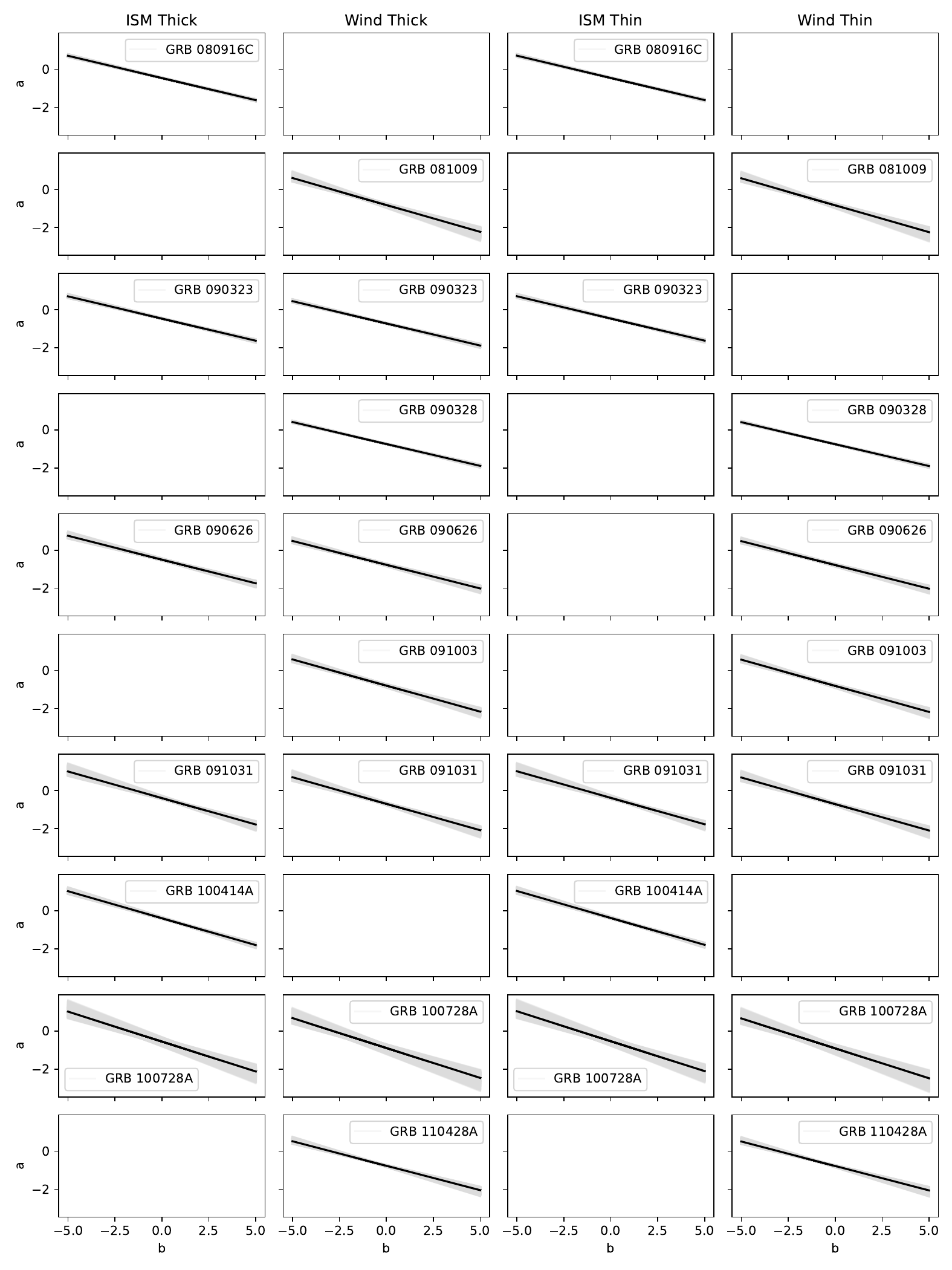}}
}
\caption{Range of microphysical parameters ($a$ and $b$) for which the CRs of bursts reported in 2FLGC are satisfied when the SSC model lies in the cooling condition ${\rm \nu_m^{ssc} < \nu_{\rm LAT} < \nu_{\rm cut}^{ssc} }$. Panels from left to right correspond to the evolution of the RS in the thick-shell regime for ISM and stellar wind, and in the thin-shell regime for ISM and stellar wind, respectively. Panels from top to bottom are for GRBs 080916C, 081009, 090323, 090328, 090626, 091003, 091031, 100414A, 100728A and 110428A.}
 \label{fig:cr_bur_1}
\end{figure}

\begin{figure}
{ \centering
\resizebox*{\textwidth}{0.9\textheight}
{\includegraphics{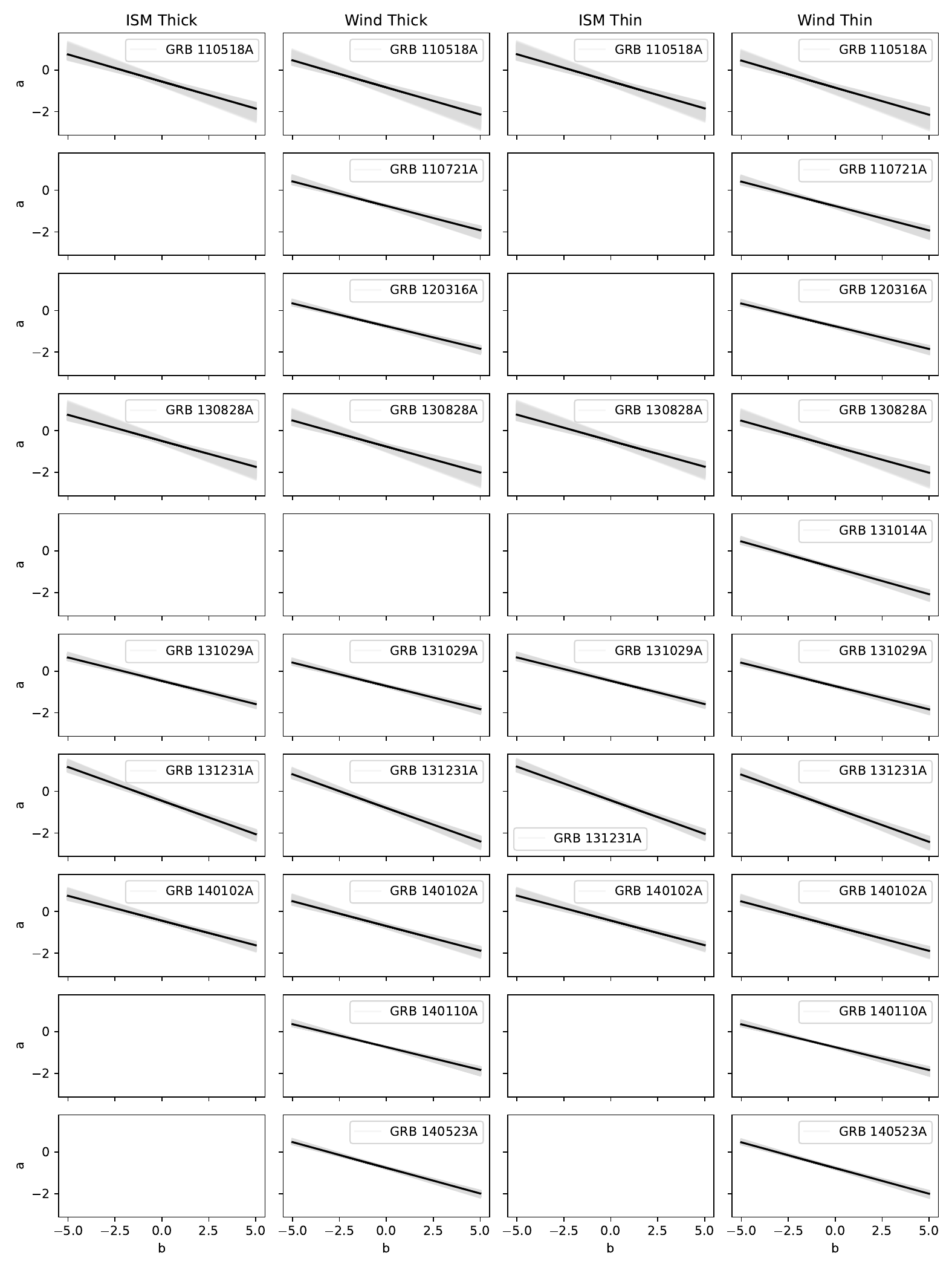}}
}
\caption{The same of Figure \ref{fig:cr_bur_1}, but panels from top to bottom are for GRBs 110518A, 110721A, 120316A, 130828A, 131014A, 131029A, 131231A, 140102A,  140110A, 140523A.}
 \label{fig:cr_bur_2}
\end{figure}

\begin{figure}
{ \centering
\resizebox*{\textwidth}{\textheight}
{\includegraphics{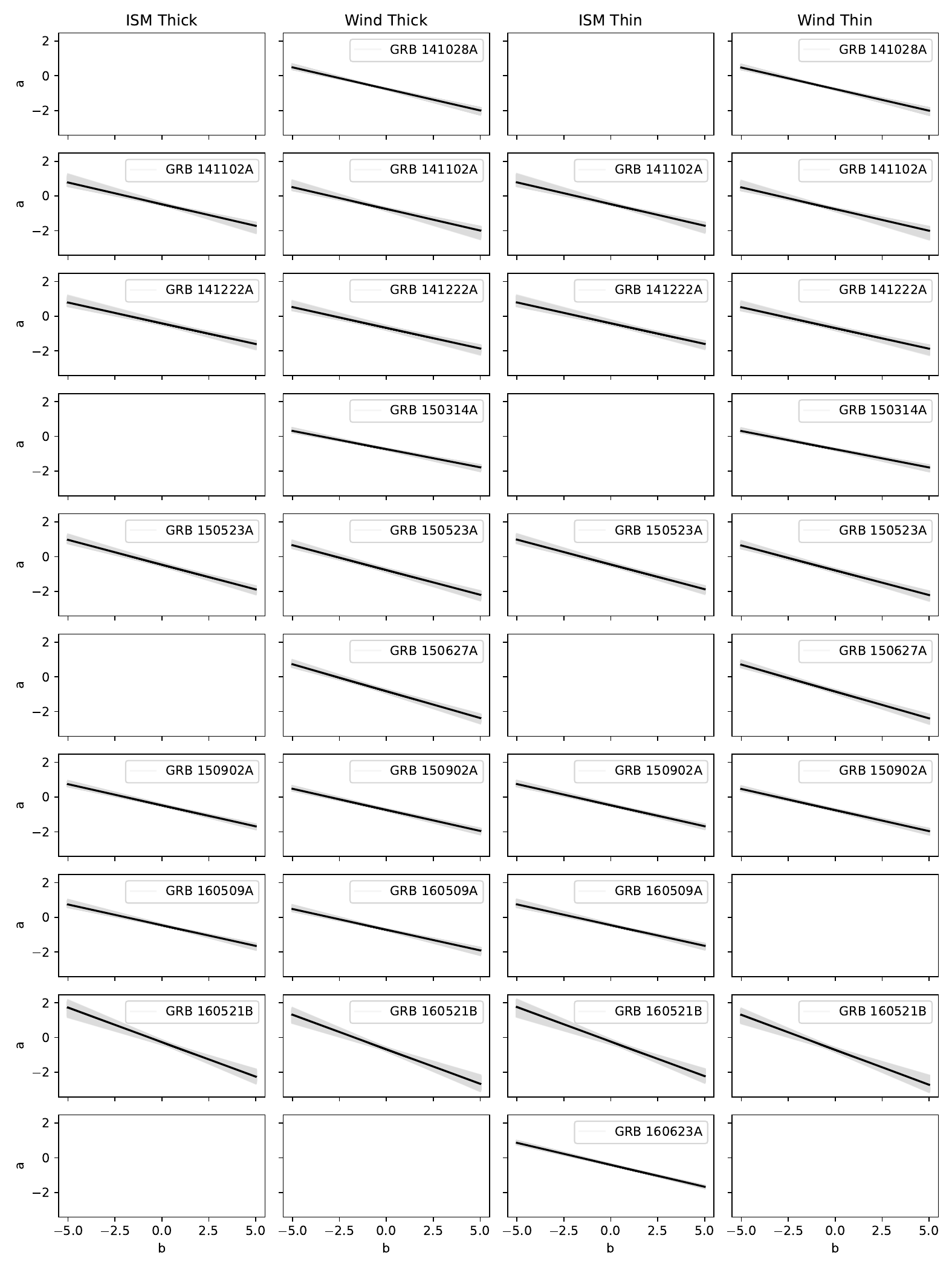}}
}
\caption{The same of Figure \ref{fig:cr_bur_1}, but panels from top to bottom are for GRBs 141028A, 141102A, 141222A, 150314A, 150523A, 150627A, 150902A, 160509A, 160521B and 160623A.
}
 \label{fig:cr_bur_3}
\end{figure}

\begin{figure}
{ \centering
\resizebox*{\textwidth}{0.9\textheight}
{\includegraphics{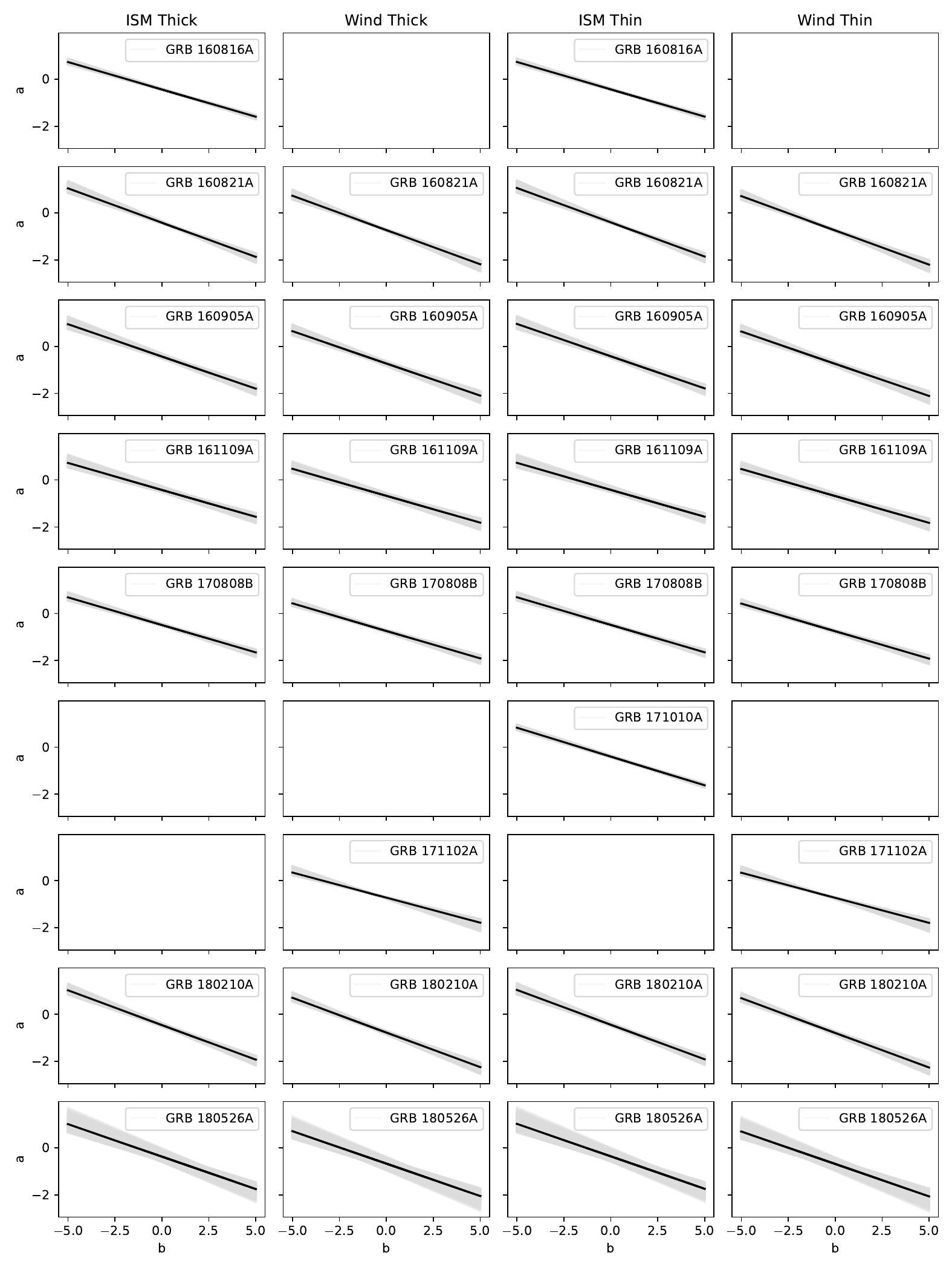}}
}
\caption{The same of Figure \ref{fig:cr_bur_1}, but panels from top to bottom are for GRBs 160816A, 160821A, 160905A, 161109A, 170808B, 171010A, 171102A, 180210A, and 180526A.}
 \label{fig:cr_bur_4}
\end{figure}

\begin{figure}
{ \centering
\resizebox*{\textwidth}{0.5\textheight}
{\includegraphics{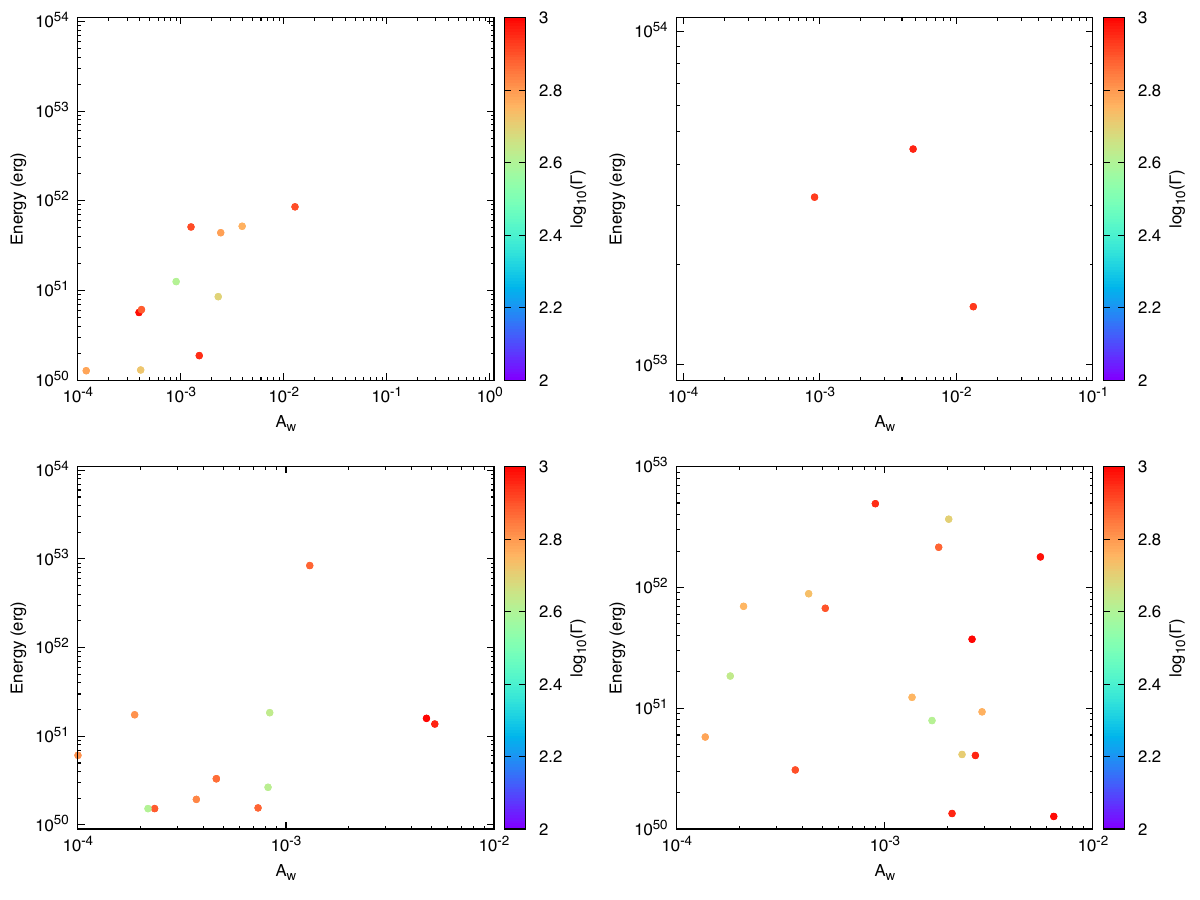}}   
}
\caption{Variation of microphysical parameters in the SSC RS mechanism evolving in a thick-shell regime and stellar-wind environment for which the atypical spectral index $\beta\approx 0$
is satisfied. The values used in each panel are:  $\rm a=0.1$ and $\rm b=1.0$ (top left), $\rm a=0.5$ and $\rm b=1.0$ (top right), $\rm a=0.1$ and $\rm b=2.0$ (bottom left), $\rm a=0.5$ and $\rm b=2.0$ (bottom right).}
 \label{fig:SPThick}
\end{figure}

\begin{figure}
{ \centering
\resizebox*{\textwidth}{0.5\textheight}
{\includegraphics{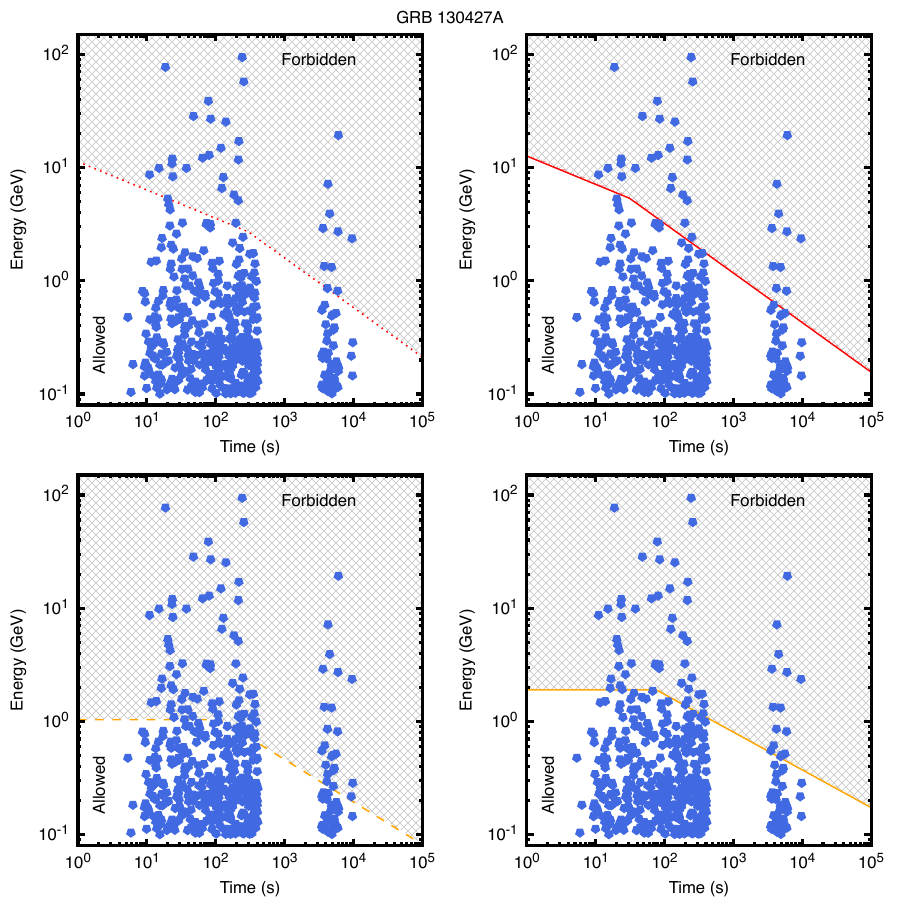}}   
}
\caption{All photons over 100 MeV are shown with a probability exceeding $90\%$ of being associated with GRB130427A. Upper panels are shown for constant medium; $n=1\,$ (left) and $ 10^{-2}\,{\rm cm^{-3}}$ (right) and lower panels for stellar wind; $A_W=1$ (left) and $10^{-2}$ (right).}
 \label{fig:Emax}
\end{figure}

\begin{figure}
{ \centering
\resizebox*{\textwidth}{0.95\textheight}
{\includegraphics{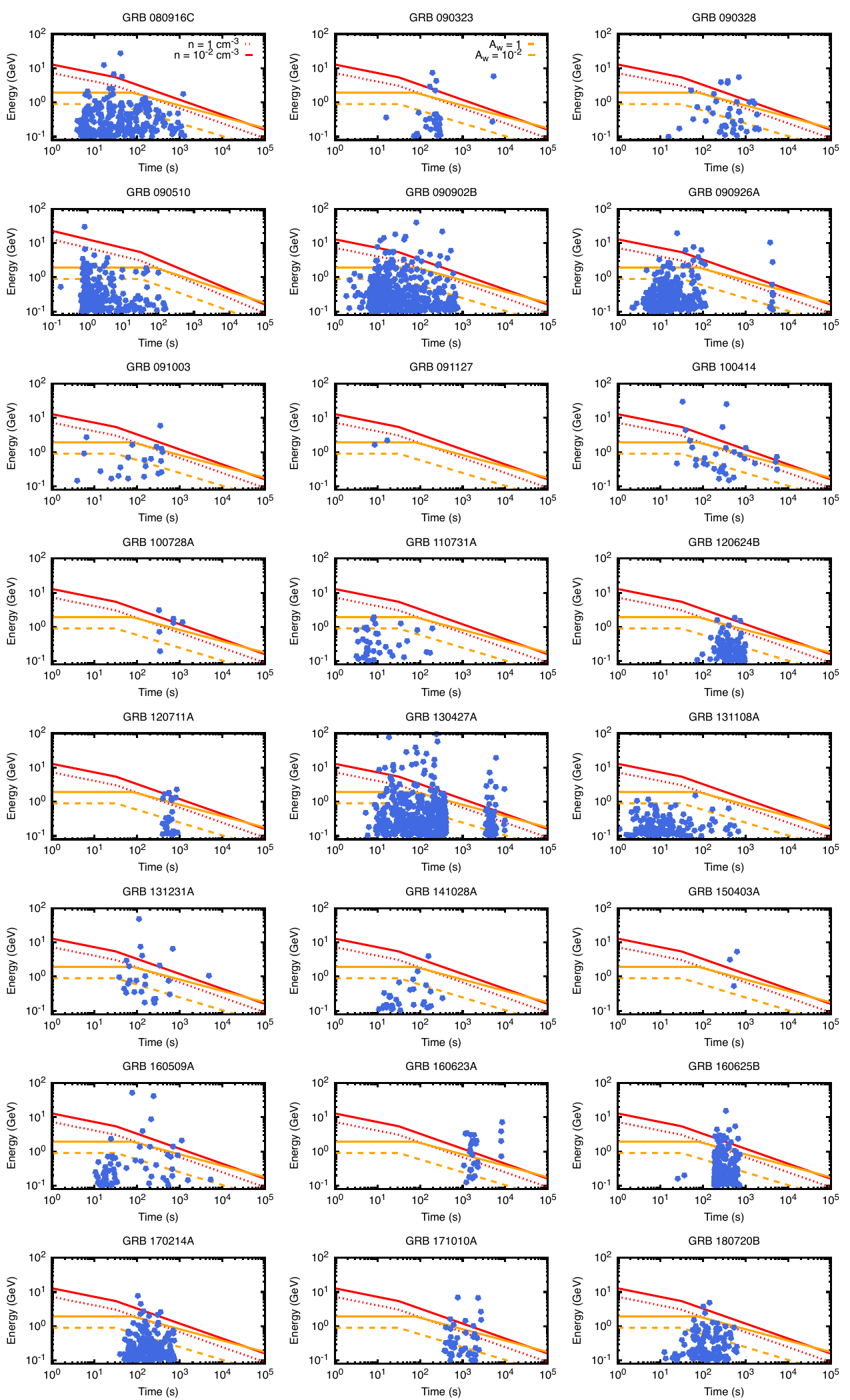}}
}
\caption{For every burst in our sample, all photons over 100 MeV are exhibited with a likelihood of more than $90\%$ that will be associated. Each panel also displays the synchrotron limit from the RS in the thick-shell regime. The evolution in a constant-density medium wih $n=1\,$ (dotted red) and $ 10^{-2}\,{\rm cm^{-3}}$ (solid red) and in a stellar wind with parameter $A_W=1$ (dashed golden) and $10^{-2}$ (solid golden) are considered. The values used are $E=10^{53}\,{\rm erg}$, and $\Gamma=600$.}
\label{fig_thick:max_en}
\end{figure}



\begin{figure}
{ \centering
\resizebox*{\textwidth}{\textheight}
{\includegraphics{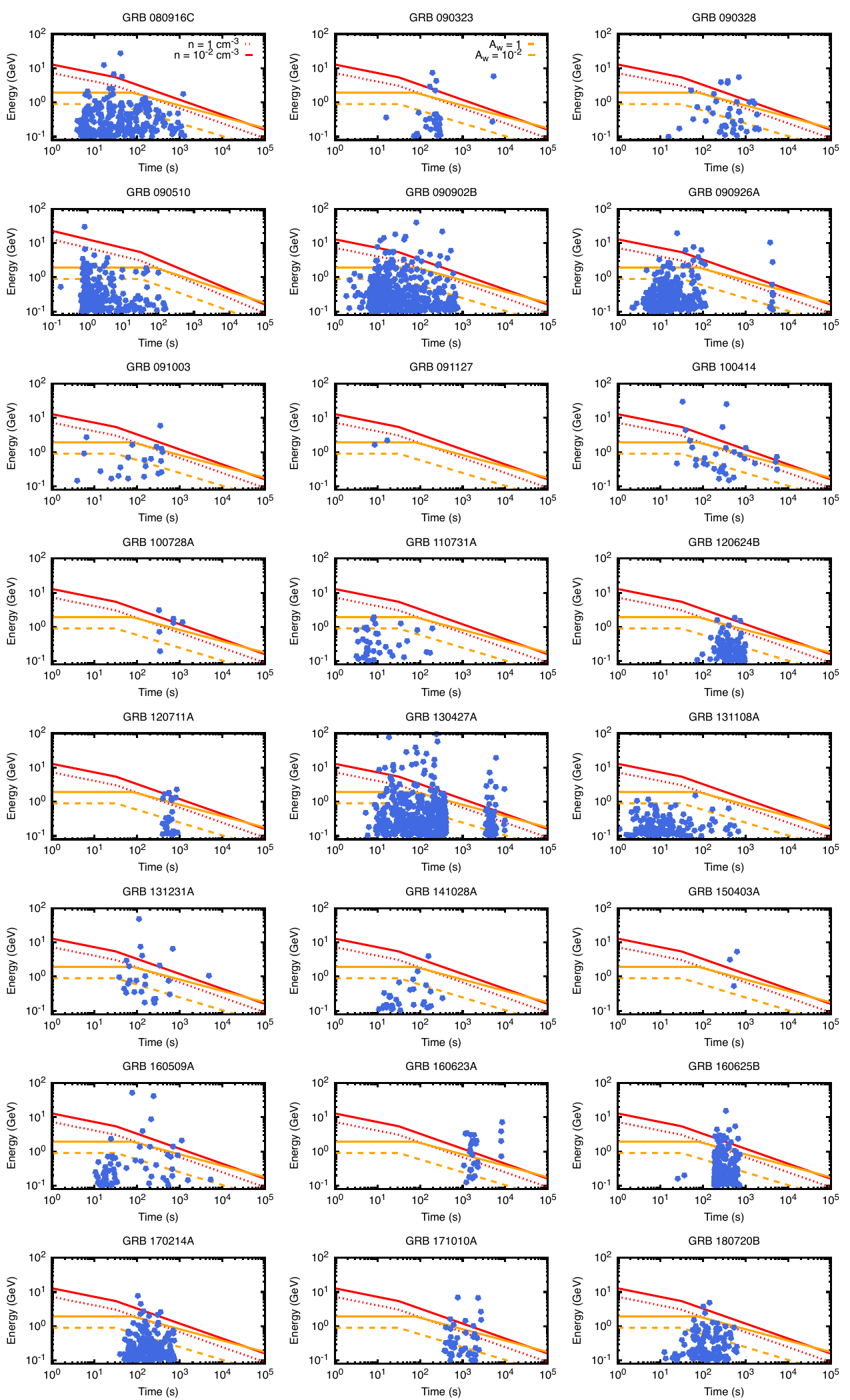}}
}
\caption{The same as Figure \ref{fig_thick:max_en}, but for thin-shell regime.}
 \label{fig_thin:max_en}
\end{figure}


\begin{figure}
{ \centering
\resizebox*{\textwidth}{0.35\textheight}
{\includegraphics{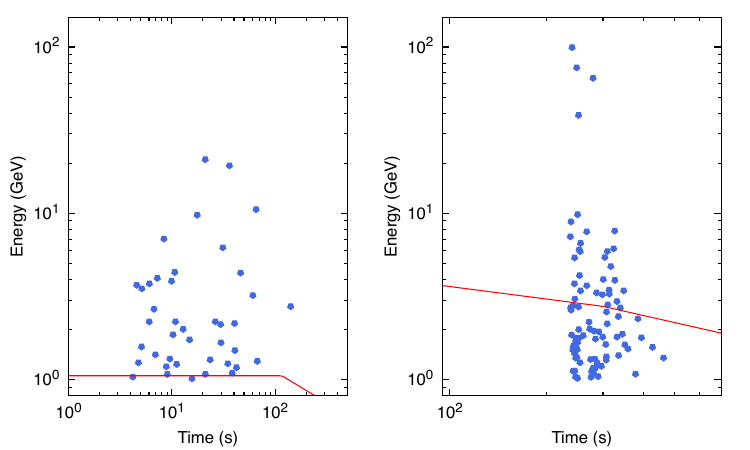}}   
}
\caption{Similar to Figures \ref{fig_thick:max_en} and \ref{fig_thin:max_en}, but focused on GRB 190114C and GRB 221009A, LAT-detected bursts (not in 2FLGC) with VHE emission. For GRB 190114C, we consider that the RS evolves in a stellar-wind medium with the parameters: $\Gamma=600$, $E=10^{54}\,{\rm erg}$, $A_W=1$ and $z=0.42$, and for GRB 221009A, we use a constant-density medium with the parameters: $\Gamma=600$, $E=10^{55}\,{\rm erg}$, $n=1\,{\rm cm^{-3}}$ and $z=0.1$.}
 \label{fig_vhe:max_en}
\end{figure}

\begin{figure}
{ \centering
\resizebox*{\textwidth}{0.5\textheight}
{\includegraphics{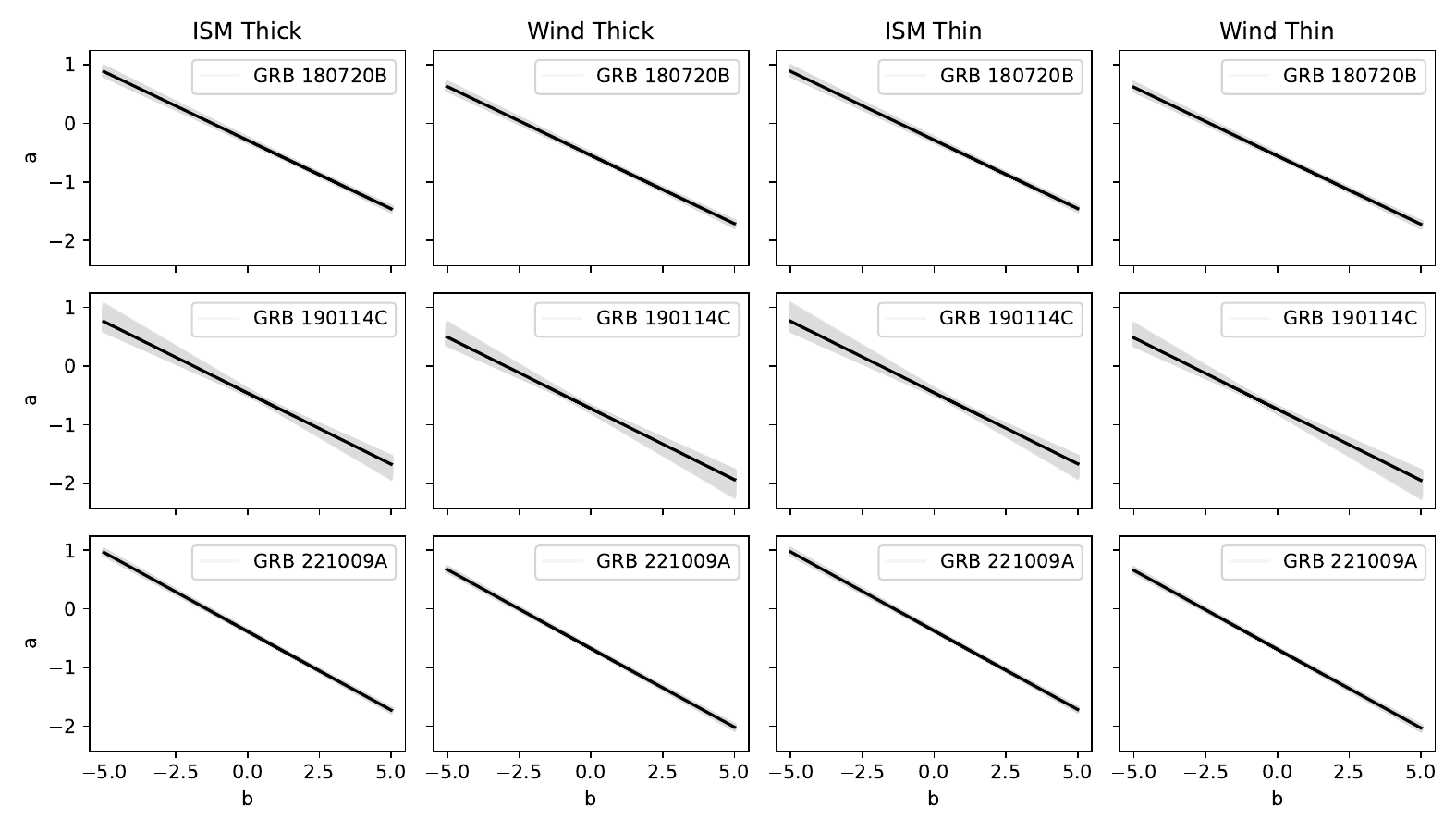}}
}
\caption{The same of Figure \ref{fig:cr_bur_1}, but panels from top to bottom are for GRBs 180720B, 190114C and 221009A.}
 \label{fig:cr_bur_5}
\end{figure}


\bsp	
\label{lastpage}
\end{document}